%% file: Main.tex
\documentclass{jfm}
\usepackage{graphicx}
\usepackage{epstopdf, epsfig}
\pdfoutput=1
\usepackage{lineno,hyperref}
\usepackage{amsmath}
\usepackage{nicefrac}
\usepackage{MnSymbol}
\usepackage{wasysym}
\usepackage{graphics}
\usepackage{graphicx}
\usepackage{float}
\usepackage{pgfplots}
\usepackage{upgreek}
\pgfplotsset{compat=newest}
\AppendGraphicsExtensions{.tif}
\usepackage{xcolor}
\usepackage{array}
\usepackage[utf8]{inputenc}
\usepackage{booktabs, makecell}
\usepackage{caption}

\usepackage[para]{threeparttable}
\usepackage{notoccite}
\usepackage{tikz}
\usepackage[T1]{fontenc}
\usetikzlibrary{backgrounds}
\usepackage{pgfplots}
\usepackage{grffile}
\usepgfplotslibrary{colormaps}
\pgfplotsset{compat=newest}
\usetikzlibrary{plotmarks,babel,quotes,angles,arrows.meta,backgrounds}
\usepgfplotslibrary{patchplots}
\usepackage{adjustbox}
\usepackage{pgfplotstable}
\usetikzlibrary{calc}
\usepackage{multirow}
\usepackage{siunitx}

\DeclareSIUnit\pixel{px} 

\newcommand{\G}{GW}
\newcommand{\Hexa}{GW/Hexa} 
\newcommand{\SO}{GW/SO} 
\newcommand{\EtOh}{GW/EtOH}

\newcommand{\dropdia}{$D_{d}$}
\newcommand{\dropvel}{$V_{d}$}
\newcommand{\filmthick}{$h_{f}$}
\newcommand{\filmres}{$h_{res}$}
\newcommand{\wft}{$\delta$}

\newcommand{\Rrim}{$R_{R}$}
\newcommand{\Hrim}{$H_{R}$}
\newcommand{\Hrimmax}{$H_{R,max}$}
\newcommand{\Rbase}{$R_{B}$}
\newcommand{\Rbasemax}{$R_{B,max}$}

\newcommand{\CS}{$\Sigma_{c}$}
\newcommand{\CSmax}{$\Sigma_{c,max}$}

\newcommand{\CSE}{$E_{\sigma,c}$}
\newcommand{\CSEmax}{$E_{\sigma,c,max}$}
\newcommand{\DSE}{$E_{\sigma,d,0}$}
\newcommand{\FSE}{$E_{\sigma,f,0}$}
\newcommand{\DKE}{$E_{k,d,0}$}
\def \spreadparam {$S$}

\newcommand{\FW}{(F.W.)}
\newcommand{\PW}{(P.W.)}

\newcommand{\tSmax}{$t_{\Sigma_{c,max}}$}
\newcommand{\tHmax}{$t_{H_{max}}$}
\newcommand{\tcapHmax}{$\tau_{cap,H_{max}}$}
\newcommand{\tcaptime}{$\tau_{cap}$}
\newcommand{\tcapHmaxmod}{$\tau_{cap,H_{max},m}$}

\newcommand{\tcapvisc}{$t_{cap,\nu}$}
\newcommand{\omegacap}{$\omega_{cap}$}
\newcommand{\dampratio}{$\eta$}

\newcommand{\TimeInert}{$\tau_{in}$}
\newcommand{\TimeHmax}{$\tau_{in,H_{max}}$}
\newcommand{\tcap}{$t_{cap}$}

\def \Re {$\mathit{Re}$}
\def \We {$\mathit{We}$}
\def \Oh {$\mathit{Oh}$}

\def \impactparam {$K$}

\def \Kavg {$K_{avg}$}

\def  \dropWe {$\mathit{We}_{d}$} 

\def \avgWed {$\mathit{We}_{avg}$} 

\def \dropRe {$\mathit{Re}_{d}$}
\def \filmRe {$\mathit{Re}_{f}$}
\def \avgRed {$\mathit{Re}_{avg}$}

\def \dropFr {$\mathit{Fr}_{d}$}

\def \rhod {$\rho_d$}
\def \rhof {$\rho_f$}
\def \rhoavg {$\rho_{avg}$}
\def \sigmad {$\sigma_d$}
\def \sigmaf {$\sigma_f$}
\def \sigmaavg {$\sigma_{avg}$}
\def \mud {$\mu_d$}

\def \nud {$\nu_d$}
\def \nuf {$\nu_f$}
\def \nuavg {$\nu_{avg}$}
\def \sigmadf {$\sigma_{d/f}$}
\newcommand{\sigmadfstar}{$\sigma_{d/f}^{*}$}

\newcommand{\md}{$M_d$}
\newcommand{\mf}{$M_f$}

\newcommand{\spreadparamstar}{$S^*$}

\shorttitle{How interfacial tension influences droplet impact on thin wall films}
\shortauthor{R. Bernard, D. Baumgartner, G. Brenn, C. Planchette, B. Weigand, G. Lamanna}

\title{Miscibility and wettability: how interfacial tension influences droplet impact onto thin wall films}

\author{Ronan Bernard\aff{1}
  \corresp{\email{ronan.bernard@itlr.uni-stuttgart.de}},
    David Baumgartner\aff{2},
  G\"unter Brenn\aff{2},
  Carole Planchette\aff{2},
  Bernhard Weigand\aff{1}
  \and Grazia Lamanna\aff{1}}

\affiliation{\aff{1}Institute of Aerospace Thermodynamics, University of Stuttgart, 70569 Stuttgart, Germany
\aff{2}Institute of Fluid Mechanics and Heat Transfer, Graz University of Technology, 8010 Graz, Austria}

\begin{document}

\maketitle
\begin{abstract} 
The influence of miscibility and liquid wettability during droplet impact onto thin wall films is investigated experimentally. Despite similar liquid properties and impact conditions, differences in the splashing limit, the crown extension and the duration of the ascending phase are observed. These differences are related to the interfacial tension of the droplet/wall-film liquid pairs, which is linked to their miscibility and wettability. More precisely, by calculating the crown surface energy, we show that the energy stored in the interface between droplet and wall-film (if any) is not negligible and leads to smaller crown extensions and the need of more kinetic energy to initiate splashing. Similarly, by calculating a modified capillary time taking into account all surface and interfacial tensions, we show that the interfacial tension acts as a non-negligible recoiling force, which reduces the duration of the ascending phase. The dynamics of this ascending phase is well captured for different wall-film thicknesses if accounting for the variations of the liquid masses in movement. Overall, droplet/wall-film interactions can be seen as inertio-capillary systems where the interfacial tension between droplet and wall film plays a significant role in the storage of energy and in the crown kinetics during the impact process. Besides, this analysis highlights that viscous losses have already a significant effect during the crown extension phase, by dissipating almost half of the initial energies for droplet impact onto thin wall films, and most likely by influencing the capillary time scale through damping.
\end{abstract}


\input{Intro}
	
\input{ExpCond_Intro}
	
\input{ExpObs_Intro}

\input{ExpObs_Splashing}
			
\input{ExpObs_Extension}
			
\input{ExpObs_Duration}

\input{IFT_Intro}

\input{IFT_nrj}
		
\input{IFT_spring}

\input{Conclusion}
		
\smallskip
	
The authors gratefully acknowledge the financial support by the Deutsche Forschungsgemeinschaft (DFG) within the international research training group GRK2160/1 \textit{Droplet Interaction Technologies} (DROPIT) in the frame of the Summer School 2018.	
\newline
Declaration of Interests. The authors report no conflict of interest.	
		
\bibliographystyle{jfm}
\bibliography{MyReferences}

\appendix

\section{Schematic of the experimental setup}\label{appExpSetup}
\begin{figure}
\centering
\input{Figures/Appendix/Exp_Setup}%
\caption{Schematic of the experimental setup allowing a two-perspective high-speed shadowgraphy of the impact area by an arrangement of lenses and mirrors. Reprinted from \cite{Bernard2020b} with permission from Elsevier Inc. (2020).}
\end{figure}

\section{Crown surface energies at $\text{\wft{}}\approx0.26$}\label{appA}
			\input{App_surf_nrj_delta2}

\end{document}

%% file: Intro.tex
\section{Introduction}\label{sec:Intro}

Droplet impact processes can be observed in many natural events and technical applications, such as rain, combustion chamber with lubricating film, spray coating, cosmetic and pharmaceutical production. In the last decade, the interest in binary systems (i.e. with different liquids for droplet and impacted liquid substrate) has increased. Besides their technical relevance, they offer the possibility to get a deeper understanding of the interaction between droplet and impacted liquid. By investigating this interaction, the question of the influence of miscibility and wettability between the droplet and the impacted liquid substrate arises. 

Few studies on miscibility or wettability can be found for various droplet impact configurations, but barely for droplet impact onto wall-films. For droplet impact onto deep pools, the effect of wettability (in terms of spreading parameter) has been studied numerically with water droplets impacting into an oil bath \citep{Wang2020}. The authors analysed the horizontal and vertical penetration into the oil cavity and found no significant effects of the wettability compared to those of the Weber number and the viscosity ratio for this bulky impact configuration where the droplet spreading is rather small and thus, does not favor interfacial forces in the process. Additional studies on deep pool considered different miscible and immiscible pairs with pure water, ethanol and silicone oil impacting onto water \citep{Hasegawa2019}. However, since the liquid properties were very different, the effects of miscibility alone could not be quantified. Only a qualitative observation of a Worthington jet during the recoiling phase could clearly be attributed to the difference in miscibility.
For droplet impact onto a continuous jet, the effects of wettability and miscibility were studied independently of the liquid properties by \cite{Baumgartner2019}. This study was carried out jointly with the present one in order to investigate the importance of the impact configuration in similar physical droplet impact processes with the same liquids. Droplet impact onto a continuous jet exhibits generally lower droplet spreading. Furthermore, the absence of a solid wall and the bulkier impact configuration leads to an impact kinetics fixed by the encapsulated drop only, and most of the effects of wettability and miscibility were observed in the recoiling phase or on the phenomenology of the impact (e.g. whether the droplet merges, is encapsulated or spreads around the jet).
In the case of droplet impact onto thin wall films, which is the focus of this work, the liquid structure formed is a lamella expanding from the impact point, the so-called \textit{crown}, as shown in the images of Fig.~\ref{fig:HighSpeed}.
Under certain impact conditions, the expanding crown can destabilize at the rim, leading to ejection of secondary droplets, referred to as \textit{crown-type splashing}, which needs particular attention in the above mentioned applications.
While the features of droplet impact onto thin wall films with similar liquids, miscible per definition, have been quite extensively studied (see e.g. the review article of \citet{Liang2016}), studies on binary droplet/wall-film systems are rarely found in the literature. 
Originally studied in the pioneering work of \citet{Worthington1897} with water and milk, most of the studies on binary systems are with miscible liquid pairs \citep{Geppert2017, Bernard2018, Geppert2019, Geppert2016, Bernard2017, Kittel2018, Thoroddsen2006} and focus on splashing limit, crown dynamic or on the viscosities.
In contrast, binary systems with exclusively immiscible droplet/wall-film systems have been investigated mainly with respect to the characteristics of the secondary droplets \citep{Shaikh2018} and to the repartition of the droplet and wall-film liquids in the crown depending on the viscosity ratio \citep{Kittel2018a}.

Only few studies report	directly or indirectly an effect of miscibility and wettability for droplet impacts onto wall-film.
\cite{Aljedaani2018} focused on the occurrence of holes in the crown, similar to some extend to those observed in the right-hand column of Fig.~\ref{fig:HighSpeed}. The holes formed on the crown wall grow till they join themselves and form a net-like structure which finally disintegrates into secondary droplets (see the video in the supplementary material of the impact process corresponding to the experiments in Fig.~\ref{fig:HighSpeed}).
This particular type of splashing event was first observed for water/glycerol droplet impacting onto very thin wall-films of ethanol \citep{Thoroddsen2006} and is a unique feature of binary droplet/wall-film systems. 
The hole formation was attributed to Marangoni driven flows initiated by small secondary droplets impacting onto the inside part of the crown wall \citep{Thoroddsen2006, Aljedaani2018, Kittel2019}. This phenomena was also observed with very similar surface tensions for droplet and wall-film (hyspin and hexadecane liquids) and was thus attributed more generally to inhomogeneities in viscosities and/or surface tensions in the crown wall \citep{Geppert2016}. Although \cite{Aljedaani2018} noticed only a weak reduction in the growth rate of the holes for increasing droplet viscosity, a large viscosity ratio between droplet and wall-film influences the liquid repartition in the crown. For droplets with much higher viscosity than the wall-film, a two-stage crown was observed with the wall-film liquid at the upper part of the crown. Similar crown morphologies with high viscosity ratios have been observed for immiscible droplet/wall-film systems by \cite{Kittel2018a}, and can be explained by the strong differences in the time scale of the ejecta sheets coming from the droplet and the wall-film \citep{Marcotte2019}. By influencing the liquid repartition inside the crown, the viscosity ratio also influences the occurrence of the holes in the crown wall. In the present study instead, the droplet and wall-film viscosities are quite similar. Hence, only one stage crowns are observed, where the droplet is expected to cover uniformly the wall-film liquid in the crown wall.
Besides the miscible configurations leading to the formation of holes, \cite{Aljedaani2018} studied the impact of immiscible liquid pairs. They did not report any major differences in the crown dynamics (looking especially at the crown angle), but the hole formation vanishes despite differences in surface tension between droplet and wall-film. They observed that the patch formed by the immiscible small droplets sitting on the crown wall do not grow. Holes in the crown were indeed observed only for miscible liquid pairs till now in the literature \citep{Geppert2019, Kittel2019, Thoroddsen2006}, although \cite{Aljedaani2018} suggested that the hole formation could be influenced by liquid wettability of immiscible pairs.
The hole formation process is not investigated in the present study, but it is encountered while studying miscibility and wettability effects with a miscible droplet/wall-film system involving glycerol, water and ethanol.

The wetting behaviour between droplet and wall film was studied by \cite{Che2018} with combinations of glycerol/water droplet impacting onto silicone oil and vice versa.
While a Worthington jet was formed at the end of the recoiling phase for the water droplet onto the oil, it was not the case for the reversed combination. The shape of the crowns were also slightly different because of the different droplet spreading speed in the wall-film liquid. The authors state that immiscible combinations share similar features with that on miscible films without drastic differences in the impact process, which is a qualitative observation though.

Only \citet{Chen2017} studied explicitly the effect of miscibility on crown formation and splashing for droplet impact onto \textit{very} thin films (maximum film height of \SI{50}{\micro\meter}, i.e. 0.017 times the droplet diameter).
They reported a significant role of the interfacial tension (assuming a value of zero for miscible pair, Sec.~\ref{sec:ExpCond}) in the receding phase, as for \cite{Che2018} for the wettability. They did not mention any influence for the extension phase, which is crucial for the ejection of secondary droplets, but did observe a small shift in the splashing limit.
They reported that, in case of miscible liquids, a larger Weber number (based on droplet properties) and film thickness are required to form a crown and to eject secondary droplets, however without distinction between crown-type splashing and \textit{prompt splash} (i.e. formed as the droplet hits the surface, within the first \SI{100}{\micro\second} \citep{Thoroddsen2011}). They explained this phenomena by the attenuation of the kinematic discontinuity due to the absence of interface between droplet and wall-film. 
To the contrary, \cite{Banks2013} noted that the immiscible liquid pair FC-72 on water required a far higher Weber number than the other miscible pairs studied for crown formation, despite having lower viscosity and surface tension, which should rather promote splashing. They could not find a clear explanation for this, attributing the observations to either the miscibility effects or some indirect effects of viscosities on the impact morphology. The effects of the interfacial tension found on the onset of splashing in the present study could explain the observation of \cite{Banks2013}.
	
The present review highlights that the question of the influence of miscibility and liquid wettability for droplet impact onto wall films remains open. 
To what extent can miscibility and wettability influence droplet impact onto thin wall films, especially during the extension phase? 

Since droplet impact onto wall-films leads to an extreme spreading of liquids, for example from 2 to 5 times the droplet diameter for the rim displacement \citep{Bernard2020b}, the interface between droplet and wall-film liquids increases drastically during the extension process. Thus, the interfacial tension acting on this interface might have a significant influence on the crown dynamics. Does the interfacial tension, linked to miscibility and wettability, influence the impact process, and if yes, how? 
To address these questions, three droplet/wall-film liquid pairs have been chosen to vary the miscibility and wettability behaviour, as presented in Sec.~\ref{sec:ExpCond}, leading to a variation of their interfacial tension. The differences observed between these three droplet/wall-film pairs in terms of impact outcome, crown morphology and crown kinetics are reported in Sec.~\ref{sec:ExpObs}. From these observations, the role of the interfacial tension during crown extension is discussed in Sec.~\ref{sec:IFT}. The conclusions are summarized in Sec.~\ref{sec:Conclusion}.

\begin{figure}
\centerline{%
\input{Figures/Fig1_CrownParameter_All.tex}
}
  \caption{High-speed images of droplet impact onto wall films for three droplet/wall-film liquid pairs at similar impact conditions (see the last row of Table \ref{tab:c3s1_exprangeavg} for $\text{\dropWe}=363\pm5$, and $\text{\wft{}} = \text{\filmthick{}}/\text{\dropdia{}} = 0.122\pm0.011$) at different time instants from the impact $\text{\TimeInert{}}=t \text{\dropvel{}}/\text{\dropdia{}}$. Only the relative importance of the interfacial tension \sigmadfstar{}=\sigmadf{}/(\sigmad{}+\sigmaf{}+\sigmadf{}) varies significantly between the liquid pairs, linked to a variation of wettability and miscibility.}
\label{fig:HighSpeed}
\end{figure}

%% file: Figures/Fig1_CrownParameter_All.tex
\newcommand{\SP}{0.3\linewidth} 
\newcommand{\SPL}{0.48\textwidth}
\newcommand{\BotHexa}{5}
\newcommand{\BotSO}{5}
\newcommand{\BotEtOh}{5}

\begin{minipage}{0.13\linewidth}
\vspace{\SP}
$\text{\TimeInert{}}=-0.53$\\
\vspace{\SPL}\\
$\text{\TimeInert{}}=1.28$\\
\vspace{\SPL}\\
$\text{\TimeInert{}}=2.25$\\
\vspace{\SPL}\\
$\text{\TimeInert{}}=4.50$\\
\vspace{\SPL}\\
$\text{\TimeInert{}}=6.75$\\
\vspace{\SPL}\\
$\text{\TimeInert{}}=9.00$\\
\end{minipage}
\begin{minipage}{\SP}
\centering
\begin{tikzpicture}
\node[anchor=south west,inner sep=0] (image) at (0,0) {\includegraphics[width=0.9\linewidth]{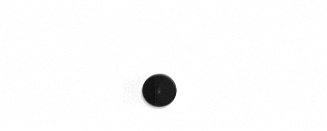}};    
\begin{scope}[x={(image.south east)},y={(image.north west)}]          
   \node[anchor=mid, align = center] at (0.5,1.3) {\Hexa{}\\ $\text{\sigmadfstar{}} \approx{} 0.35$};    
     \path[draw=black, line width=0.051pt, anchor=north east] (0,0) rectangle (1,1);
     \draw[](0,\BotHexa{}*0.01)--(1,\BotHexa{}*0.01); 
     \draw[->] (0.08,0.9)--(0.08,0.6) node[midway,right] {$g$};;    
		 \draw[thick,|-|] (0.75,0.7)--(0.95657,0.7) node[midway,sloped,above] {$4$ mm};;
\end{scope}  
\end{tikzpicture}
\begin{tikzpicture}
\node[anchor=south west,inner sep=0] (image) at (0,0) {\includegraphics[width=0.9\linewidth]{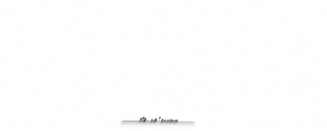}};    
    \begin{scope}[x={(image.south east)},y={(image.north west)}]  
     \path[draw=black, line width=0.051pt, anchor=north east] (0,0) rectangle (1,1);
        \draw[](0,\BotHexa{}*0.01)--(1,\BotHexa{}*0.01); 
     \end{scope}  
\end{tikzpicture}
\begin{tikzpicture}
\node[anchor=south west,inner sep=0] (image) at (0,0) {\includegraphics[width=0.9\linewidth]{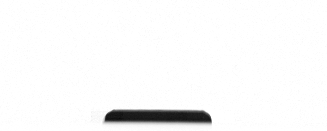}};    
    \begin{scope}[x={(image.south east)},y={(image.north west)}]  
     \path[draw=black, line width=0.051pt, anchor=north east] (0,0) rectangle (1,1);      
        \draw[](0,\BotHexa{}*0.01)--(1,\BotHexa{}*0.01);
     \end{scope}  
\end{tikzpicture}
\begin{tikzpicture}
\node[anchor=south west,inner sep=0] (image) at (0,0) {\includegraphics[width=0.9\linewidth]{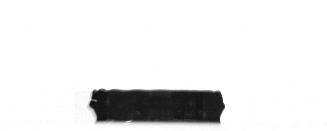}};    
    \begin{scope}[x={(image.south east)},y={(image.north west)}]  
     \path[draw=black, line width=0.051pt, anchor=north east] (0,0) rectangle (1,1);   
		
        \draw[](0,\BotHexa{}*0.01)--(1,\BotHexa{}*0.01);
     \end{scope}  
\end{tikzpicture}
\begin{tikzpicture}
\node[anchor=south west,inner sep=0] (image) at (0,0) {\includegraphics[width=0.9\linewidth]{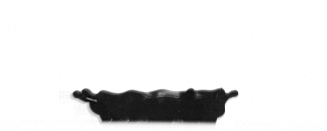}};    
    \begin{scope}[x={(image.south east)},y={(image.north west)}] 
     \path[draw=black, line width=0.051pt, anchor=north east] (0,0) rectangle (1,1);    
        \draw[](0,\BotHexa{}*0.01)--(1,\BotHexa{}*0.01);
     \end{scope}  
\end{tikzpicture}
\begin{tikzpicture}
\node[anchor=south west,inner sep=0] (image) at (0,0) {\includegraphics[width=0.9\linewidth]{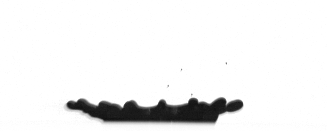}};    
    \begin{scope}[x={(image.south east)},y={(image.north west)}]   
     \path[draw=black, line width=0.051pt, anchor=north east] (0,0) rectangle (1,1); 
        \draw[](0,\BotHexa{}*0.01)--(1,\BotHexa{}*0.01);
     \end{scope}  
\end{tikzpicture}
\end{minipage}%
\hspace{-0.03\linewidth}
\begin{minipage}{\SP}
\centering
\begin{tikzpicture}
\node[anchor=south west,inner sep=0] (image) at (0,0) {\includegraphics[width=0.9\linewidth]{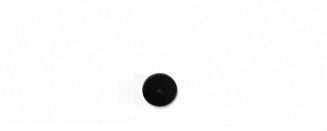}};    
    \begin{scope}[x={(image.south east)},y={(image.north west)}]    
    
       \node[anchor=mid, align = center] at (0.5,1.3) {\SO{} \\ $\text{\sigmadfstar{}}\approx{} 0.28$};  
     \path[draw=black, line width=0.051pt, anchor=north east] (0,0) rectangle (1,1); 
        \draw[](0,\BotSO{}*0.01)--(1,\BotSO{}*0.01); 
     \end{scope}  
\end{tikzpicture}
\begin{tikzpicture}
\node[anchor=south west,inner sep=0] (image) at (0,0) {\includegraphics[width=0.9\linewidth]{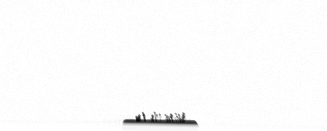}};    
    \begin{scope}[x={(image.south east)},y={(image.north west)}] 
     \path[draw=black, line width=0.051pt, anchor=north east] (0,0) rectangle (1,1); 
        \draw[](0,\BotSO{}*0.01)--(1,\BotSO{}*0.01); 
     \end{scope}  
\end{tikzpicture}

\begin{tikzpicture}
\node[anchor=south west,inner sep=0] (image) at (0,0) {\includegraphics[width=0.9\linewidth]{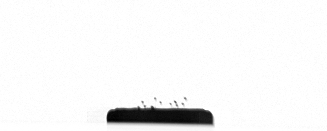}};    
    \begin{scope}[x={(image.south east)},y={(image.north west)}] 
     \path[draw=black, line width=0.051pt, anchor=north east] (0,0) rectangle (1,1); 
        \draw[](0,\BotSO{}*0.01)--(1,\BotSO{}*0.01); 
     \end{scope}  
\end{tikzpicture}
\begin{tikzpicture}
\node[anchor=south west,inner sep=0] (image) at (0,0) {\includegraphics[width=0.9\linewidth]{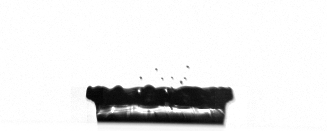}};    
    \begin{scope}[x={(image.south east)},y={(image.north west)}]     
     \path[draw=black, line width=0.051pt, anchor=north east]
  (0,0) rectangle (1,1);
        \draw[](0,\BotSO{}*0.01)--(1,\BotSO{}*0.01); 
     \end{scope}  
\end{tikzpicture}
\begin{tikzpicture}
\node[anchor=south west,inner sep=0] (image) at (0,0) {\includegraphics[width=0.9\linewidth]{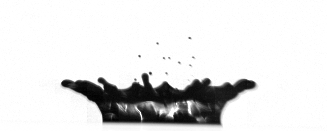}};    
    \begin{scope}[x={(image.south east)},y={(image.north west)}] 
     \path[draw=black, line width=0.051pt, anchor=north east]
  (0,0) rectangle (1,1);   
        \draw[](0,\BotSO{}*0.01)--(1,\BotSO{}*0.01);
     \end{scope}  
\end{tikzpicture}
\begin{tikzpicture}
\node[anchor=south west,inner sep=0] (image) at (0,0) {\includegraphics[width=0.9\linewidth]{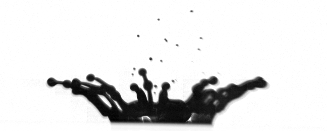}};    
    \begin{scope}[x={(image.south east)},y={(image.north west)}]
     \path[draw=black, line width=0.051pt, anchor=north east]
  (0,0)  rectangle (1,1);   
        \draw[](0,\BotSO{}*0.01)--(1,\BotSO{}*0.01); 
     \end{scope}  
\end{tikzpicture}
\end{minipage}
\hspace{-0.04\linewidth}
\begin{minipage}{\SP}
\centering
\begin{tikzpicture}
\node[anchor=south west,inner sep=0] (image) at (0,0) {\includegraphics[width=0.9\linewidth]{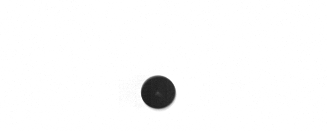}};    
    \begin{scope}[x={(image.south east)},y={(image.north west)}]
           \node[anchor=mid, align = center] at (0.5,1.3) {\EtOh{} \\ $\text{\sigmadfstar{}} \approx{} 0$};  
     \path[draw=black, line width=0.051pt, anchor=north east] (0,0) rectangle (1,1);
        \draw[](0,\BotEtOh{}*0.01)--(1,\BotEtOh{}*0.01);
     \end{scope}  
\end{tikzpicture}

\begin{tikzpicture}
\node[anchor=south west,inner sep=0] (image) at (0,0) {\includegraphics[width=0.9\linewidth]{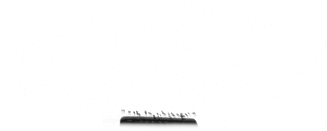}};    
    \begin{scope}[x={(image.south east)},y={(image.north west)}] 
     \path[draw=black, line width=0.051pt, anchor=north east] (0,0) rectangle (1,1);   
        \draw[](0,\BotEtOh{}*0.01)--(1,\BotEtOh{}*0.01); 
     \end{scope}  
\end{tikzpicture}
\begin{tikzpicture}
\node[anchor=south west,inner sep=0] (image) at (0,0) {\includegraphics[width=0.9\linewidth]{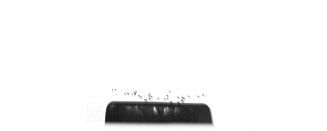}};    
    \begin{scope}[x={(image.south east)},y={(image.north west)}] 
     \path[draw=black, line width=0.051pt, anchor=north east]
  (0,0)
    rectangle (1,1);      
        \draw[](0,\BotEtOh{}*0.01)--(1,\BotEtOh{}*0.01);
     \end{scope}  
\end{tikzpicture}
\begin{tikzpicture}
\node[anchor=south west,inner sep=0] (image) at (0,0) {\includegraphics[width=0.9\linewidth]{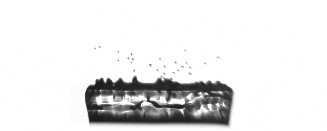}};    
    \begin{scope}[x={(image.south east)},y={(image.north west)}] 
     \path[draw=black, line width=0.051pt, anchor=north east] (0,0) rectangle (1,1);       
        \draw[](0,\BotEtOh{}*0.01)--(1,\BotEtOh{}*0.01);
     \end{scope}  
\end{tikzpicture}
\begin{tikzpicture}
\node[anchor=south west,inner sep=0] (image) at (0,0) {\includegraphics[width=0.9\linewidth]{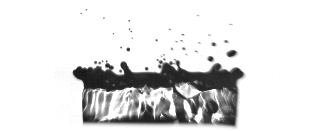}};    
    \begin{scope}[x={(image.south east)},y={(image.north west)}] 
     \path[draw=black, line width=0.051pt, anchor=north east]
  (0,0)
    rectangle (1,1);    
        \draw[](0,\BotEtOh{}*0.01)--(1,\BotEtOh{}*0.01);
     \end{scope}  
\end{tikzpicture}
\begin{tikzpicture}
\node[anchor=south west,inner sep=0] (image) at (0,0) {\includegraphics[width=0.9\linewidth]{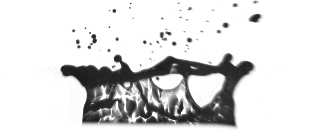}};    
    \begin{scope}[x={(image.south east)},y={(image.north west)}] 
     \path[draw=black, line width=0.051pt, anchor=north east] (0,0) rectangle (1,1);      
        \draw[](0,\BotEtOh{}*0.01)--(1,\BotEtOh{}*0.01);
     \end{scope}  
\end{tikzpicture}
\end{minipage}

%% file: ExpCond_Intro.tex
\section{Experimental setup and conditions}\label{sec:ExpCond}

Three different droplet/wall-film liquid pairs listed in Table \ref{tab:prop} are used to vary miscibility and liquid wettability. All the binary droplet/wall-film systems are similar, only their respective interfacial tension \sigmadf{} is changing significantly so that it corresponds to immiscible liquids with partial and total wetting conditions as well as to miscible liquids. The experiments are recorded by high-speed shadowgraphy.

\subsection{Experimental setup} \label{expsetup}

The experimental setup is composed of a dropper, an impact area, and a high-speed imaging system. A schematic is provided in Appendix \ref{appExpSetup}.
The dropper consists of a needle (Gauge 21), connected through flexible tubes (PTFE) to a liquid reservoir. The liquid is set in motion by a peristaltic pump with a flow rate
kept low enough to let a single droplet fall by gravity. The height from the wall-film surface to the needle tip is varied to obtain different droplet impact velocities \dropvel{}.
The impact area is a shallow pool in which the liquid is progressively added with a micro-pipette, bounded by a metallic ring of \SI{0.6}{\milli\meter} height and \SI{60}{\milli\meter} diameter fixed on a sapphire plate. Sapphire is used for its high refractive index compared to the liquids used. This allows a reliable use of a confocal chromatic sensor (Micro-Epsilon, IFS2405-3) to measure the film thickness \filmthick{} with an accuracy of about 1\% for the values investigated in this work (principle described in \citet{Lel2008}).

The imaging system is designed for high-speed shadowgraphy, recording simultaneously two orthogonal perspectives to track asymmetric features such as holes in the crown. Typical high-speed images taken with a high-speed digital video camera (Photron FASTCAM SA1.1) and LED lights operated in backlit mode can be seen in Fig.~\ref{fig:HighSpeed} (only one perspective is shown). 
The spatial and temporal resolutions of the high-speed imaging were \SI{80}{\micro\meter\per\pixel} and \SI{0.05}{\milli\second}, respectively combined with an exposure time of 1/92 ms.

The high-speed images are post-processed with an in-house MATLAB routine that extracts the primary geometrical parameters of the crown (rim radius \Rrim{}, base radius \Rbase{} and height \Hrim{}).
The high-speed images are first processed by subtracting the background image (first image before the droplet enters the frame).
This step, combined with an increase of contrast (grey-scale spread over the full dynamic range), improves significantly the observation of the liquid structures. Thus, the processed high-speed images, as shown in Fig.~\ref{fig:HighSpeed} are preferred to the raw images (as in the Appendix \ref{appA} for example) to illustrate the impact process. Finally, the images are binarized in order to detect the four edges of the crown like the four orange dots in Fig.~\ref{fig:CrownSchematic}. The distance between the detected crown edges enables the evaluation of the crown geometrical parameters: crown rim radius \Rrim{} and crown base radius \Rbase{} for the horizontal distances, and the crown rim height \Hrim{} in the vertical.
To detect these edges, the image is analysed pixel line by pixel line. At the bottom, a search for the maximum radius is carried out within the last five lines belonging to the crown.
The detection of the crown rim is more difficult, especially when the rim has corrugations or fingers.
For each pixel line at the crown top, the presence of fingers (alternation of black and white pixels) is checked for. The height of the crown \Hrim{} is set at the finger base, meaning that the corrugations due to the fingers are not taken into account.
The values of the geometrical parameters are averaged between the two perspectives of the experimental setup to reduce the measurement uncertainty. The typical errors associated with the measurement of the crown geometrical parameters are between 1.5\% and 4.3\% for the radii, and between 1.6 and 5.2\% for the crown height \citep{Geppert2019}. The droplet diameter \dropdia{} and impact velocity \dropvel{} are also measured from the high-speed images shortly before impact ($\sim$ \SI{1.5}{\milli\second}). By adding up the systematic and random uncertainties, which maximises the error, the overall uncertainty for the droplet diameter remains smaller than 2.8\%, and below 3\% for the impact velocity \citep{Geppert2019}.
A detailed description of the setup (which has been used with the same operating conditions), the post-processing routine and the associated measurement errors can be found in \cite{Geppert2016, Geppert2017, Geppert2019}.


\subsection{Variation of droplet/wall-film liquids}

The liquids of the droplet/wall-film pairs have been chosen to vary the wettability and miscibility behaviours, keeping other liquid properties almost unchanged. The liquid properties are summarized in Table \ref{tab:prop}, together with the abbreviations used for each liquid. 

The droplet liquid is systematically composed of a mixture of glycerol and water (abbreviation \G{}), which is coloured with Indigotin 85 (E 132, BASF, Germany) similar to the liquids used by \cite{Baumgartner2019}.
The only parameter varying significantly is the interfacial tension with the wall-film liquids $\sigma_{d/f}$ (see \cite{Baumgartner2019} for more details on their determination), the subscript $d$ standing for droplet and $f$ for wall-film. The interfacial tension is associated with the miscibility behaviour ($\sigma_{d/f}=0$ for miscible liquids) and the wettability behaviour through the spreading parameter $S$. The spreading parameter is defined as $\text{\spreadparam{}}=\text{\sigmad}-\text{\sigmaf}-\text{\sigmadf}$, which takes negative values for partial wetting \PW{} and positive values for full wetting \FW{} \citep{DeGennes2013}). 
For the wettability, n-hexadecane (abbreviation Hexa, third column in Table \ref{tab:prop}) provides partial wetting with aqueous solutions as many alkanes. This is confirmed by the estimation of \spreadparam{} for \Hexa{} which is negative with approximately \SI{-9}{\milli\newton\per\meter}, and the observation of n-hexadecane lenses on top of \G{} \citep{Baumgartner2019}.
To the contrary, silicone oil (abbreviation S.O., fourth column in Table \ref{tab:prop}) provides full wetting with aqueous solutions \citep{Ross1992, DeGennes2013}. The spreading parameter \spreadparam{} is positive and equal to approximately 15. The full wetting behaviour is confirmed by the observation of an oil drop spreading on the surface of a liquid bath of \G{} \citep{Baumgartner2019}.

The last wall-film liquid is a mixture of ethanol, water and glycerol (abbreviation EtOH, fifth column in Table \ref{tab:prop}), which has been tuned to approach the two other wall-film liquid properties. The mixture of ethanol is miscible with the droplet liquid \G{}. In the case of miscible liquids, the interfacial tension is assumed to be zero because no interface, in a strict sense, exists between the liquids. Note that for miscible liquids with composition gradients, \sigmadf{} may transiently differ from zero before diffusion takes place \cite{truzzolillo2017}, which may be the case for droplet impact processes of a few milliseconds. Nevertheless, the measurements of interfacial tensions for water and glycerol mixtures show values of \SI{1}{\milli\newton\per\meter} or lower \citep{Petitjeans1996}, and around \SI{2}{\milli\newton\per\meter} for water and alcohols like butanol \citep{Enders2008}. These values remain within the experimental uncertainties, thus, the interfacial tension can be approximated to be zero for \EtOh{}.
\begin{table}
  \begin{center}
  \begin{tabular}{l | c | c c c }
	\hline
       		 & Droplet &   \multicolumn{3}{c}{Wall-film}     \\
       		 \hline
       Abbreviation   		 & \G{}   & Hexa & S.O. & EtOH \\
       Composition   		 & G50 W50     & Hexa100 & SO100 & EtOH55 G30 W15 \\
       $\rho \, \left(\si{\kilogram\per\cubic\meter}\right)$ & $1116\pm2$ & $767\pm10$ & $908\pm5$ & $936\pm10$\\
       $\mu \, \left(\si{\milli\pascal\second}\right)$& $4.97\pm0.10$ & $3.50\pm 0.30$ & $5.10\pm0.05$ & $4.58\pm0.40$\\
			 $\nu \, \left(\si{\milli\meter\squared\per\second}\right)$ & $4.45\pm0.10$ & $4.57\pm 0.45$ & $5.62\pm0.09$ & $4.90\pm0.48$\\
       $\sigma \left(\si{\milli\newton\per\meter}\right)$ & $68\pm2$  & $26.5\pm1$ & $19.5\pm0.5$ & $25.7 \pm 0.7$\\
       $\text{\sigmadf{}} \left(\si{\milli\newton\per\meter}\right)$ & / & $50\pm2$ & $34\pm1$ & 0\\
       $\text{\spreadparam{}} \left(\si{\milli\newton\per\meter}\right)$ & / & $\approx-9$ \PW{} & $\approx15$ \FW{} & miscible\\
			$\text{\sigmadfstar{}} (-)$  & / & $\approx0.35$  & $\approx0.28$ & $\approx{}0$\\
			\hline
  \end{tabular}
\caption{Liquid properties of droplet and wall-film liquids. In the line \textit{Composition}, the numbers indicate the mass percentage of G: glycerol,  W: water, Hexa: n-hexadecane, S.O.: silicone oil, EtOH: ethanol. The spreading parameter \spreadparam{} is associated with the wettability behaviour: partial wetting \PW{} for $\text{\spreadparam{}}<0$ or full wetting \FW{} for $\text{\spreadparam{}}>0$. The non-dimensionalised interfacial tension \sigmadfstar{} quantifies the importance of \sigmadf{} in the droplet/wall-film system.}
  \label{tab:prop}
  \end{center}
\end{table}
In order to quantify the importance of the interfacial tension in the droplet/wall-film system, the non-dimensionalised interfacial tension \sigmadfstar{}=\sigmadf{}/(\sigmad{}+\sigmaf{}+\sigmadf{}) is introduced. It corresponds to the ratio of the interfacial tension to the sum of the surface tensions of droplet and wall film. Thus, it is zero for miscible liquids as \EtOh{}, and increases with increasing importance of the interfacial tension in the droplet/wall-film system as for \SO{} and \Hexa{} in the last row of Table \ref{tab:prop}. 
Note that in some cases, it could be interesting to have a unified parameter combining the importance of the interfacial tension with the spreading parameter, e.g. as \spreadparamstar{}=\sigmadf{}/\spreadparam{} in the case where the spreading parameter is different from zero. This parameter could help to understand the kinetics of the receding where the wettability has a strong influence for example in the formation of a Worthington jet \citep{Che2018}, whose magnitude could be related to \spreadparamstar{}. In the present study, we rather focus on the crown extension phase where the value of the interfacial tension seems to play a significant role independently on the wetting behaviour of the liquid pair, probably because of the large Weber numbers involved (see Sec.~\ref{exprange}).

\subsection{Experimental range} \label{exprange}

The single droplets of the glycerol/water mixture exhibit diameters of \dropdia{} = \SI[parse-numbers=false]{2.20\pm0.07}{\milli\meter} for the full database. The impact velocity is varied from \dropvel{}=\SIrange{2.5}{3.7}{\meter\per\second} with four different fall heights. For each impact velocity investigated, all three liquid pairs are used as wall-film. The dimensionless wall-film thickness \wft{}=\filmthick{}/\dropdia{} is kept quasi-constant at two investigated ranges, i.e. at \text{\wft}=\SI[parse-numbers=false]{0.122\pm0.011}, and at \text{\wft}=\SI[parse-numbers=false]{0.259\pm0.008} .
In order to describe accurately the impact process, the initial state of the droplet/wall-film system needs to be quantified.
Therefore, the range of Weber and Reynolds numbers based on droplet properties, as $\text{\dropWe{}}=\text{\rhod} \text{\dropvel{}}^2 \text{\dropdia{}}/\text{\sigmad}$ and $\text{\dropRe}=\rho\text{\dropvel{}}\text{\dropdia{}}/\text{\mud}$ respectively, are given in the two first columns of Table \ref{tab:c3s1_exprangeavg}.
Note that the range indicated entails the experimental conditions of all three liquids pairs of \Hexa{}, \SO{} and \EtOh{}.
However, these numbers are not representative of the full binary droplet/wall-film systems since the liquid properties of droplet and wall-film differ. 
A proper way of taking both droplet and wall-film liquid properties into account in these non-dimensional numbers is still under discussion in literature \citep{geppert2014, Kittel2017}. Recently, averaged liquid properties (subscript \textit{avg}) as $\text{\rhoavg{}}=\left(\text{\rhod{}+\rhof{}}\right)/2$, $\text{\sigmaavg{}}=\left(\text{\sigmad{}+\sigmaf{}}\right)/2$ and $\text{\nuavg{}}=\sqrt{\text{\nud\nuf}}$ have shown good ability to capture the crown rim dynamic for binary droplet/wall-film systems \citep{Bernard2020b}. Note that this approach was developed for miscible droplet/wall-film liquid pairs made of silicone oils. Hence, it does not consider wettability and miscibility effects which are considered separately in the present study (e.g. in terms of spreading parameter \spreadparam{} and dimensionless interfacial tension \sigmadfstar{} summarized in Table \ref{tab:prop}).
These averaged properties can be used in the dimensionless numbers to form \avgWed{} and \avgRed{} summarized in Table \ref{tab:c3s1_exprangeavg} for each reference value of \dropWe{} and all liquid pairs combined. 
Furthermore, they can be combined in a single impact parameter $\text{\impactparam{}}=\text{\We}^{0.5}\text{\Re}^{0.25}$ (also summarized in the last column of Table \ref{tab:c3s1_exprangeavg}), which, considered together with \wft{}, is fully representative of the droplet wall-film system for the onset of splashing \citep{Cossali1997}. 
These non-dimensional impact parameters between the three liquid pairs indicate very similar droplet/wall-film systems for each set of experiments.
The variation does not exceed $\pm8\%$ in general at a given \dropWe{}, while the non-dimensionalised interfacial tension \sigmadfstar{} increases by more than 25\% between \SO{} and \Hexa{}, and is zero for \EtOh{}.
Hence, the major varying parameter between the liquid pairs investigated is \sigmadfstar{}, which, combined to the surface tensions of droplet and wall-films defined the miscibility and wettability of the liquid pairs.
   
\begin{table}
\begin{center}
\begin{tabular}{c | l l | l l l }
\hline
\wft{} & \dropWe{} & \dropRe{} & \avgWed{}  & \avgRed{}   &  \Kavg{} \\
\hline
\multirow{4}{*}{$0.122\pm0.011$} & $218\pm2$ & $1212\pm10$ & $286\pm18$ & $1152\pm76$ & $98\pm2$ \\
 & $269\pm2$ & $1341\pm8$  & $354\pm30$ & $1274\pm66$ & $112\pm3$ \\
 & $318\pm3$ & $1448\pm10$ & $417\pm35$ & $1375\pm71$ & $124\pm4$ \\
 & $363\pm5$ & $1547\pm16$ & $477\pm42$ & $1469\pm70$ & $135\pm4$ \\
\hline
\multirow{4}{*}{$0.259\pm0.008$} & $217\pm1$ & $1205\pm2$ & $284\pm21$ & $1144\pm65$ & $98\pm2$ \\
 & $268\pm2$ & $1336\pm7$  & $353\pm28$ & $1272\pm70$ & $112\pm3$ \\
 & $320\pm2$ & $1453\pm7$ & $420\pm29$ & $1383\pm85$ & $125\pm2$ \\
 & $363\pm7$ & $1546\pm24$ & $481\pm40$ & $1482\pm76$ & $136\pm4$ \\
\end{tabular}
\caption{The investigation range of typical non-dimensional parameters. The subscript \textit{avg} refers to averaged liquid properties between droplet and wall-film.}
\label{tab:c3s1_exprangeavg}
\end{center}
\end{table}

%% file: ExpObs_Intro.tex
\section{Experimental observations}\label{sec:ExpObs}

The temporal evolutions of the three liquid pairs are shown in Fig.~\ref{fig:HighSpeed}. Despite similar impact conditions (see Table \ref{tab:c3s1_exprangeavg}, row $\text{\wft{}}= 0.130\pm0.001$ with $\text{\dropWe}=363\pm5$), differences between the liquid pairs can be observed in the impact outcome, the crown extension and the crown kinetics.

%% file: ExpObs_Splashing.tex
\subsection{Shift in the splashing limit}\label{sec:ShiftSplashing}

The first differences that can be observed qualitatively between the three droplet/wall-film pairs in Fig.~\ref{fig:HighSpeed} are their impact outcomes. The full temporal evolution of the impact process for each liquid pair side-by-side can be observed in the video of the supplementary material. 
For \Hexa{}, the crown rim stays relatively stable, undulations become significant only in the receding phase (the maximum crown height is reached at about \TimeHmax{} = \tHmax{} \dropvel{}/\dropdia{} = 5.40) and fingers can barely be observed at the end of the impact process. Thus, the outcome of \Hexa{} can be categorised as a \textit{transition} case (i.e. fingers formed) close to \textit{deposition} (i.e. no fingers). 
In the case of \SO{}, undulations on the rim can already be observed in the ascending phase (up to \TimeHmax{} = 6.53), and at the end of the extension process, long fingers with droplets about to detach are formed. The long fingers only disintegrate into secondary droplets at the end of the impact process, leaving the droplets formed in the vicinity of the impact location. Thus, the outcome of \SO{} can again be categorised as a \textit{transition} case, but close to \textit{splashing} (i.e. secondary droplets are ejected far beyond the crown dimensions). Note that the tiny droplets observed already on the picture at \TimeInert{} = 2.25 are associated with \textit{prompt splash}, i.e. they are not considered as part of the crown-type outcome. 
The prompt splashing consists in the fragmentation of a thin and fast liquid lamella called \textit{ejecta sheet} \citep{Thoroddsen2002} formed at the base of the impacting droplet. The distinction between prompt and crown-type splashing relies first on their different time scales, since prompt splashing occurs quasi-immediately after the droplet has impacted the liquid surface (within the first \SI{100}{\micro\second}, \cite{Liang2016}). Second, the ejecta sheet and the crown wall arise from different dynamics and can be observed dinstinctly \citep{Zhang2012}. Third, prompt and crown-type splashing can be observed independently of each other in a splashing regime map \citep{Deegan2007}. Last, prompt splashing leads to the formation of much smaller secondary droplets than crown-type splashing \citep{Cossali1997, Motzkus2009}. Hence, there is most likely no or small interdependence between the prompt splashing observed and the crown-type splashing studied in the present experiments. In the video of the impact experiments similar to those of Fig.~\ref{fig:HighSpeed} provided in the supplementary material, one can see that prompt splashing occurs in all three cases as the droplets penetrate the wall-film, although the tiny droplets produced are ejected beyond the crown size and thus observable later only for \SO{} and \EtOh{}. Furthermore, the video shows clearly that the crown wall arises far  beyond the location where these tiny droplets are form, in the vicinity of the droplet/wall-film neck region.
For the last liquid pair \EtOh{}, fingers are already formed in the ascending phase (e.g. at \TimeInert{} = 4.50), and some droplets are ejected from the fingers.
Towards the end of the crown extension, holes are formed on the crown wall, as already reported in literature for droplet impacts with aqueous and ethanol solutions \citep{Thoroddsen2006, Aljedaani2018}. As explained in the introduction, mixture gradients in the crown wall and/or fine droplets that cannot be observed with the current setup impacting on the crown wall could cause these holes.
Despite the formation of holes in the crown (the first occurrence could be back-tracked at \TimeInert{} = 6.30), it can be observed that a few secondary droplets (the bigger ones) are ejected from the fingers. Thus, the outcome of \EtOh{} can be categorised as a \textit{splashing} case, but close to \textit{transition} since only few droplets are ejected.
Hence, a small shift in the splashing limit is observed between the three cases despite similar liquid properties and impact conditions: the miscible pair \EtOh{} starts to splash, the full wetting pair \SO{} is close to splash and the partially wetting pair \Hexa{} is closer to deposition. 
Note that the smallest viscosity (despite the small differences between the pairs) is the one of n-hexadecane. Since less viscous losses are expected, splashing with less kinetic energy should be observed (see the review article of \cite{Liang2016}). Yet it is not the case, hence, small viscosity variations are not expected to modify the trend observed between the liquid pairs.
This shift in the fragmentation limit, classically derived based on empirical observation of the outcome, is very small. Thus, it is hard to evaluate it in terms of critical Weber or Reynolds numbers.
Furthermore, there is a smooth transition from deposition to splashing \citep{Cossali1997}, which lead to the introduction of the \textit{transition} outcome \citep{Geppert2016} as described above to refine the splashing criteria. Qualitatively though, one could modify the averaged Weber number \avgWed{} defined in Sec.\ref{sec:ExpCond} taking into account all surface and interfacial tensions with $\left(\text{\sigmad{}}+\text{\sigmaf{}}+\text{\sigmadf{}}\right)/3$ instead of $\left(\text{\sigmad{}}+\text{\sigmaf{}}\right)/2$. This would lead to a characteristic Weber number of the impact multiplied by 0.98 for \Hexa{}, 1.08 for \SO{} and 1.5 for \EtOh{}, highlighting that \EtOh{} would cross the splashing limit earlier than \SO{} and finally than \Hexa{}.
This difficulty to assert the splashing limit for such a narrow difference highlights the need to consider another parameter to quantify the differences observed between the liquid pairs.

%% file: ExpObs_Extension.tex
\subsection{Shift in crown extension}\label{sec:CrownExt}

Besides the small shift in the fragmentation observed in Fig.~\ref{fig:HighSpeed}, the liquid pairs show differences in the size of the crown.
\begin{figure}
\centerline{\input{Figures/Crown_Repere_tikz_review.tex}}
  \caption{Schematic of a meridional section of a crown with droplet (green) and wall-film (blue) liquids, illustrating the full coverage of the droplet liquid on the inner part of the crown. The interface between them is marked by a red line. Note that the crown thickness is not representative of the reality. The crown edges (orange dots) extracted from the post-processing are used to define the crown geometrical parameters: rim radius \Rrim{}, rim height \Hrim{} and base radius \Rbase{}.}
\label{fig:CrownSchematic}
\end{figure}
To quantify this observation, the crown surface is calculated from the high-speed images by extracting the geometrical parameters of the crown as shown in the schematic of Fig.~\ref{fig:CrownSchematic}. The crown rim height \Hrim{}, the rim radius \Rrim{}, and the base radius \Rbase{} are extracted from the crown edges marked with orange circles. The crown surface, expressed in Eq.~\ref{eq:cs}, is calculated by summing the surface of the disc at the crown base, and the surface of the crown wall counted twice for the inner and outer parts. The crown wall is approximated by a conical frustum formed between the base and rim radii. 
This assumes having a straight line between the base and rim radii, although sometimes the crown wall is slightly bended inward. This bending and its influence on the determination of the crown surface is discussed in Appendix \ref{appA}. However, the difference in crown morphologies (between the crown edges) is taken indirectly into account via the calculation of \CS{}. Besides, the surface area at the rim due to the crown thickness is neglected.

\begin{equation}
  \text{\CS{}}=\upi \text{\Rbase{}}^2 + 2 \pi \left(\text{\Rbase{}}+\text{\Rrim{}}\right) \sqrt{\text{\Hrim{}}^2+\left(\text{\Rrim{}}-\text{\Rbase{}}\right)^2}
\label{eq:cs}
\end{equation}

The propagation of uncertainty from the measurement of \Hrim{}, \Rrim{} and \Rbase{} given in Sec.~\ref{expsetup} to \CS{} can be calculated based on the formula of Gauss-Laplace as $\Delta_{\text{\CS{}}}=\sqrt{\sum_{i} \left[\left(\partial \text{\CS{}}/\partial x_i\right) \; \Delta  x_i \right]^2 }$, where $x_i$ are the different variables of \CS{} (here the geometrical parameters) and $\Delta x_i$ their respective absolute uncertainty. The relative measurement uncertainty $\Delta_{\text{\CS{}}}/\text{\CS}$ is dependent upon the crown morphology, i.e. \Hrim{}/\Rrim{} and \Rbase{}/\Rrim{}.
Note that these ratios vary during the impact process, and between different impact conditions. A typical crown configuration though in the present experimental conditions is $\text{\Hrim{}}/\text{\Rrim{}} \approx{} 0.5$ and $\text{\Rbase{}}/\text{\Rrim{}} \approx{} 1$. In this case, the relative error $\Delta_{\text{\CS{}}}/\text{\CS{}}$ is equal to 5.7\%. In the present database, the ratio $\text{\Hrim{}}/\text{\Rrim{}}$ varies from 0.05 to 0.65, and $\text{\Rbase{}}/\text{\Rrim{}}$ from 0.65 to 1.30. By considering these extreme values, propagation of uncertainties up to 8.27\% can be found, tendentially for high \Rbase{} and/or small \Hrim{}.

The temporal evolution of \CS{} is shown in Fig.~\ref{fig:CrownSurfTime}(a) for the same experiments as in Fig.~\ref{fig:HighSpeed} that have similar impact conditions. Except at the early stage of impact where kinetic energy dominates, the curves separate progressively. As observed qualitatively in the high-speed images, the crown surface of \Hexa{} is smaller than the one of \SO{}, which is smaller than the one of \EtOh{}. 
Note that this ranking of crown surface extensions correlates with the values of \sigmadfstar{} reported in Table \ref{tab:prop}: The larger is \sigmadfstar{} (tendentially for immiscible liquid pairs with partial wetting), the smaller becomes the crown surface.
A similar ranking in the spreading between miscible and immiscible droplet/wall-films during the crown extension phase has also been measured by \citet{Chen2017}, without being explained. The formation of holes in the crown of \EtOh{} (grey symbols in Fig.~\ref{fig:CrownSurfTime}) does not influence this trend since the values of \CS{} for \EtOh{} are systematically above, but adds noise which prevents the exact determination of the maximum crown surface \CSmax{} (filled symbols in Fig.~\ref{fig:CrownSurfTime}(a,b)). \CSmax{} for \EtOh{} is lying between the value where the first hole appears (filled grey symbols and grey dashed line), and the maximum value calculated before the destruction of the crown (filled red symbols an red dashed line). Furthermore, the determination of the crown parameters during extension remains repeatable despite stochastic holes \citep{Geppert2017}.

\begin{figure}
\centerline{\input{Figures/CrownEvolution-B2W-G50W50-123-j3-k1-s5-t1-Hole2.tex}
\input{Figures/MaxCrownSurf_B2W_We_d_delta_1_Dim.tex}
\input{Figures/TimeHmax_We_delta_1_1_tex.tex}}
  \caption{(a) Temporal evolution of the crown surface \CS{} for the same experiments as in Fig.~\ref{fig:HighSpeed}. The filled symbols with yellow edge correspond to \CSmax{}. The dashed lines correspond to \tHmax{}, the time at which \Hrimmax{} is reached. (b) Maximum crown surface \CSmax{} at different Weber numbers \dropWe{}. (c) Time duration \tHmax{} of the crown ascending phase (marked with dashed lines in (a)) at different Weber numbers \dropWe{}. The dimensionless wall-film thickness is kept constant for the data of this figure at $\text{\wft}=0.122\pm0.011$. The grey symbols correspond to the measurements where holes in the crown are observed (only for \EtOh{}).}
\label{fig:CrownSurfTime}
\end{figure}

The values of \CSmax{} are reported for different Weber numbers in Fig.~\ref{fig:CrownSurfTime}(b). It is clear that the trend observed between the liquid pairs observed in (a) remains the same: at a given \dropWe{}, the values of \CSmax{} are systematically smaller for \Hexa{} than for \SO{}, which are smaller than \EtOh{}. 
For a given liquid pair, \CSmax{} increases with growing \dropWe{}, i.e. with increasing droplet kinetic energy. 
This corroborates the larger rim expansions observed at higher droplet kinetic energy for droplet impact onto wall-films \citep{Bernard2020b}, or for other impact configurations, e.g. onto dry surfaces \citep{Huang2018}, with another droplet \citep{Roisman2012}, or with a continuous jet \citep{Baumgartner2019}.

%% file: Figures/Crown_Repere_tikz_review.tex
\definecolor{c0000ff}{RGB}{0,0,255}
\definecolor{c44aa00}{RGB}{68,170,0}
\definecolor{c668500}{RGB}{102,133,0}
\definecolor{c55d400}{RGB}{85,212,0}
\definecolor{cff0000}{RGB}{255,0,0}
\definecolor{cff8300}{RGB}{255,131,0}
\newcommand{\WidthDashed}{0.5pt}

\begin{tikzpicture}    
	 \node[anchor=south west,inner sep=0] (image) at (0,0) {\includegraphics[scale=0.55,trim={3.3cm 12cm 2.7cm 9.5cm},clip]{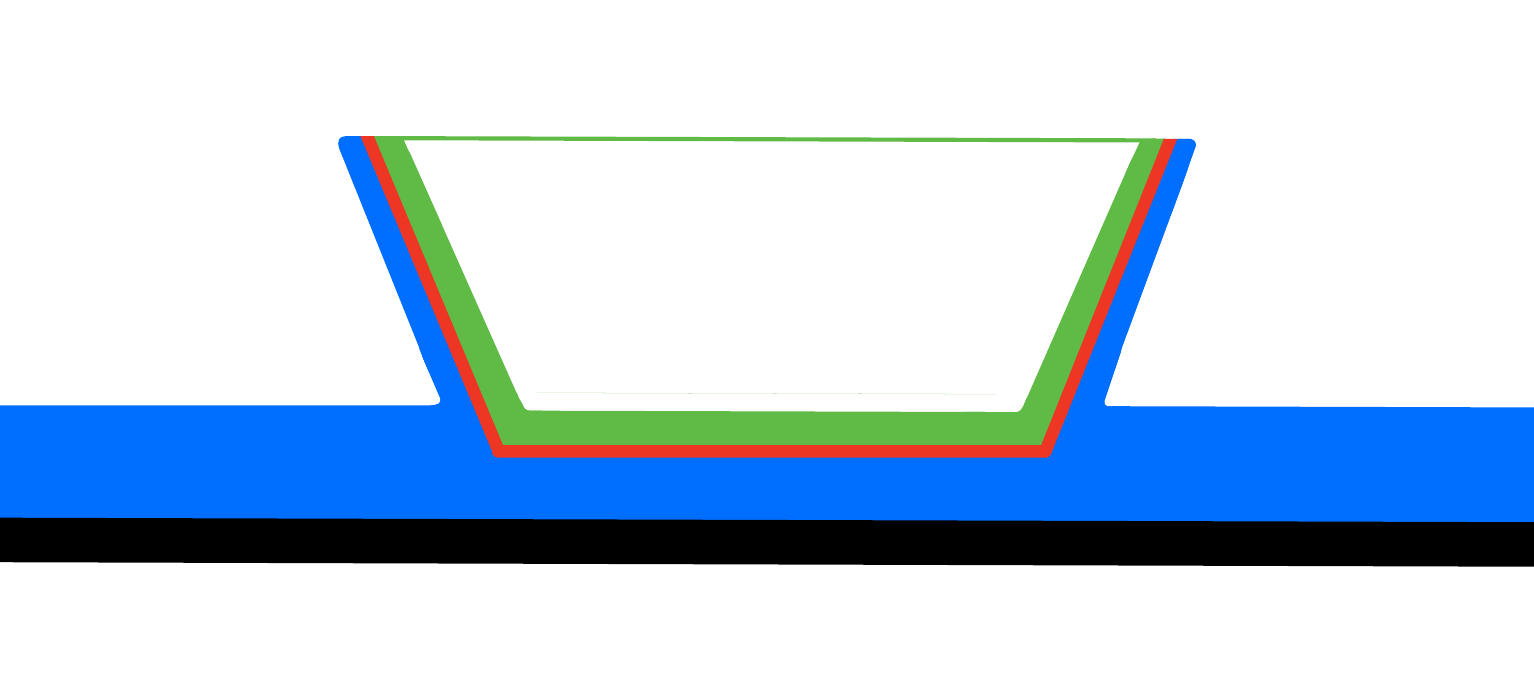}};  
		
    \begin{scope}[x={(image.south east)},y={(image.north west)}] 		
		
\draw[-,thick, c668500] (0.9,0.9) node[fill=white, anchor=mid]{$\sigma_d$, $\rho_d$, $\mu_d$} -- (0.725,0.65) ;;	
\draw[-,thick, c0000ff] (0.9,0.7) node[fill=white, anchor=mid]{$\sigma_f$, $\rho_f$, $\mu_f$} -- (0.73,0.55) ;;			
\draw[-,thick, cff0000] (0.9,0.5) node[fill=white, anchor=mid]{$\sigma_{d/f}$} -- (0.7,0.45) ;;	

        \node[orange] (R1) at (0.22,0.79) {$\bullet$}; 
        \node[orange] (R2) at (0.78,0.79) {$\bullet$}; 
        
        \node[orange] (B1) at (0.29,0.4) {$\bullet$}; 
        \node[orange] (B2) at (0.72,0.4) {$\bullet$}; 
				\draw[orange, dashed](R1)--(0.08,0.79); 
				\draw[orange, dashed](B1)--(0.08,0.4);
        \draw[orange, thick, arrows={Triangle[angle=90:5pt,orange] - Triangle[angle=90:5pt,orange]}](0.15,0.4)--(0.15,0.79) node[midway,sloped,left,rotate=270,orange] {\Hrim};;			                
        \draw[orange, dashed](R1)--(R1|-,0.95); 
        \draw[orange, dashed](R2)--(R2|-,0.95);
        \draw[orange,thick, arrows={Triangle[angle=90:5pt,orange] - Triangle[angle=90:5pt,orange]}](R1|-,0.875)--(R2|-,0.875) node[midway,sloped,above,orange] {2\Rrim};;
        
        \draw[orange, dashed](B1|-,0.035)--(B1); 
        \draw[orange, dashed](B2|-,0.035)--(B2);
        \draw[orange, thick, arrows={Triangle[angle=90:5pt,orange] - Triangle[angle=90:5pt,orange]}](B1|-,0.1)--(B2|-,0.1) node[midway,sloped,below,orange] {2\Rbase};;

     \end{scope}  
\end{tikzpicture}

%% file: Figures/CrownEvolution-B2W-G50W50-123-j3-k1-s5-t1-Hole2.tex
\begin{tikzpicture}[baseline]

\begin{axis}[%
width=0.245\linewidth,
height=0.245\linewidth,
scale only axis,
xmin=0,
xmax=8,
xlabel={$t$ $(\si{\milli\second})$},
ymin=0,
ymax=350,
ylabel={\CS{} (\si{\milli\meter\squared})},
axis background/.style={fill=white},
xmajorgrids,
ymajorgrids,
title={$\text{\dropWe}=363\pm5$},
legend style={at={(0.02,0.98)}, anchor=north west, legend cell align=left, align=left, draw=white!15!black},
clip mode=individual
]

\draw[green, dashed, thick] (4.35,0)--(4.35,204.377021163015);
\draw[blue, dashed, thick] (3.6,0)--(3.6,155.498563007783);
\draw[black!30!white, dashed, thick] (4.2,0)--(4.2,252.238774822162);
\draw[red, dashed, thick] (5.8,0)--(5.8,307.61951207877);

\node[anchor=south west] at (4.6,15) {\tHmax{}};

\addplot[only marks, mark=square, mark options={}, mark size=2.0pt, draw=blue] table[row sep=crcr]{%
x	y\\
0	nan\\
0.05	nan\\
0.1	nan\\
0.15	nan\\
0.2	nan\\
0.25	nan\\
0.3	nan\\
0.35	nan\\
0.4	nan\\
0.45	nan\\
0.5	nan\\
0.55	nan\\
0.6	nan\\
0.65	nan\\
0.7	nan\\
0.75	nan\\
0.8	nan\\
0.85	nan\\
0.9	nan\\
0.95	nan\\
1	23.886574910699\\
1.05	28.3663493379032\\
1.1	36.887874544837\\
1.15	40.5177784980729\\
1.2	46.5416479933272\\
1.25	48.4705372484394\\
1.3	50.5356621380225\\
1.35	54.966105999252\\
1.4	59.5021784488429\\
1.45	62.2836707043345\\
1.5	65.3655740669344\\
1.55	68.7483791055767\\
1.6	71.4651750021254\\
1.65	73.1978302813939\\
1.7	76.7423274387473\\
1.75	73.7952869691075\\
1.8	81.1582359337388\\
1.85	82.6897212431757\\
1.9	85.2776123810508\\
1.95	89.0487235766641\\
2	91.0643548759572\\
2.05	94.7371512683276\\
2.1	96.8602079733137\\
2.15	99.6629301614719\\
2.2	102.454961881879\\
2.25	103.144291800547\\
2.3	105.682950383281\\
2.35	108.880633365768\\
2.4	112.073992681337\\
2.45	113.23317686874\\
2.5	115.901779705761\\
2.55	118.599939213771\\
2.6	121.329456232483\\
2.65	122.788875358416\\
2.7	126.899918679007\\
2.75	128.486910238168\\
2.8	130.202349512971\\
2.85	132.797280639111\\
2.9	134.341908025188\\
2.95	136.208811330329\\
3	138.570008917964\\
3.05	140.547696220255\\
3.1	141.297311439685\\
3.15	143.96709674343\\
3.2	145.922447160616\\
3.25	147.647670058705\\
3.3	148.707559624823\\
3.35	150.044552753962\\
3.4	150.80342001029\\
3.45	152.136573737673\\
3.5	153.20134115009\\
3.55	153.942207716072\\
3.6	155.498563007783\\
3.65	155.188407413393\\
3.7	156.388843382804\\
3.75	155.381332857373\\
3.8	157.525127917328\\
3.85	158.470243309927\\
3.9	158.952303524039\\
3.95	158.903155100154\\
4	159.411804814276\\
4.05	159.952288177261\\
4.1	161.0091610436\\
4.15	160.018527887404\\
4.2	161.117583667737\\
4.25	160.15864282244\\
4.3	159.791951715029\\
4.35	160.374878729382\\
4.4	158.73836777156\\
4.45	159.341003157091\\
4.5	159.950088863578\\
4.55	157.12729071246\\
4.6	158.235623390673\\
4.65	155.312054145987\\
4.7	156.063412852521\\
4.75	153.861617884625\\
4.8	153.724327947795\\
4.85	152.224896953944\\
4.9	150.881247735117\\
4.95	150.934802498337\\
5	148.180903264314\\
5.05	145.389831543993\\
5.1	145.377652623567\\
5.15	148.135566467838\\
5.2	146.054398280673\\
5.25	145.420296711073\\
5.3	146.189534903904\\
5.35	147.86838736855\\
5.4	145.072703415805\\
5.45	146.977950583398\\
5.5	149.375480848994\\
5.55	144.556080791489\\
5.6	147.094819056835\\
5.65	143.137532777556\\
5.7	nan\\
5.75	nan\\
5.8	nan\\
5.85	nan\\
5.9	nan\\
5.95	nan\\
6	nan\\
6.05	nan\\
6.1	nan\\
6.15	nan\\
6.2	nan\\
6.25	nan\\
6.3	nan\\
6.35	nan\\
6.4	nan\\
6.45	nan\\
6.5	nan\\
6.55	nan\\
6.6	nan\\
6.65	nan\\
6.7	nan\\
6.75	nan\\
6.8	nan\\
6.85	nan\\
6.9	nan\\
};

\addplot [color=yellow, mark=square*, mark options={solid, fill=blue}, forget plot]
  table[row sep=crcr]{%
4.2	161.117583667737\\
};

\addplot[only marks, mark=o, mark options={}, mark size=2.0pt, draw=green] table[row sep=crcr]{%
x	y\\
0	nan\\
0.05	nan\\
0.1	nan\\
0.15	nan\\
0.2	nan\\
0.25	nan\\
0.3	nan\\
0.35	nan\\
0.4	nan\\
0.45	nan\\
0.5	nan\\
0.55	nan\\
0.6	nan\\
0.65	nan\\
0.7	nan\\
0.75	nan\\
0.8	nan\\
0.85	nan\\
0.9	36.5703238110898\\
0.95	33.1279970256748\\
1	37.2791679544423\\
1.05	41.0874315472547\\
1.1	45.6083675809696\\
1.15	49.8977072681385\\
1.2	54.3306553951384\\
1.25	70.3052931479222\\
1.3	73.1358778101401\\
1.35	77.6364124502373\\
1.4	77.0424891582067\\
1.45	67.6874755669434\\
1.5	71.4803416289326\\
1.55	74.1174704959334\\
1.6	78.7107740804491\\
1.65	80.6005324959571\\
1.7	85.4531496493743\\
1.75	89.6270288701767\\
1.8	91.5416741319184\\
1.85	95.9291696316945\\
1.9	98.2211037621049\\
1.95	101.452242680685\\
2	105.999540838845\\
2.05	106.251291630683\\
2.1	111.011779518447\\
2.15	113.340801595271\\
2.2	118.177180067716\\
2.25	120.26663720692\\
2.3	124.182930177855\\
2.35	127.645326125087\\
2.4	129.768738959253\\
2.45	133.469353046275\\
2.5	136.071287436873\\
2.55	138.908846554812\\
2.6	141.831928464029\\
2.65	145.163829009425\\
2.7	147.647670058705\\
2.75	149.699624925389\\
2.8	152.652745866528\\
2.85	155.909721914617\\
2.9	158.115560989522\\
2.95	161.411973248535\\
3	162.910411828892\\
3.05	166.071850843999\\
3.1	167.762203584241\\
3.15	168.642920134582\\
3.2	170.475426188123\\
3.25	171.911923998697\\
3.3	173.401545562342\\
3.35	174.640153422712\\
3.4	175.707394632456\\
3.45	177.351884978615\\
3.5	179.625553464438\\
3.55	180.796906999359\\
3.6	182.361778426346\\
3.65	183.376857165993\\
3.7	182.082720484966\\
3.75	183.38212629774\\
3.8	185.372642815806\\
3.85	186.047125936964\\
3.9	187.78300184543\\
3.95	189.048932813986\\
4	190.354294033814\\
4.05	191.824096703525\\
4.1	198.342246494906\\
4.15	199.124938287673\\
4.2	198.927825516521\\
4.25	200.520762642255\\
4.3	200.347324728963\\
4.35	204.377021163015\\
4.4	200.463475820771\\
4.45	194.157708711227\\
4.5	194.085092437627\\
4.55	195.578629583388\\
4.6	197.228660994703\\
4.65	198.897538434912\\
4.7	193.379578656743\\
4.75	194.105225957204\\
4.8	187.894017321915\\
4.85	191.938498891196\\
4.9	186.544752499714\\
4.95	188.175501539069\\
5	183.641846086537\\
5.05	180.71329088221\\
5.1	181.409994490142\\
5.15	183.044880358719\\
5.2	174.098734752346\\
5.25	175.660169683073\\
5.3	170.462035542733\\
5.35	167.606910030896\\
5.4	163.294407630576\\
5.45	156.787209647096\\
5.5	148.582912826999\\
5.55	145.211281145863\\
5.6	141.92937082144\\
5.65	138.628484455894\\
5.7	131.613215603992\\
5.75	125.420970392285\\
5.8	120.486436999581\\
5.85	118.082508777863\\
5.9	115.464869495867\\
5.95	111.4062747128\\
6	106.498673643293\\
6.05	104.232073032742\\
6.1	102.116181453238\\
6.15	99.8968253156707\\
6.2	95.6769355414976\\
6.25	91.8120310461247\\
6.3	89.1558279261341\\
6.35	88.0598531569209\\
};

\addplot [color=yellow, mark=*, mark options={solid, fill=green}, forget plot]
  table[row sep=crcr]{%
4.35	204.377021163015\\
};

\addplot[only marks, mark=triangle, mark options={}, mark size=2.0pt, draw=red] table[row sep=crcr]{%
x	y\\
0	nan\\
0.05	nan\\
0.1	nan\\
0.15	nan\\
0.2	nan\\
0.25	nan\\
0.3	nan\\
0.35	nan\\
0.4	nan\\
0.45	nan\\
0.5	nan\\
0.55	nan\\
0.6	nan\\
0.65	nan\\
0.7	nan\\
0.75	nan\\
0.8	37.9597802727136\\
0.85	42.6211535446929\\
0.9	33.0409230336213\\
0.95	33.2424261053916\\
1	39.4488249441078\\
1.05	42.8786549616634\\
1.1	46.1319644547608\\
1.15	50.6427533490967\\
1.2	72.5995003092223\\
1.25	60.2980272438414\\
1.3	80.9379888283441\\
1.35	70.0998794100184\\
1.4	74.3081285160656\\
1.45	77.5045204302662\\
1.5	81.9780864022271\\
1.55	102.072298324688\\
1.6	90.0953548507072\\
1.65	91.1298472593244\\
1.7	97.2330931918406\\
1.75	100.962867112182\\
1.8	106.015703765575\\
1.85	123.029698148553\\
1.9	104.765112332558\\
1.95	108.625965821178\\
2	113.254549955662\\
2.05	126.09138531605\\
2.1	129.058380439253\\
2.15	132.833778725579\\
2.2	136.430695101245\\
2.25	142.363809667892\\
2.3	143.783618229076\\
2.35	148.588937183753\\
2.4	149.255495199135\\
2.45	155.717173086387\\
2.5	155.660174402771\\
2.55	160.097339138233\\
2.6	163.089119295627\\
2.65	165.135880976333\\
2.7	167.724241364704\\
2.75	172.591083687525\\
2.8	174.607592586476\\
2.85	179.269009980522\\
2.9	181.073624925491\\
2.95	184.700100160936\\
3	187.809411560706\\
3.05	192.522115175206\\
3.1	193.636631768693\\
3.15	197.773654017343\\
3.2	201.77859343404\\
3.25	206.187546639117\\
3.3	206.785217216294\\
3.35	212.020772803009\\
3.4	215.848322999496\\
3.45	218.706068405418\\
3.5	222.36169256086\\
3.55	225.133846363754\\
3.6	229.133327181654\\
3.65	229.698449900209\\
3.7	232.335210460973\\
3.75	234.27566420142\\
3.8	237.059554803196\\
3.85	239.417997832633\\
3.9	240.416265090253\\
3.95	241.457584320079\\
4	245.38044706013\\
4.05	248.231923551264\\
4.1	249.375026893558\\
4.15	250.803393453331\\
4.2	252.238774822162\\
};

\addplot[only marks, mark=triangle, mark options={}, mark size=2.0pt, draw=black!30!white] table[row sep=crcr]{%
x	y\\
4.25	255.533776444057\\
4.3	256.773691077673\\
4.35	257.196881493407\\
4.4	257.625036334147\\
4.45	261.863714801804\\
4.5	262.313823804901\\
4.55	265.710859040025\\
4.6	269.476826718074\\
4.65	269.960136561378\\
4.7	273.333477890574\\
4.75	274.861624024205\\
4.8	275.380957750367\\
4.85	278.335287131519\\
4.9	282.960979551562\\
4.95	283.512688674846\\
5	284.630984032834\\
5.05	288.261468374118\\
5.1	290.761245840695\\
5.15	291.94169880255\\
5.2	294.460005265177\\
5.25	296.283975001474\\
5.3	296.90175307849\\
5.35	297.524430580388\\
5.4	298.784495052845\\
5.45	298.289844835864\\
5.5	298.931254714117\\
5.55	302.149877517655\\
5.6	304.729262709408\\
5.65	304.909690651632\\
5.7	305.390735441747\\
5.75	306.728201879568\\
5.8	307.61951207877\\
5.85	304.336730049633\\
5.9	305.016473504543\\
5.95	305.016473504543\\
6	304.336730049633\\
6.05	301.666387254417\\
6.1	299.873516836027\\
6.15	293.488547650176\\
6.2	292.859048567427\\
6.25	292.859048567427\\
6.3	294.122952993268\\
6.35	293.001048608669\\
6.4	294.282775999982\\
6.45	294.930988757833\\
6.5	295.11708834633\\
6.55	293.030253889476\\
6.6	293.030253889476\\
6.65	293.030253889476\\
6.7	293.030253889476\\
6.75	292.564748928533\\
6.8	292.564748928533\\
6.85	290.429607255289\\
6.9	290.429607255289\\
6.95	289.987038237285\\
7	289.987038237285\\
7.05	286.120891620428\\
};

\addplot [color=yellow, mark=triangle*,mark size=2.0pt, mark options={solid, fill=black!30!white}, forget plot]
  table[row sep=crcr]{%
4.2	252.238774822162\\
};

\addplot [draw=yellow, mark=triangle*, mark size=2.0pt, mark options={solid, fill=red}]
  table[row sep=crcr]{%
5.8	307.61951207877\\
};

\end{axis}
\end{tikzpicture}%

%% file: Figures/MaxCrownSurf_B2W_We_d_delta_1_Dim.tex
\begin{tikzpicture}[baseline]

\begin{axis}[%
width=0.245\linewidth,
height=0.245\linewidth,
scale only axis,
xmin=200,
xmax=400,
xlabel style={font=\color{white!15!black}},
xlabel={\dropWe{} (-)},
ymin=0,
ymax=350,
ylabel style={font=\color{white!15!black}},
ylabel={\CSmax{} (\si{\milli\meter\squared})},
axis background/.style={fill=white},
title style={font=\bfseries},
title={$\text{\wft{}} \approx{} 0.12$},
xmajorgrids,
ymajorgrids,
clip mode=individual
]

\addplot [color=black, mark=square*, mark options={solid, fill=blue}]
  table[row sep=crcr]{%
508.471029883192	200.291006763379\\
508.471029883192	200.291006763379\\
};

\addplot [color=black, mark=*, mark options={solid, fill=green}]
  table[row sep=crcr]{%
508.48504088168	236.145101384781\\
508.48504088168	236.145101384781\\
};

\addplot [color=black, mark=triangle*, mark options={solid, fill=red}]
  table[row sep=crcr]{%
498.867960163284	350.029084940422\\
498.867960163284	288.82839917649\\
};

\addplot [color=black, mark=square*, mark options={solid, fill=blue}, forget plot]
  table[row sep=crcr]{%
440.840954053392	151.023613883439\\
440.840954053392	151.023613883439\\
};
\addplot [color=black, mark=*, mark options={solid, fill=green}, forget plot]
  table[row sep=crcr]{%
445.036209274399	192.717190260402\\
445.036209274399	192.717190260402\\
};
\addplot [color=black, mark=triangle*, mark options={solid, fill=red}, forget plot]
  table[row sep=crcr]{%
434.940594142564	304.477989460389\\
434.940594142564	258.73358576654\\
};
\addplot [color=black, mark=square*, mark options={solid, fill=blue}, forget plot]
  table[row sep=crcr]{%
358.310179270558	161.117583667737\\
358.310179270558	161.117583667737\\
};
\addplot [color=black, mark=*, mark options={solid, fill=green}, forget plot]
  table[row sep=crcr]{%
368.354660425129	204.377021163015\\
368.354660425129	204.377021163015\\
};
\addplot [color=black, mark=triangle*, mark options={solid, fill=red}, forget plot]
  table[row sep=crcr]{%
362.804222243713	307.61951207877\\
362.804222243713	252.238774822162\\
};
\addplot [color=black, mark=triangle*, mark options={solid, fill=black!30!white}, forget plot] 
  table[row sep=crcr]{%
362.804222243713	252.238774822162\\
};

\addplot [color=black, mark=square*, mark options={solid, fill=blue}, forget plot]
  table[row sep=crcr]{%
314.864350862924	131.959989693274\\
314.864350862924	131.959989693274\\
};
\addplot [color=black, mark=*, mark options={solid, fill=green}, forget plot]
  table[row sep=crcr]{%
320.545648797683	174.406790546171\\
320.545648797683	174.406790546171\\
};
\addplot [color=black, mark=triangle*, mark options={solid, fill=red}, forget plot]
  table[row sep=crcr]{%
316.28831821484	284.236815534747\\
316.28831821484	226.15697807761\\
};
\addplot [color=black, mark=triangle*, mark options={solid, fill=black!30!white}, forget plot] 
  table[row sep=crcr]{%
316.28831821484	226.15697807761\\
};

\addplot [color=black, mark=square*, mark options={solid, fill=blue}, forget plot]
  table[row sep=crcr]{%
266.853404078276	102.634521070003\\
266.853404078276	102.634521070003\\
};
\addplot [color=black, mark=*, mark options={solid, fill=green}, forget plot]
  table[row sep=crcr]{%
271.76583073723	152.210988952478\\
271.76583073723	152.210988952478\\
};
\addplot [color=black, mark=triangle*, mark options={solid, fill=red}, forget plot]
  table[row sep=crcr]{%
269.170235106682	225.231143770399\\
269.170235106682	225.231143770399\\
};
\addplot [color=black, mark=square*, mark options={solid, fill=blue}, forget plot]
  table[row sep=crcr]{%
220.427335475403	115.700373802581\\
220.427335475403	115.700373802581\\
};
\addplot [color=black, mark=*, mark options={solid, fill=green}, forget plot]
  table[row sep=crcr]{%
215.741494675615	114.955522487748\\
215.741494675615	114.955522487748\\
};
\addplot [color=black, mark=triangle*, mark options={solid, fill=red}, forget plot]
  table[row sep=crcr]{%
218.583938591942	175.635477396474\\
218.583938591942	175.635477396474\\
};
\end{axis}
\end{tikzpicture}%

%% file: Figures/TimeHmax_We_delta_1_1_tex.tex
\begin{tikzpicture}[baseline]

\begin{axis}[%
width=0.245\linewidth,
height=0.245\linewidth,
scale only axis,
xmin=200,
xmax=400,
xlabel style={font=\color{white!15!black}},
xlabel={\dropWe{} (-)},
ymin=0,
ymax=7,
ylabel style={font=\color{white!15!black}},
ylabel={\tHmax{} $\left( \si{\milli\second} \right)$},
axis background/.style={fill=white},
title style={font=\bfseries},
title={$\text{\wft{}} \approx{} 0.12$},
xmajorgrids,
ymajorgrids,
legend style={at={(0.98,0.02)}, anchor=south east, legend cell align=left, align=left, draw=white!15!black},
clip mode=individual
]
\node at (220,6.5) {\textbf{(c)}};

\addplot [color=black, mark=square*, mark options={solid, fill=blue}, only marks]
  table[row sep=crcr]{%
508.471029883192	3.25\\
508.471029883192	3.25\\
};
\addlegendentry{\Hexa{}}

\addplot [color=black, mark=*, mark options={solid, fill=green}, only marks]
  table[row sep=crcr]{%
508.48504088168	3.25\\
508.48504088168	3.25\\
};
\addlegendentry{\SO{}}

\addplot [color=black, mark=triangle*, mark options={solid, fill=red}, only marks]
  table[row sep=crcr]{%
218.583938591942	3.8\\
218.583938591942	3.8\\
};
\addlegendentry{\EtOh{}}

\addplot [color=black, mark=triangle*, mark options={solid, fill=red}]
  table[row sep=crcr]{%
498.867960163284	4.1\\
498.867960163284	3\\
};
\addplot [draw=black, mark=triangle*, mark options={solid, fill=black!30!white}, only marks]
  table[row sep=crcr]{%
498.867960163284	3\\
};

\addplot [color=black, mark=square*, mark options={solid, fill=blue}, forget plot]
  table[row sep=crcr]{%
440.840954053392	2.2\\
440.840954053392	2.2\\
};
\addplot [color=black, mark=*, mark options={solid, fill=green}, forget plot]
  table[row sep=crcr]{%
445.036209274399	3.15\\
445.036209274399	3.15\\
};
\addplot [color=black, mark=triangle*, mark options={solid, fill=red}, forget plot]
  table[row sep=crcr]{%
434.940594142564	4.25\\
434.940594142564	3.2\\
};
\addplot [draw=black, mark=triangle*, mark options={solid, fill=black!30!white}, only marks]
  table[row sep=crcr]{%
434.940594142564	3.2\\
};

\addplot [color=black, mark=square*, mark options={solid, fill=blue}, forget plot]
  table[row sep=crcr]{%
358.310179270558	3.6\\
358.310179270558	3.6\\
};
\addplot [color=black, mark=*, mark options={solid, fill=green}, forget plot]
  table[row sep=crcr]{%
368.354660425129	4.35\\
368.354660425129	4.35\\
};
\addplot [color=black, mark=triangle*, mark options={solid, fill=red}, forget plot]
  table[row sep=crcr]{%
362.804222243713	5.8\\
362.804222243713	4.2\\
};
\addplot [draw=black, mark=triangle*, mark options={solid, fill=black!30!white}, only marks]
  table[row sep=crcr]{%
362.804222243713	4.2\\
};

\addplot [color=black, mark=square*, mark options={solid, fill=blue}, forget plot]
  table[row sep=crcr]{%
314.864350862924	3.35\\
314.864350862924	3.35\\
};
\addplot [color=black, mark=*, mark options={solid, fill=green}, forget plot]
  table[row sep=crcr]{%
320.545648797683	3.35\\
320.545648797683	3.35\\
};
\addplot [color=black, mark=triangle*, mark options={solid, fill=red}, forget plot]
  table[row sep=crcr]{%
316.28831821484	5.8\\
316.28831821484	3.9\\
};
\addplot [draw=black, mark=triangle*, mark options={solid, fill=black!30!white}, only marks]
  table[row sep=crcr]{%
316.28831821484	3.9\\
};

\addplot [color=black, mark=square*, mark options={solid, fill=blue}, forget plot]
  table[row sep=crcr]{%
266.853404078276	3.5\\
266.853404078276	3.5\\
};
\addplot [color=black, mark=*, mark options={solid, fill=green}, forget plot]
  table[row sep=crcr]{%
271.76583073723	3.8\\
271.76583073723	3.8\\
};
\addplot [color=black, mark=triangle*, mark options={solid, fill=red}, forget plot]
  table[row sep=crcr]{%
269.170235106682	4.3\\
269.170235106682	4.3\\
};
\addplot [color=black, mark=square*, mark options={solid, fill=blue}, forget plot]
  table[row sep=crcr]{%
220.427335475403	3.15\\
220.427335475403	3.15\\
};
\addplot [color=black, mark=*, mark options={solid, fill=green}, forget plot]
  table[row sep=crcr]{%
215.741494675615	3.35\\
215.741494675615	3.35\\
};
\end{axis}
\end{tikzpicture}%

%% file: ExpObs_Duration.tex
\subsection{Shift in duration of crown ascending phase}\label{sec:CrownDur}

It is interesting to note that the duration of the crown ascending phase (i.e. the time at which \Hrimmax{} is reached, marked by dashed lines in Fig.~\ref{fig:CrownSurfTime}(a)) also exhibits a systematic ranking between the liquid pairs, although in smaller amplitude than the differences in crown surfaces. The ascending duration of \Hexa{} is smaller than the one of \SO{}, which may be smaller than the one of \EtOh{} (the grey dashed line corresponds to the time at which the first hole could be back-tracked). Here again, this ranking correlates with the values of \sigmadfstar{} reported in Table \ref{tab:prop}. The higher is \sigmadfstar{}, the shorter becomes the ascending duration.
 Note that a similar observation could not be made with \tSmax{} (the time at which the maximum crown surface is reached). In fact, the exact time of crown surface receding is unclear because it has, in some cases, a plateau at its maximum value. 
This effect is known as the stabilization phase of the crown at the end of the surface extension \citep{Zhang2019}.
This happens when the radial extension of the crown is longer than the axial one \citep{Bernard2018}. In this case, the decrease of \Hrim{} is compensated by a continuous increase of \Rrim{} and/or \Rbase{}. This effect shifts \tSmax{} compared to \tHmax{} (as in Fig.~\ref{fig:CrownSurfTime}(a) for \Hexa{} between the blue filled symbol and the blue dashed line) and leads to scattered data. Hence, \tHmax{} is preferred to \tSmax{}.
The durations of the crown ascending phase \tHmax{} are given in Fig.~\ref{fig:CrownSurfTime}(c) for different \dropWe{}. 
A trend between the liquid pairs is observed, similar to the maximum crown surface \CSmax{}: the ascending phase of \Hexa{} is slightly shorter than the one of \SO{}, which is (not systematically) slightly smaller than \EtOh{}. These durations are slightly increasing with growing \dropWe{}, as already observed for example with droplet impact onto dry surfaces \citep{Huang2018}, but not for droplet head-on collisions or impacts on a continuous jet where the kinetics is fixed by the encapsulated drop only \citep{Baumgartner2019}.

In summary, slight differences between the liquid pairs were observed for the onset of splashing, but more prominently, differences in maximum crown surface, and in the duration of the crown ascending phase despite similar impact conditions and liquid properties. Thus, these differences can be attributed to their wettability and miscibility behaviours.

%% file: IFT_Intro.tex
\section{Importance of the interfacial tension}\label{sec:IFT}

The miscibility and wettability behaviours responsible for the differences observed in Sec.~\ref{sec:ExpObs} are linked to the interfacial tension \sigmadfstar{} characteristic of each liquid droplet/wall-film pair (see Table \ref{tab:prop}). During droplet impact onto thin wall-films, the initial kinetic energy of the droplet is partially converted into surface energy and partially dissipated due to impact losses (initial droplet deformation) and viscous losses during the extension phase. This energy transfer leads to an important crown extension and a significant role of the surface forces. Although the role of surface tension is quite clear, the importance of interfacial tension for droplet impact onto wall-films during extension remains poorly documented and understood. 
In the following, we investigate the role of the interfacial tension, starting with energetical considerations followed by kinetic ones.

%% file: IFT_nrj.tex
		\subsection{Energy storage} \label{sec:nrj_storage}

In order to estimate if the differences in the crown extension observed in Sec.~\ref{sec:CrownExt} can be explained by the differences in interfacial tension $\sigma_{d/f}$, the temporal evolution of the crown surface energy \CSE{} is calculated. Each portion of the crown (i.e. base disc and conical frustum) is multiplied by the corresponding surface/interfacial tension (i.e. $\sigma_d$, $\sigma_f$ and/or $\sigma_{d/f}$). The droplet liquid is assumed to completely cover the wall-film liquid on the inside of the crown, as represented in Fig.~\ref{fig:CrownSchematic}. This full coverage by the droplet liquid on the inner crown wall has been observed numerically for aqueous droplet/wall-film systems \cite{Zhang2019}, and experimentally for dyed water droplet impacting on silicone oil films \citep{Shaikh2018}, for $Re_{f}=\rho_{f}\text{\dropvel{}\filmthick}/\mu_{f}$ below 400 \citep{Kittel2018a}.  
In the current work, the maximum value of \filmRe{} is 196. Hence, it can reasonably be assumed that the droplet liquid completely covers the inner part of the crown, which leads to the following expression for the crown surface energy:

\begin{equation}
  \text{\CSE{}}= \left( \sigma_d+\sigma_{d/f} \right) \upi \text{\Rbase{}}^2 + \left( \sigma_d+\sigma_{d/f}+\sigma_f \right)\upi \left(\text{\Rbase{}}+\text{\Rrim{}}\right) \sqrt{\text{\Hrim{}}^2+\left(\text{\Rrim{}}-\text{\Rbase{}} \right)^2}
\label{eq:cse}
\end{equation}

The propagation of uncertainty for \CSE{} can be calculated similar to that of the crown surface \CS{}, by considering additionally the measurement uncertainty associated with the surface and interfacial tensions given in Table \ref{tab:prop}. This provides a propagation of error for the reference geometrical configuration ($\text{\Hrim{}}/\text{\Rrim{}} \approx{} 0.5$ and $\text{\Rbase{}}/\text{\Rrim{}} \approx{} 1$) of 6.1\% for \Hexa{}, 7.2\% for \SO{} and 7.6\% for \EtOh{}. By considering the extreme values of the crown geometrical configuration as presented for $\Delta_{\text{\CS{}}}$, the maximum propagation of uncertainty to \CSE{} becomes 9.6\% (obtained for \EtOh{}).

The temporal evolution of \CSE{} is shown in Fig.~\ref{fig:CrownSurfNrj-time}(a) for the same cases as in Fig.~\ref{fig:CrownSurfTime}(a). In contrast to \CS{}, the maxima are now comparable (note that at a constant Weber number, \DKE{} is constant for the three liquid pairs). Indeed, the overall surface energy of \Hexa{}, which has the smallest crown surface, is relatively increased due to the high value of $\sigma_{d/f}$ for this liquid pair. Likewise, the overall surface energy of \EtOh{}, which has the biggest crown surface, is relatively decreased since the interfacial tension is zero (miscible liquid pair). In between lies the case of \SO{}. For all \dropWe{} investigated in our study,  all \CSEmax{} normalized with the droplet kinetic energy $\text{\DKE{}}=\rho_d \upi \text{\dropdia{}}^3 \text{\dropvel{}}^2/12$ are systematically coming together. Hence, the interfacial tension appears as a good candidate to explain the differences in maximum crown extensions. This means that, for immiscible liquid pairs, the interface between droplet and wall film stores a non-negligible amount of energy during the extension phase, proportionally to \sigmadfstar{}. This stored energy neither participates in the crown extension, nor in the splashing process. Hence, liquid pairs with smaller interfacial tension (the limit being zero for miscible cases) expand more, and splash at lower droplet kinetic energy as observed in Fig.~\ref{fig:HighSpeed}.
For impact processes with smaller extensions (e.g. at very small \dropWe{}, or onto another impacted substrates), the interfacial energy could be less significant and its effect may not be noticeable.

\begin{figure}
\centerline{\input{Figures/CrownEvolution-G50W50-123-j3-k1-s4-t4-Hole2.tex}
\input{Figures/MaxSurfNrj_FW_KinEn_Wed_delta_1_withFit.tex}
\input{Figures/TimeHmax_We_delta_1_4_tex.tex}
}
  \caption{(a) Temporal evolution of the crown surface energy \CSE{} normalized with the initial droplet kinetic energy \DKE{} in function of \tcaptime{} (see Eq.~\ref{eq:captime}) for the same experiments as in Fig.~\ref{fig:HighSpeed} and Fig.~\ref{fig:CrownSurfTime}(a). The filled symbols with yellow edge correspond to \CSEmax{}/\DKE{}. The dashed lines correspond to \tcapHmax{}, the time at which \Hrimmax{} is reached normalized with \tcap{}. (b) Maximum crown surface energy \CSEmax{} normalized with the initial droplet kinetic energy \DKE{} at different Weber numbers \dropWe{}. (c) Capillary time duration \tcapHmax{} of the crown ascending phase (highlighted with dashed lines in (a)) at different Weber numbers \dropWe{}. The capillary time scale \tcap{} used has a typical spring constant of \sigmad{}+\sigmadf{}+\sigmaf{} and a characteristic mass $\text{\md{}}+\text{\mf{}}=\left( \text{\rhod}+\text{\rhof} \right) \pi \text{\dropdia{}}^3/6$. The dimensionless wall-film thickness is kept constant for the data of this figure at $\text{\wft}=0.122\pm0.011$. The grey symbols correspond to the measurements where holes in the crown are observed (only for \EtOh{}).}
\label{fig:CrownSurfNrj-time}
\end{figure}

The ratio of \CSEmax{} with \DKE{} in Fig.~\ref{fig:CrownSurfNrj-time}(b) indicates how much the incoming kinetic energy has been transferred into surface energy. 
On average, this ratio is lying around 50\% of the initial kinetic energy in our experimental range. This amount corroborates numerical simulations at similar \dropWe{} and wall-film thickness of aqueous droplet/wall-film systems \citep{Zhang2019}.
Similar values have also been found in numerical simulations of droplet impact onto dry surface with slip condition \citep{Wildeman2016}, i.e. with negligible surface friction which can be assimilated to a liquid interface.
At the end of crown extension, the quantity $1-\text{\CSEmax{}}/\text{\DKE{}}$ is representative of the remaining energy and losses before the onset of retraction. Some kinetic energy might remain due to internal flow motions and the complex shape of the crown. Furthermore, the motion of a liquid mass in the vertical direction of a liquid mass leads to a conversion of kinetic energy into gravitational energy. 
Numerical studies of water droplet impact onto wall-films at \wft{}=0.1 and \dropWe{}=250 \citep{Zhang2019} show that at the end of the expansion, the remaining kinetic energy is about 8\% of \DKE{}, and the gravitational energy of 2\%. 
Combined with a surface energy of about 50\%, this corresponds to an energy loss of about 40\% of the droplet kinetic energy during the extension phase. This corroborates the study on dry surface with slip condition at the wall \citep{Wildeman2016}, where they found 10\% remaining energy (no gravitation energy since the spreading is horizontal), and 40\% of \DKE{} converted into heat.
Hence, in the present droplet/wall-film experiments, the remaining kinetic and gravitational energies at the end of crown extension are expected to be about 10\% of \DKE{}. Note that the gravitational energy is thus minor for \wft{}=0.1, but for higher wall-film thickness like \wft{}=0.3 it can represent up to 15\% of the crown surface energy \citep{Cossali2004}.

A full energy balance of the impact process (always assuming a zero kinetic energy at maximum crown surface) should also consider the initial surface energies of droplet and wall-film. The initial surface energy of the droplet can be calculated with $\text{\DSE}=\upi \text{\dropdia{}}^2 \sigma_{d}$.
For the wall-film, the initial surface energy can be approximated with the free surface of the wall-film covered by the crown base at maximum extension, i.e. $\text{\FSE}\approx\upi\text{\Rbasemax{}}^2 \sigma_{f}$ (see e.g. similar approaches for droplet impact onto dry surfaces by \citet{Vaikuntanathan2016}).
This surface corresponds to the free surface of the wall-film covered by the crown base at maximum extension and is therefore an impact dependent variable that can be estimated \textit{a posteriori} in the present database.
The values of \CSEmax{}/(\DKE+\DSE+\FSE) taking all initial energies into account would be diminished from 5 to 10\% compared to \CSEmax{}/\DKE{} alone reported in Fig.~\ref{fig:CrownSurfNrj-time}(b).
Since this shift roughly corresponds to the value of the remaining kinetic and gravitational energies of approximately 10\% of \DKE{} at the end of the crown extension discussed previously, this means that \CSEmax{}/\DKE{} gives an estimation of the losses at the end of the crown extension, of about 50\% of all initial energies for these impact conditions.

Looking at the influence of the droplet kinetic energy, the maximum crown surface energy normalized with the droplet kinetic energy slightly decreases with increasing \dropWe{}. Hence, the percentage of total losses is increasing with growing \dropWe{}, which can be explained by higher deformation losses of the impacting droplet \citep{Gao2015} and a more pronounced influence of the boundary layer because of a higher shear due to the higher velocity gradient, and a deeper penetration of the droplet bringing the flow closer to the wall \citep{Lamanna2019}.

The extension of the experimental database to $\text{\wft{}} \approx{} 0.26$ given in Appendix \ref{appA} shows a similar data reduction of the normalized crown surface energies between the different liquid pairs at a given Weber number. However, the values are now higher compared to $\text{\wft{}} \approx{} 0.12$, lying around 0.6 and above. This corroborates the influence of shear during the extension process since it is expected to be reduced for higher wall-film thickness because of a decreased influence of the solid wall. However, a more refined quantitative analysis is not possible here since the gravitational energies and remaining kinetic energy at the end of crown extension might be larger for wall-film thickness above 0.2 \citep{Zhang2019}.

To conclude, the interface (if any) between the droplet and wall-film  stores a non-negligible amount of energy during the extension.
Taking into account the interfacial energy generalizes the estimation of viscous losses made for miscible liquid pairs to immiscible ones with different interfacial tensions. 
With about 50\% loss of the initial surface and kinetic energies for \wft{} $\approx{}$ 0.1, the viscous effects prevent the crown extension as much as the capillary forces for thin wall-films. Hence, the droplet/wall-film system should be considered as a inertio-capillary-\textit{viscous} system.

%% file: Figures/CrownEvolution-G50W50-123-j3-k1-s4-t4-Hole2.tex
\definecolor{mycolor1}{rgb}{0.00000,0.0,0.0}%
\definecolor{mycolor2}{rgb}{0.0,0.0,0.00}%
\definecolor{mycolor3}{rgb}{0.0,0.0,0.0}%
\definecolor{mycolor4}{rgb}{0.0,0.0,0.0}%
\begin{tikzpicture}[baseline]

\begin{axis}[%
width=0.245\linewidth,
height=0.245\linewidth,
scale only axis,
xmin=0,
xmax=0.8,
xlabel={\tcaptime{} (-)},
ymin=0,
ymax=0.6,
ylabel={\CSE{}/\DKE{} $(-)$},
axis background/.style={fill=white},
xmajorgrids,
ymajorgrids,
title={$\text{\dropWe}=363\pm5$},
clip mode=individual
]

\node at (0.07,0.55) {\textbf{(a)}};

\addplot[only marks, mark=square, mark options={}, mark size=2.0pt, draw=blue] table[row sep=crcr]{%
x	y\\
0	nan\\
0.00616139874589175	nan\\
0.0123227974917835	nan\\
0.0184841962376753	nan\\
0.024645594983567	nan\\
0.0308069937294588	nan\\
0.0369683924753505	nan\\
0.0431297912212423	nan\\
0.049291189967134	nan\\
0.0554525887130258	nan\\
0.0616139874589175	nan\\
0.0677753862048093	nan\\
0.0739367849507011	nan\\
0.0800981836965928	nan\\
0.0862595824424846	nan\\
0.0924209811883763	nan\\
0.0985823799342681	nan\\
0.10474377868016	nan\\
0.110905177426052	nan\\
0.117066576171943	nan\\
0.123227974917835	0.0879480448162668\\
0.129389373663727	0.102134234286983\\
0.135550772409619	0.127379232747878\\
0.14171217115551	0.138513687701716\\
0.147873569901402	0.156554461414947\\
0.154034968647294	0.163134016233323\\
0.160196367393186	0.1700916194721\\
0.166357766139077	0.182996913163555\\
0.172519164884969	0.196203779762965\\
0.178680563630861	0.205064256203714\\
0.184841962376753	0.214712993907884\\
0.191003361122644	0.225151218622988\\
0.197164759868536	0.23345357594252\\
0.203326158614428	0.2382922381494\\
0.20948755736032	0.249209764216715\\
0.215648956106211	0.243416791860917\\
0.221810354852103	0.262342266660054\\
0.227971753597995	0.268305325669812\\
0.234133152343887	0.274771493630461\\
0.240294551089778	0.286368183281707\\
0.24645594983567	0.291404488842332\\
0.252617348581562	0.302793173465553\\
0.258778747327454	0.308097894766811\\
0.264940146073345	0.315659663614625\\
0.271101544819237	0.324326476088843\\
0.277262943565129	0.32604885264834\\
0.283424342311021	0.332392006945079\\
0.289585741056912	0.340950046158288\\
0.295747139802804	0.350647860611341\\
0.301908538548696	0.353544226514594\\
0.308069937294588	0.360212062696689\\
0.314231336040479	0.366953749898315\\
0.320392734786371	0.374931424964627\\
0.326554133532263	0.378577965213229\\
0.332715532278155	0.390016967301547\\
0.338876931024046	0.393982263145518\\
0.345038329769938	0.398268501586862\\
0.35119972851583	0.404752259834016\\
0.357361127261722	0.409198757356816\\
0.363522526007613	0.413863447521198\\
0.369683924753505	0.420352600201106\\
0.375845323499397	0.425294097964709\\
0.382006722245289	0.427167104848097\\
0.388168120991181	0.434429655248471\\
0.394329519737072	0.439315341525767\\
0.400490918482964	0.444220138166292\\
0.406652317228856	0.446868404105286\\
0.412813715974747	0.449614935170959\\
0.418975114720639	0.450919299741495\\
0.425136513466531	0.454250350183478\\
0.431297912212423	0.45691080402829\\
0.437459310958315	0.45876195131264\\
0.443620709704206	0.46265069863101\\
0.449782108450098	0.461875736292793\\
0.45594350719599	0.465466934583483\\
0.462104905941881	0.46235778451592\\
0.468266304687773	0.468306082855134\\
0.474427703433665	0.470667571242842\\
0.480589102179557	0.471872058659631\\
0.486750500925449	0.471749255208129\\
0.49291189967134	0.473020179798626\\
0.499073298417232	0.475561222840826\\
0.505234697163124	0.478201951180121\\
0.511396095909016	0.475726730997324\\
0.517557494654907	0.4784728585951\\
0.523718893400799	0.476076825593839\\
0.529880292146691	0.475160602242904\\
0.536041690892583	0.476617117908668\\
0.542203089638474	0.473126910128183\\
0.548364488384366	0.474632669611201\\
0.554525887130258	0.476154546023594\\
0.56068728587615	0.469702605029003\\
0.566848684622041	0.471870740747443\\
0.573010083367933	0.464565839276103\\
0.579171482113825	0.467044373841724\\
0.585332880859717	0.460941743832165\\
0.591494279605608	0.460598707845117\\
0.5976556783515	0.456253374525131\\
0.603817077097392	0.453494918233675\\
0.609978475843284	0.454833427118979\\
0.616139874589175	0.447952467277477\\
0.622301273335067	0.440375103070441\\
0.628462672080959	0.440948196806596\\
0.634624070826851	0.447839187651252\\
0.640785469572742	0.44203560573902\\
0.646946868318634	0.44045122408048\\
0.653108267064526	0.440576806079296\\
0.659269665810418	0.445368093344228\\
0.665431064556309	0.438382729184673\\
0.671592463302201	0.442546760739216\\
0.677753862048093	0.448537288378185\\
0.683915260793985	0.436495417765738\\
0.690076659539877	0.442244658560146\\
0.696238058285768	0.432356886475882\\
0.70239945703166	nan\\
0.708560855777552	nan\\
0.714722254523443	nan\\
0.720883653269335	nan\\
0.727045052015227	nan\\
0.733206450761119	nan\\
0.73936784950701	nan\\
0.745529248252902	nan\\
0.751690646998794	nan\\
0.757852045744686	nan\\
0.764013444490577	nan\\
0.770174843236469	nan\\
0.776336241982361	nan\\
0.782497640728253	nan\\
0.788659039474145	nan\\
0.794820438220036	nan\\
0.800981836965928	nan\\
0.80714323571182	nan\\
0.813304634457712	nan\\
0.819466033203603	nan\\
0.825627431949495	nan\\
0.831788830695387	nan\\
0.837950229441279	nan\\
0.84411162818717	nan\\
0.850273026933062	nan\\
};

\addplot [color=yellow, mark=square*, mark options={solid, fill=blue}, forget plot]
  table[row sep=crcr]{%
0.517557494654907	0.4784728585951\\
};
\addplot [color=blue, dashed, thick, forget plot]
  table[row sep=crcr]{%
0.443620709704206	0\\
0.443620709704206	0.46265069863101\\
};

\addplot[only marks, mark=o, mark options={}, mark size=2.000pt, draw=green] table[row sep=crcr]{%
x	y\\
0	nan\\
0.00533271136836316	nan\\
0.0106654227367263	nan\\
0.0159981341050895	nan\\
0.0213308454734527	nan\\
0.0266635568418158	nan\\
0.031996268210179	nan\\
0.0373289795785421	nan\\
0.0426616909469053	nan\\
0.0479944023152685	nan\\
0.0533271136836316	nan\\
0.0586598250519948	nan\\
0.063992536420358	nan\\
0.0693252477887211	nan\\
0.0746579591570843	nan\\
0.0799906705254474	nan\\
0.0853233818938106	nan\\
0.0906560932621738	nan\\
0.0959888046305369	0.0957696488378465\\
0.1013215159989	0.0907995917251491\\
0.106654227367263	0.101308129627258\\
0.111986938735626	0.110548401806402\\
0.11731965010399	0.121605641570516\\
0.122652361472353	0.131911342280139\\
0.127985072840716	0.142552294807345\\
0.133317784209079	0.176158973316653\\
0.138650495577442	0.183718736674851\\
0.143983206945805	0.194242264796816\\
0.149315918314169	0.194683011043643\\
0.154648629682532	0.1777610106826\\
0.159981341050895	0.186551687509508\\
0.165314052419258	0.19306573442092\\
0.170646763787621	0.203917804701746\\
0.175979475155984	0.208547997294721\\
0.181312186524348	0.21996304280358\\
0.186644897892711	0.229606442554181\\
0.191977609261074	0.234322135273429\\
0.197310320629437	0.24441973585014\\
0.2026430319978	0.249904570847882\\
0.207975743366163	0.256790028302936\\
0.213308454734527	0.267240999426515\\
0.21864116610289	0.2677408414348\\
0.223973877471253	0.278633137085801\\
0.229306588839616	0.283741508779393\\
0.234639300207979	0.294808541217233\\
0.239972011576342	0.298957082101896\\
0.245304722944706	0.307224941820648\\
0.250637434313069	0.314593610466168\\
0.255970145681432	0.31980399958366\\
0.261302857049795	0.327651656351007\\
0.266635568418158	0.332817702112883\\
0.271968279786521	0.338953799640719\\
0.277300991154884	0.345261704449135\\
0.282633702523248	0.352383308637877\\
0.287966413891611	0.357314884380974\\
0.293299125259974	0.361388965872382\\
0.298631836628337	0.367252279712041\\
0.3039645479967	0.374227128946204\\
0.309297259365063	0.378606741916259\\
0.314629970733427	0.385151648137416\\
0.31996268210179	0.388126743245677\\
0.325295393470153	0.394403665017585\\
0.330628104838516	0.397759798685426\\
0.335960816206879	0.399508429249957\\
0.341293527575242	0.403146803132755\\
0.346626238943606	0.40599891710345\\
0.351958950311969	0.40895650634685\\
0.357291661680332	0.41090747487403\\
0.362624373048695	0.41302644334165\\
0.367957084417058	0.41629151890002\\
0.373289795785421	0.420805804689676\\
0.378622507153785	0.423131484378833\\
0.383955218522148	0.425732242208984\\
0.389287929890511	0.427243411625945\\
0.394620641258874	0.423671500441593\\
0.399953352627237	0.426251423254517\\
0.4052860639956	0.430203521147246\\
0.410618775363964	0.431542682765656\\
0.415951486732327	0.433994770582546\\
0.42128419810069	0.436014022444296\\
0.426616909469053	0.438113566530821\\
0.431949620837416	0.441031806083707\\
0.437282332205779	0.452994963342674\\
0.442615043574143	0.454548969328946\\
0.447947754942506	0.453187237477543\\
0.453280466310869	0.456349955998452\\
0.458613177679232	0.455043249313517\\
0.463945889047595	0.46256589583016\\
0.469278600415958	0.454319530628527\\
0.474611311784322	0.441325501324146\\
0.479944023152685	0.440709170394481\\
0.485276734521048	0.442736240645638\\
0.490609445889411	0.446012317786625\\
0.495942157257774	0.4493258130301\\
0.501274868626137	0.4379039660275\\
0.5066075799945	0.438418448419292\\
0.511940291362864	0.426086320383163\\
0.517273002731227	0.433656366346868\\
0.52260571409959	0.422947279758432\\
0.527938425467953	0.426185072453698\\
0.533271136836316	0.417183665031211\\
0.53860384820468	0.410911005896977\\
0.543936559573043	0.410931955876217\\
0.549269270941406	0.414177962094909\\
0.554601982309769	0.396415716527949\\
0.559934693678132	0.399065788522457\\
0.565267405046495	0.388745082918813\\
0.570600116414859	0.382182149991351\\
0.575932827783222	0.37361983388194\\
0.581265539151585	0.360255944430793\\
0.586598250519948	0.343966612551488\\
0.591930961888311	0.336830280445495\\
0.597263673256674	0.330314167286144\\
0.602596384625038	0.323320302322129\\
0.607929095993401	0.309391742074131\\
0.613261807361764	0.297097265223361\\
0.618594518730127	0.286861858615152\\
0.62392723009849	0.282088946874359\\
0.629259941466853	0.278656031694571\\
0.634592652835216	0.271043930159466\\
0.63992536420358	0.262198262599505\\
0.645258075571943	0.257698009818221\\
0.650590786940306	0.253949089285327\\
0.655923498308669	0.24954263868073\\
0.661256209677032	0.241618311428375\\
0.666588921045395	0.233944684621903\\
0.671921632413759	0.228216780410424\\
0.677254343782122	0.228331360851079\\
};
\addplot [color=yellow, mark=*, mark options={solid, fill=green}, forget plot]
  table[row sep=crcr]{%
0.463945889047595	0.46256589583016\\
};
\addplot [color=green, dashed,  thick, forget plot]
  table[row sep=crcr]{%
0.463945889047595	0\\
0.463945889047595	0.46256589583016\\
};

\addplot[only marks, mark=triangle, mark options={}, mark size=2.0pt, draw=red] table[row sep=crcr]{%
x	y\\
0	nan\\
0.00470323938798851	nan\\
0.00940647877597702	nan\\
0.0141097181639655	nan\\
0.018812957551954	nan\\
0.0235161969399426	nan\\
0.0282194363279311	nan\\
0.0329226757159196	nan\\
0.0376259151039081	nan\\
0.0423291544918966	nan\\
0.0470323938798851	nan\\
0.0517356332678736	nan\\
0.0564388726558621	nan\\
0.0611421120438506	nan\\
0.0658453514318392	nan\\
0.0705485908198277	nan\\
0.0752518302078162	0.0698392390842567\\
0.0799550695958047	0.0783894933902302\\
0.0846583089837932	0.0642132997400362\\
0.0893615483717817	0.0653443101668279\\
0.0940647877597702	0.0763203186153922\\
0.0987680271477587	0.082431511062903\\
0.103471266535747	0.0882811167540324\\
0.108174505923736	0.0965047743390638\\
0.112877745311724	0.132295007240825\\
0.117580984699713	0.113668108759003\\
0.122284224087701	0.147037320831942\\
0.12698746347569	0.130960505348711\\
0.131690702863678	0.138439820453541\\
0.136393942251667	0.1443394489831\\
0.141097181639655	0.152271313882147\\
0.145800421027644	0.184868358779471\\
0.150503660415632	0.166876429066445\\
0.155206899803621	0.169432537782061\\
0.159910139191609	0.180003716379366\\
0.164613378579598	0.18684668036697\\
0.169316617967586	0.195304868608413\\
0.174019857355575	0.222641121369488\\
0.178723096743563	0.194812502758452\\
0.183426336131552	0.201913149605724\\
0.18812957551954	0.209984216142998\\
0.192832814907529	0.230497750589745\\
0.197536054295517	0.236221853978027\\
0.202239293683506	0.242972080473201\\
0.206942533071494	0.248914177217635\\
0.211645772459483	0.259081833530344\\
0.216349011847472	0.261592851690104\\
0.22105225123546	0.269993641716446\\
0.225755490623449	0.271045421236792\\
0.230458730011437	0.282347438157198\\
0.235161969399426	0.282257498426927\\
0.239865208787414	0.290099614586789\\
0.244568448175403	0.295102741183383\\
0.249271687563391	0.298900193757684\\
0.25397492695138	0.303556486667858\\
0.258678166339368	0.311523620208233\\
0.263381405727357	0.314994196611776\\
0.268084645115345	0.322930095819251\\
0.272787884503334	0.326069497207928\\
0.277491123891322	0.332084719483057\\
0.282194363279311	0.337579984418698\\
0.286897602667299	0.345609521745331\\
0.291600842055288	0.347368146952837\\
0.296304081443276	0.354194271626918\\
0.301007320831265	0.36111336522601\\
0.305710560219253	0.368371755006652\\
0.310413799607242	0.369314835216451\\
0.31511703899523	0.378182106733034\\
0.319820278383219	0.384831886719952\\
0.324523517771207	0.389341198772245\\
0.329226757159196	0.395109504804206\\
0.333929996547184	0.400098184322393\\
0.338633235935173	0.40702773300746\\
0.343336475323161	0.407919455090373\\
0.34803971471115	0.412080069311299\\
0.352742954099138	0.415452882681146\\
0.357446193487127	0.420157636733087\\
0.362149432875115	0.424506223316372\\
0.366852672263104	0.42608141562247\\
0.371555911651092	0.427724540774069\\
0.376259151039081	0.434229687143288\\
0.380962390427069	0.439045324169293\\
0.385665629815058	0.44084905716375\\
0.390368869203046	0.443420190746283\\
0.395072108591035	0.44600345253451\\
};

\addplot[only marks, mark=triangle, mark options={}, mark size=2.0pt, draw=black!30!white] table[row sep=crcr]{%
x	y\\
0.399775347979023	0.451522117652757\\
0.404478587367012	0.453478611744077\\
0.409181826755	0.454146375092809\\
0.413885066142989	0.454821971939048\\
0.418588305530977	0.461830749061784\\
0.423291544918966	0.46254098796141\\
0.427994784306954	0.468222773322297\\
0.432698023694943	0.474165193316738\\
0.437401263082932	0.474927820700159\\
0.44210450247092	0.480250705195301\\
0.446807741858908	0.4826620073931\\
0.451510981246897	0.483481477813975\\
0.456214220634886	0.488143192275056\\
0.460917460022874	0.495764767654211\\
0.465620699410863	0.496635324070409\\
0.470323938798851	0.498399911890355\\
0.47502717818684	0.504128549144707\\
0.479730417574828	0.508073014118643\\
0.484433656962817	0.509935682065559\\
0.489136896350805	0.513909384432384\\
0.493840135738794	0.516787474516216\\
0.498543375126782	0.517762282881775\\
0.503246614514771	0.518744822179426\\
0.507949853902759	0.520733111234337\\
0.512653093290748	0.519630016597447\\
0.517356332678736	0.520642114207147\\
0.522059572066725	0.525720864248948\\
0.526762811454713	0.529790944357755\\
0.531466050842702	0.529754132754441\\
0.53616929023069	0.530834699675503\\
0.540872529618679	0.532945123338544\\
0.545575769006667	0.534030031648596\\
0.550279008394656	0.52885004306796\\
0.554982247782644	0.529922628242042\\
0.559685487170633	0.529922628242042\\
0.564388726558621	0.52885004306796\\
0.56909196594661	0.524636438551172\\
0.573795205334598	0.521486966631517\\
0.578498444722587	0.511411954895476\\
0.583201684110575	0.510418651644996\\
0.587904923498564	0.510418651644996\\
0.592608162886552	0.512412999863904\\
0.597311402274541	0.510323322346383\\
0.602014641662529	0.512345793893857\\
0.606717881050518	0.513368625947185\\
0.611421120438506	0.513343941843414\\
0.616124359826495	0.510369406131335\\
0.620827599214483	0.510369406131335\\
0.625530838602472	0.510369406131335\\
0.63023407799046	0.510369406131335\\
0.634937317378449	0.509316537976588\\
0.639640556766437	0.509316537976588\\
0.644343796154426	0.505313948390903\\
0.649047035542414	0.505313948390903\\
0.653750274930403	0.504300449525982\\
0.658453514318391	0.504300449525982\\
0.66315675370638	0.49819995453465\\
};

\addplot [color=yellow, mark=triangle*, mark options={solid, fill=black!30!white}, forget plot]
  table[row sep=crcr]{%
0.395072108591035	0.44600345253451\\
};

\addplot [color=red, dashed,  thick, forget plot]
  table[row sep=crcr]{%
0.545575769006667	0\\
0.545575769006667	0.534030031648596\\
};
\addplot [draw=yellow, mark=triangle*, mark options={solid, fill=red}]
  table[row sep=crcr]{%
0.545575769006667	0.534030031648596\\
};
\addplot [color=black!30!white, dashed,  thick, forget plot ]
  table[row sep=crcr]{%
0.395072108591035	0\\
0.395072108591035	0.44600345253451\\
};

\node[] at (0.635,0.04) {\tcapHmax{}};

\end{axis}
\end{tikzpicture}%

%% file: Figures/MaxSurfNrj_FW_KinEn_Wed_delta_1_withFit.tex
\begin{tikzpicture}[baseline]

\begin{axis}[%
width=0.245\linewidth,
height=0.245\linewidth,
scale only axis,
xmin=200,
xmax=400,
xlabel style={font=\color{white!15!black}},
xlabel={\dropWe{} (-)},
ymin=0,
ymax=1,
ylabel style={font=\color{white!15!black}},
ylabel={\CSEmax{}/\DKE{} $(-)$},
axis background/.style={fill=white},
title style={font=\bfseries},
title={$\text{\wft{}} \approx{} 0.12$},
xmajorgrids,
ymajorgrids,
legend style={at={(0.98,0.98)}, anchor=north east, legend cell align=left, align=left, draw=white!15!black},
clip mode=individual
]
\node at (220,0.93) {\textbf{(b)}};

\addplot [color=black, mark=square*, mark options={solid, fill=blue}]
  table[row sep=crcr]{%
508.471029883192	0.368398257376051\\
508.471029883192	0.368398257376051\\
};

\addplot [color=black, mark=*, mark options={solid, fill=green}]
  table[row sep=crcr]{%
508.48504088168	0.363350328546628\\
508.48504088168	0.363350328546628\\
};

\addplot [color=black, mark=triangle*, mark options={solid, fill=red}]
  table[row sep=crcr]{%
498.867960163284	0.4115101751182\\
498.867960163284	0.341421872230432\\
};
\addplot [draw=black, mark=triangle*, mark options={solid, fill=black!30!white}, only marks]
  table[row sep=crcr]{%
498.867960163284	0.341421872230432\\
};

\addplot [color=black, mark=square*, mark options={solid, fill=blue}, forget plot]
  table[row sep=crcr]{%
440.840954053392	0.344967582323257\\
440.840954053392	0.344967582323257\\
};
\addplot [color=black, mark=*, mark options={solid, fill=green}, forget plot]
  table[row sep=crcr]{%
445.036209274399	0.358799910694997\\
445.036209274399	0.358799910694997\\
};
\addplot [color=black, mark=triangle*, mark options={solid, fill=red}, forget plot]
  table[row sep=crcr]{%
434.940594142564	0.418030566787486\\
434.940594142564	0.358007152906086\\
};
\addplot [draw=black, mark=triangle*, mark options={solid, fill=black!30!white}, only marks]
  table[row sep=crcr]{%
434.940594142564	0.358007152906086\\
};

\addplot [color=black, mark=square*, mark options={solid, fill=blue}, forget plot]
  table[row sep=crcr]{%
358.310179270558	0.4784728585951\\
358.310179270558	0.4784728585951\\
};
\addplot [color=black, mark=*, mark options={solid, fill=green}, forget plot]
  table[row sep=crcr]{%
368.354660425129	0.46256589583016\\
368.354660425129	0.46256589583016\\
};
\addplot [color=black, mark=triangle*, mark options={solid, fill=red}, forget plot]
  table[row sep=crcr]{%
362.804222243713	0.534030031648596\\
362.804222243713	0.44600345253451\\
};
\addplot [draw=black, mark=triangle*, mark options={solid, fill=black!30!white}, only marks]
  table[row sep=crcr]{%
362.804222243713	0.44600345253451\\
};

\addplot [color=black, mark=square*, mark options={solid, fill=blue}, forget plot]
  table[row sep=crcr]{%
314.864350862924	0.462341848235315\\
314.864350862924	0.462341848235315\\
};
\addplot [color=black, mark=*, mark options={solid, fill=green}, forget plot]
  table[row sep=crcr]{%
320.545648797683	0.462646470304137\\
320.545648797683	0.462646470304137\\
};
\addplot [color=black, mark=triangle*, mark options={solid, fill=red}, forget plot]
  table[row sep=crcr]{%
316.28831821484	0.564733598546298\\
316.28831821484	0.457039859009651\\
};
\addplot [draw=black, mark=triangle*, mark options={solid, fill=black!30!white}, only marks]
  table[row sep=crcr]{%
316.28831821484	0.457039859009651\\
};

\addplot [color=black, mark=square*, mark options={solid, fill=blue}, forget plot]
  table[row sep=crcr]{%
266.853404078276	0.450345912510613\\
266.853404078276	0.450345912510613\\
};
\addplot [color=black, mark=*, mark options={solid, fill=green}, forget plot]
  table[row sep=crcr]{%
271.76583073723	0.481010305008296\\
271.76583073723	0.481010305008296\\
};
\addplot [color=black, mark=triangle*, mark options={solid, fill=red}, forget plot]
  table[row sep=crcr]{%
269.170235106682	0.51435804539953\\
269.170235106682	0.51435804539953\\
};
\addplot [color=black, mark=square*, mark options={solid, fill=blue}, forget plot]
  table[row sep=crcr]{%
220.427335475403	0.553777592538329\\
220.427335475403	0.553777592538329\\
};
\addplot [color=black, mark=*, mark options={solid, fill=green}, forget plot]
  table[row sep=crcr]{%
215.741494675615	0.498895644580862\\
215.741494675615	0.498895644580862\\
};
\addplot [color=black, mark=triangle*, mark options={solid, fill=red}, forget plot]
  table[row sep=crcr]{%
218.583938591942	0.498884850755158\\
218.583938591942	0.498884850755158\\
};
\end{axis}
\end{tikzpicture}%

%% file: Figures/TimeHmax_We_delta_1_4_tex.tex
\begin{tikzpicture}[baseline]

\begin{axis}[%
width=0.245\linewidth,
height=0.245\linewidth,
scale only axis,
xmin=200,
xmax=400,
xlabel style={font=\color{white!15!black}},
xlabel={\dropWe{} (-)},
ymin=0,
ymax=0.6,
ylabel style={font=\color{white!15!black}},
ylabel={\tcapHmax{} (-)},
axis background/.style={fill=white},
title style={font=\bfseries},
title={$\text{\wft{}} \approx{} 0.12$},
xmajorgrids,
ymajorgrids,
legend style={at={(0.98,0.02)}, anchor=south east, legend cell align=left, align=left, draw=white!15!black},
clip mode=individual
]
\node at (220,0.55) {\textbf{(c)}};
\addplot [color=black, mark=square*, mark options={solid, fill=blue}, only marks]
  table[row sep=crcr]{%
508.471029883192	0.365728213220436\\
508.471029883192	0.365728213220436\\
};
\addlegendentry{\Hexa{}}

\addplot [color=black, mark=*, mark options={solid, fill=green}, only marks]
  table[row sep=crcr]{%
508.48504088168	0.324605987975313\\
508.48504088168	0.324605987975313\\
};
\addlegendentry{\SO{}}

\addplot [color=black, mark=triangle*, mark options={solid, fill=red}, only marks]
  table[row sep=crcr]{%
217.94292997437	0.348889416910854\\
217.94292997437	0.348889416910854\\
};
\addlegendentry{\EtOh{}}

\addplot [color=black, mark=triangle*, mark options={solid, fill=red}]
  table[row sep=crcr]{%
498.867960163284	0.362627000746312\\
498.867960163284	0.265336829814375\\
};
\addplot [draw=black, mark=triangle*, mark options={solid, fill=black!30!white}, only marks]
  table[row sep=crcr]{%
498.867960163284	0.265336829814375\\
};

\addplot [color=black, mark=square*, mark options={solid, fill=blue}, forget plot]
  table[row sep=crcr]{%
440.840954053392	0.254973381476396\\
440.840954053392	0.254973381476396\\
};
\addplot [color=black, mark=*, mark options={solid, fill=green}, forget plot]
  table[row sep=crcr]{%
445.036209274399	0.324174685003228\\
445.036209274399	0.324174685003228\\
};
\addplot [color=black, mark=triangle*, mark options={solid, fill=red}, forget plot]
  table[row sep=crcr]{%
434.940594142564	0.382255414574836\\
434.940594142564	0.287815841562229\\
};
\addplot [draw=black, mark=triangle*, mark options={solid, fill=black!30!white}, only marks]
  table[row sep=crcr]{%
434.940594142564	0.287815841562229\\
};

\addplot [color=black, mark=square*, mark options={solid, fill=blue}, forget plot]
  table[row sep=crcr]{%
358.310179270558	0.443620709704206\\
358.310179270558	0.443620709704206\\
};
\addplot [color=black, mark=*, mark options={solid, fill=green}, forget plot]
  table[row sep=crcr]{%
368.354660425129	0.463945889047595\\
368.354660425129	0.463945889047595\\
};
\addplot [color=black, mark=triangle*, mark options={solid, fill=red}, forget plot]
  table[row sep=crcr]{%
362.804222243713	0.545575769006667\\
362.804222243713	0.395072108591035\\
};
\addplot [draw=black, mark=triangle*, mark options={solid, fill=black!30!white}, only marks]
  table[row sep=crcr]{%
362.804222243713	0.395072108591035\\
};

\addplot [color=black, mark=square*, mark options={solid, fill=blue}, forget plot]
  table[row sep=crcr]{%
314.864350862924	0.410107802561169\\
314.864350862924	0.410107802561169\\
};
\addplot [color=black, mark=*, mark options={solid, fill=green}, forget plot]
  table[row sep=crcr]{%
320.545648797683	0.357932396932754\\
320.545648797683	0.357932396932754\\
};
\addplot [color=black, mark=triangle*, mark options={solid, fill=red}, forget plot]
  table[row sep=crcr]{%
316.28831821484	0.541615272661636\\
316.28831821484	0.364189579893169\\
};
\addplot [draw=black, mark=triangle*, mark options={solid, fill=black!30!white}, only marks]
  table[row sep=crcr]{%
316.28831821484	0.364189579893169\\
};

\addplot [color=black, mark=square*, mark options={solid, fill=blue}, forget plot]
  table[row sep=crcr]{%
266.853404078276	0.420581699363483\\
266.853404078276	0.420581699363483\\
};
\addplot [color=black, mark=*, mark options={solid, fill=green}, forget plot]
  table[row sep=crcr]{%
271.76583073723	0.399767578345252\\
271.76583073723	0.399767578345252\\
};
\addplot [color=black, mark=triangle*, mark options={solid, fill=red}, forget plot]
  table[row sep=crcr]{%
269.170235106682	0.397111929934686\\
269.170235106682	0.397111929934686\\
};
\addplot [color=black, mark=square*, mark options={solid, fill=blue}, forget plot]
  table[row sep=crcr]{%
220.427335475403	0.368915078919916\\
220.427335475403	0.368915078919916\\
};
\addplot [color=black, mark=*, mark options={solid, fill=green}, forget plot]
  table[row sep=crcr]{%
215.741494675615	0.352614343079758\\
215.741494675615	0.352614343079758\\
};
\addplot [color=black, mark=triangle*, mark options={solid, fill=red}, forget plot]
  table[row sep=crcr]{%
218.583938591942	0.348155258357597\\
218.583938591942	0.348155258357597\\
};
\end{axis}
\end{tikzpicture}%

%% file: IFT_spring.tex
		\subsection{Recoiling force}

Droplet impact processes can be considered as oscillating mass-spring systems \citep{Planchette2017, Okumura2003}, where surface tension is the recoiling force opposing the deformation induced by the impacting droplet. Hence, the droplet and the wall-film could be seen as springs in parallel whose spring constants are given by the surface and interfacial tensions. Following this reasoning, the system during impact could be represented by an equivalent spring constant equal to the sum of each, i.e. $\sigma_d+\sigma_f+\sigma_{d/f}$. Considering the crown as an inertio-capillary system, the capillary time scales of the crown ascending phase (i.e. increase of \Hrim{}) can be compared for the three liquid pairs to see whether the trend between the liquid pairs observed in Fig.~\ref{fig:CrownSurfTime}(c) is related to the differences in interfacial tensions. Thus,  we propose to evaluate the capillary time based on the equivalent spring constant as:

\begin{equation}
  \text{\tcaptime{}}=\frac{t}{\text{\tcap{}}}=\frac{t}{\sqrt{\frac{\text{\md{}}+\text{\mf{}}}{\sigma_d+\sigma_{d/f}+\sigma_f}}}
\label{eq:captime}
\end{equation}

The characteristic mass of the droplet/wall-film system entering \tcap{} is most likely a combination of droplet and wall-film masses as (\md{}+\mf{}). While the characteristic mass of the droplet can be easily calculated before impact as $\text{\md{}}=\text{\rhod}  \pi \text{\dropdia{}}^3/6$, the determination of wall-film mass \mf{} is more difficult. By assuming an equal volume of droplet and wall-film liquid participating to the oscillation process, it comes $\text{\md{}}+\text{\mf{}}=\left(\text{\rhod}+\text{\rhof} \right) \pi \text{\dropdia{}}^3/6$. This rough estimation has the advantage of being determined with pre-impact parameters.
In Fig.~\ref{fig:CrownSurfNrj-time}(a), the time is non-dimensionalized with \tcap{} to form \tcaptime{}. Compared to \tHmax{} in Fig.~\ref{fig:CrownSurfTime}(a), we observe that \tcapHmax{} (the non-dimensional time at which \Hrimmax{} is reached) are coming together. 
Indeed, the differences of \sigmadf{} in \tcap{} compensate the differences in extension duration for each liquid pair, similar to the compensating effect for \CSEmax{} in Fig.~\ref{fig:CrownSurfNrj-time}(b). 
Hence, the interfacial tension is responsible for the different durations of ascending phases. This means that, for immiscible liquid pairs, the interface between droplet and wall-film acts as a non-negligible force preventing the extension, proportionally to \sigmadfstar{}.
This scaling leads to a unified temporal evolution for the three liquid pairs of the normalized crown surface energy with the capillary time during the ascending phase at a given impact condition.
The role of interfacial tension is observed for all impact conditions investigated in our study as shown in Fig.~\ref{fig:CrownSurfNrj-time}(c), all data points coming together at a given \dropWe{}. \tcapHmax{} is lying around 0.4, similar to what has been also observed for much lower Weber numbers (below 10) for the droplet oscillation alone impacting a shallow and deep pool \citep{Tang2019}.

The values of \tcapHmax{} are slightly increasing with increasing \dropWe{}. 
This indicates that \tcaptime{} can not properly capture the process for all impact conditions, but only highlights the influence of the interfacial tension during the extension.
The consideration of the inertial time \TimeHmax{}=\tHmax{}\dropvel{}/\dropdia{} would lead to an even stronger dependency on the Weber number since \dropWe{} increases with \dropvel{}. Furthermore, it would not bring the experiments together at a given \dropWe{} since no surface tensions are considered. A viscous time as $\text{\dropdia{}}^2/\nu$ would also let the ranking of \tHmax{} and the trend with \dropWe{} unchanged given the similar kinematic viscosities involved for all impact conditions (see Table \ref{tab:prop}).
The extension of the experimental database to $\text{\wft{}} \approx{} 0.26$ represented in Fig.~\ref{fig:timecap_delta2} shows a similar data reduction of the ascending phase durations between the different liquid pairs at a given Weber number. However, the values are now higher compared to $\text{\wft{}} \approx{} 0.12$, lying around 0.6. Furthermore, the trend with increasing Weber number becomes weaker compared to $\text{\wft{}} \approx{} 0.12$, exhibiting almost constant values.

\begin{figure}
\centering
\input{Figures/TimeHmax_B2W_We_time1delta2.tex}
\input{Figures/TimeHmax_B2W_We_time4delta2.tex}
  \caption{(a) Time duration \tHmax{} of the crown ascending phase at different Weber numbers \dropWe{}. (b) Capillary time duration \tcapHmax{} of the crown ascending phase at different Weber numbers \dropWe{}. The dimensionless wall-film thickness is kept constant at \text{\wft}=\SI[parse-numbers=false]{0.259\pm0.008} for the data of this figure. 
}
\label{fig:timecap_delta2}
\end{figure}

These discrepancies between different Weber numbers and with higher wall-film thicknesses suggest a variation of the characteristic mass used in \tcap{} between these different experiments, since the variations of crown surfaces and with it, of the recoiling forces are already taken into account. 
Indeed, the kinetic energy of the droplet influences the droplet penetration inside the film. This could lead to an increased interacting wall-film mass, which is not captured with the current capillary scaling assuming an equal amount of droplet and wall-film liquids. Similarly, larger wall-film thicknesses increase the available amount of wall-film liquid that could enter the crown. Hence, the estimation of the interacting mass of wall-film needs to be refined to get a deeper insight into the oscillating behaviour.
Therefore, the maximal interacting wall-film mass \mf{} can be estimated by considering the mass contained in a cylinder of radius \Rbase{} at \tHmax{} and of height equal to the difference of the wall-film height \filmthick{} with the residual film thickness \filmres{}, as $\text{\mf{}}=\pi \text{\Rbase{}}\left(\text{\tHmax{}}\right)^2 \left(\text{\filmthick{}} - \text{\filmres{}} \right)$.
The residual thickness corresponds to the thickness of the wall-film which is not set in motion by the droplet impact. It can be approximated with $\text{\filmres{}} \approx (0.098\text{\wft{}}^{4.0413}+0.79)\text{\Re}^{-2/5}$ \citep{vanHinsberg2010}. Hence, the residual thickness increases with increasing initial wall-film thickness, and decreases with increasing Reynolds number due to a deeper penetration of the droplet in the wall-film. In the present database, the Reynolds number can be approximated with averaged liquid properties as introduced in the section on the experimental range, corresponding to \avgRed.
For the value of $\text{\Rbase{}}\left(\text{\tHmax{}}\right)$, it has now the drawback that it should be extracted post-impact. For predicting purposes however, one could use available theoretical models predicting \Rbase{} in the literature, e.g. the one of \cite{Roisman2008} predicting the maximum value of \Rbase{} during the impact process in function of \dropWe{}, \wft{} and the Froude number \dropFr{}.
The experimental data scaled with the modified capillary time scale \tcapHmaxmod{} based on the maximum mass of the wall-film are given in Fig.~\ref{fig:timecap_modified_delta1-delta2} for $\text{\wft{}} \approx{} 0.12$ (a) and $\text{\wft{}} \approx{} 0.26$ (b). It can be seen that the values are now comparable for both wall-film thicknesses. Hence, the modified scaling captures the oscillating dynamics for different wall-film thickness, due to the increase of the characteristic mass.
This correction also enables to reduce the dependency with \dropWe{} for $\text{\wft{}} \approx{} 0.1$. Yet, a slight slowdown of the crown dynamics with increasing \dropWe{} can still be observed. Hence, additional effects might be responsible for this remaining trend.

Considering a classical oscillating mass-spring system, the damping can increase the inviscid oscillating period \tcap{} as $\text{\tcapvisc{}} = 2 \pi / \left(\text{\omegacap}\sqrt{1-\text{\dampratio}^2}\right)$, where \omegacap{} is the inviscid pulsation and \dampratio{} the damping ratio. Applied to oscillating droplets during an impact, the damping ratio can be expressed as $\text{\dampratio{}} \approx{} 32\nu \text{\tcap{}}/(2\pi\text{\dropdia{}}^2)$ by approximating a damping coefficient of $32\nu/\text{\dropdia{}}^2$ \citep{Tang2019}.
By replacing \tcap{} in \dampratio{}, the damping ratio can be re-expressed with the impact parameters of the present study in terms of a modified Ohnesorge number as $\text{\dampratio{}}=16\pi\text{\Oh}/9$, with $\text{\Oh}=\nu{}\sqrt{(\text{\rhod{}}+\text{\rhof{}})/\left(2\text{\dropdia{}}(\text{\sigmad{}}+\text{\sigmaf{}}+\text{\sigmadf{}})\right)}$. The maximum value of \Oh{} in the present database (considering averaged kinematic viscosity as $\sqrt{\text{\nud\nuf}}$ suitable for binary droplet/wall-film system \citep{Bernard2020b}) is less than 0.015, leading to $\text{\dampratio{}}=0.084$, and an increase of \tcap{} of less than 0.5\%. Hence, the damping due to viscosity alone is negligible. Note however that this damping does not take shear into account during the extension, which has been shown to play a significant role for the crown extension dynamics at the base \citep{Geppert2020} and at the rim \citep{Bernard2020b}. This viscous losses due to shear might increase the damping, and with it the oscillation period when shear increases. Since shear losses are expected to increase with increasing \dropWe{} as highlighted by the increasing trend of $1-\text{\CSEmax{}}/\text{\DKE{}}$ in Fig.~\ref{fig:CrownSurfNrj-time}(b), this could explain the increasing trend of \tcapHmax{} with growing \dropWe{}. This also corroborates that this trend is weaker for \wft{}=0.2 where shear is smaller because the influence of the solid wall is reduced.

Another mechanism that could lead to the increase of \tcapHmax{} with growing \dropWe{} is the amplitude of the oscillation. Since droplet/wall-film systems are characterised by large extensions, the isochronism of small oscillations might be altered. In this case, the oscillation period increases with increasing amplitude \citep{Fulcher1976} which corresponds in our case to the Weber number.

\begin{figure}
\centering
\input{Figures/TimeHmax_B2W_We_time7delta1.tex}
\input{Figures/TimeHmax_B2W_We_time7delta2.tex}
  \caption{(a) Time duration \tHmax{} of the crown ascending phase at different Weber numbers \dropWe{}. (b) Capillary time duration \tcapHmax{} of the crown ascending phase at different Weber numbers \dropWe{}. The dimensionless wall-film thickness is kept constant at \text{\wft}=\SI[parse-numbers=false]{0.259\pm0.008} for the entire database.}
\label{fig:timecap_modified_delta1-delta2}
\end{figure}

To conclude, the interface (if any) between the droplet and wall-film  brings a supplementary recoiling force during the extension.
Similar to a mass-spring oscillating system, the interfacial tension influences the oscillation behaviour of immiscible droplet/wall-film systems. Furthermore, the impact configuration (especially the wall-film thickness and the droplet kinetic energy) changes the characteristic mass of the system by modifying the mass of wall-film interacting in the crown, and the influence of shear, which might influence the damping significantly.

%% file: Figures/TimeHmax_B2W_We_time1delta2.tex
\begin{tikzpicture}

\begin{axis}[%
width=0.245\linewidth,
height=0.245\linewidth,
scale only axis,
xmin=200,
xmax=400,
xlabel style={font=\color{white!15!black}},
xlabel={\dropWe{} (-)},
ymin=0,
ymax=8,
ylabel style={font=\color{white!15!black}},
ylabel={\tHmax{} $\left( \si{\milli\second} \right)$},
axis background/.style={fill=white},
title style={font=\bfseries},
xmajorgrids,
ymajorgrids,
legend style={at={(0.98,0.02)}, anchor=south east, legend cell align=left, align=left, draw=white!15!black},
clip mode=individual,
title={$\text{\wft{}} \approx{} 0.26$}
]

\node at (220,7.5) {\textbf{(a)}};

\addplot [color=black, mark=square*, mark options={solid, fill=blue}]
  table[row sep=crcr]{%
502.006156921766	4.5\\
502.006156921766	4.5\\
};

\addplot [color=black, mark=*, mark options={solid, fill=green}]
  table[row sep=crcr]{%
501.547976394186	4.6\\
501.547976394186	4.6\\
};

\addplot [color=black, mark=triangle*, mark options={solid, fill=red}]
  table[row sep=crcr]{%
495.658218686723	8.05\\
495.658218686723	4.55\\
};

\addplot [color=black, mark=square*, mark options={solid, fill=blue}, forget plot]
  table[row sep=crcr]{%
434.268229939372	4.5\\
434.268229939372	4.5\\
};
\addplot [color=black, mark=*, mark options={solid, fill=green}, forget plot]
  table[row sep=crcr]{%
434.063366381827	4.75\\
434.063366381827	4.75\\
};
\addplot [color=black, mark=triangle*, mark options={solid, fill=red}, forget plot]
  table[row sep=crcr]{%
434.092265396643	8.1\\
434.092265396643	5.7\\
};
\addplot [color=black, mark=square*, mark options={solid, fill=blue}, forget plot]
  table[row sep=crcr]{%
363.494692562247	5.3\\
363.494692562247	5.3\\
};
\addplot [color=black, mark=*, mark options={solid, fill=green}, forget plot]
  table[row sep=crcr]{%
369.324643699557	5.75\\
369.324643699557	5.75\\
};
\addplot [color=black, mark=triangle*, mark options={solid, fill=red}, forget plot]
  table[row sep=crcr]{%
355.919634927362	6.95\\
355.919634927362	6.95\\
};
\addplot [color=black, mark=square*, mark options={solid, fill=blue}, forget plot]
  table[row sep=crcr]{%
321.655135921171	5.55\\
321.655135921171	5.55\\
};
\addplot [color=black, mark=*, mark options={solid, fill=green}, forget plot]
  table[row sep=crcr]{%
318.321194011279	6.05\\
318.321194011279	6.05\\
};
\addplot [color=black, mark=triangle*, mark options={solid, fill=red}, forget plot]
  table[row sep=crcr]{%
318.36112148599	7.35\\
318.36112148599	7.35\\
};
\addplot [color=black, mark=square*, mark options={solid, fill=blue}, forget plot]
  table[row sep=crcr]{%
267.654800883446	5\\
267.654800883446	5\\
};
\addplot [color=black, mark=*, mark options={solid, fill=green}, forget plot]
  table[row sep=crcr]{%
269.636317537986	5.5\\
269.636317537986	5.5\\
};
\addplot [color=black, mark=triangle*, mark options={solid, fill=red}, forget plot]
  table[row sep=crcr]{%
266.494273655914	6.45\\
266.494273655914	6.45\\
};
\addplot [color=black, mark=square*, mark options={solid, fill=blue}, forget plot]
  table[row sep=crcr]{%
216.514512112825	5.1\\
216.514512112825	5.1\\
};
\addplot [color=black, mark=*, mark options={solid, fill=green}, forget plot]
  table[row sep=crcr]{%
216.660937992126	5.8\\
216.660937992126	5.8\\
};
\addplot [color=black, mark=triangle*, mark options={solid, fill=red}, forget plot]
  table[row sep=crcr]{%
217.038084289605	6.2\\
217.038084289605	6.2\\
};
\end{axis}
\end{tikzpicture}%

%% file: Figures/TimeHmax_B2W_We_time4delta2.tex
\begin{tikzpicture}

\begin{axis}[%
width=0.245\linewidth,
height=0.245\linewidth,
scale only axis,
xmin=200,
xmax=400,
xlabel style={font=\color{white!15!black}},
xlabel={\dropWe{} (-)},
ymin=0,
ymax=1,
ylabel style={font=\color{white!15!black}},
ylabel={\tcapHmax{} (-)},
axis background/.style={fill=white},
title style={font=\bfseries},
xmajorgrids,
ymajorgrids,
legend style={at={(0.98,0.02)}, anchor=south east, legend cell align=left, align=left, draw=white!15!black},
clip mode=individual,
title={$\text{\wft{}} \approx{} 0.26$}
]

\node at (220,0.9) {\textbf{(b)}};
\addplot [color=black, mark=square*, mark options={solid, fill=blue}, only marks]
  table[row sep=crcr]{%
502.006156921766	0.510585288430639\\
502.006156921766	0.510585288430639\\
};
\addlegendentry{\Hexa{}}

\addplot [color=black, mark=*, mark options={solid, fill=green}, only marks]
  table[row sep=crcr]{%
501.547976394186	0.46509660394855\\
501.547976394186	0.46509660394855\\
};
\addlegendentry{\SO{}}

\addplot [color=black, mark=triangle*, mark options={solid, fill=red}, only marks]
  table[row sep=crcr]{%
495.658218686723	0.718498378688667\\
495.658218686723	0.406107779258812\\
};
\addlegendentry{\EtOh{}}

\addplot [color=black, mark=square*, mark options={solid, fill=blue}, forget plot]
  table[row sep=crcr]{%
434.268229939372	0.527487524703446\\
434.268229939372	0.527487524703446\\
};
\addplot [color=black, mark=*, mark options={solid, fill=green}, forget plot]
  table[row sep=crcr]{%
434.063366381827	0.496429729041294\\
434.063366381827	0.496429729041294\\
};
\addplot [color=black, mark=triangle*, mark options={solid, fill=red}, forget plot]
  table[row sep=crcr]{%
434.092265396643	0.730293529405888\\
434.092265396643	0.513910261433773\\
};
\addplot [color=black, mark=square*, mark options={solid, fill=blue}, forget plot]
  table[row sep=crcr]{%
363.494692562247	0.643712418220729\\
363.494692562247	0.643712418220729\\
};
\addplot [color=black, mark=*, mark options={solid, fill=green}, forget plot]
  table[row sep=crcr]{%
369.324643699557	0.607944986841861\\
369.324643699557	0.607944986841861\\
};
\addplot [color=black, mark=triangle*, mark options={solid, fill=red}, forget plot]
  table[row sep=crcr]{%
355.919634927362	0.666471080647916\\
355.919634927362	0.666471080647916\\
};
\addplot [color=black, mark=square*, mark options={solid, fill=blue}, forget plot]
  table[row sep=crcr]{%
321.655135921171	0.670175122744696\\
321.655135921171	0.670175122744696\\
};
\addplot [color=black, mark=*, mark options={solid, fill=green}, forget plot]
  table[row sep=crcr]{%
318.321194011279	0.650849163499668\\
318.321194011279	0.650849163499668\\
};
\addplot [color=black, mark=triangle*, mark options={solid, fill=red}, forget plot]
  table[row sep=crcr]{%
318.36112148599	0.69429691162914\\
318.36112148599	0.69429691162914\\
};
\addplot [color=black, mark=square*, mark options={solid, fill=blue}, forget plot]
  table[row sep=crcr]{%
267.654800883446	0.599765448084762\\
267.654800883446	0.599765448084762\\
};
\addplot [color=black, mark=*, mark options={solid, fill=green}, forget plot]
  table[row sep=crcr]{%
269.636317537986	0.580004108312575\\
269.636317537986	0.580004108312575\\
};
\addplot [color=black, mark=triangle*, mark options={solid, fill=red}, forget plot]
  table[row sep=crcr]{%
266.494273655914	0.600934034209861\\
266.494273655914	0.600934034209861\\
};
\addplot [color=black, mark=square*, mark options={solid, fill=blue}, forget plot]
  table[row sep=crcr]{%
216.514512112825	0.608160359207386\\
216.514512112825	0.608160359207386\\
};
\addplot [color=black, mark=*, mark options={solid, fill=green}, forget plot]
  table[row sep=crcr]{%
216.660937992126	0.607376044335884\\
216.660937992126	0.607376044335884\\
};
\addplot [color=black, mark=triangle*, mark options={solid, fill=red}, forget plot]
  table[row sep=crcr]{%
217.038084289605	0.568814873393642\\
217.038084289605	0.568814873393642\\
};
\end{axis}
\end{tikzpicture}%

%% file: Figures/TimeHmax_B2W_We_time7delta1.tex
\begin{tikzpicture}

\begin{axis}[%
width=0.245\linewidth,
height=0.245\linewidth,
scale only axis,
xmin=200,
xmax=400,
xlabel style={font=\color{white!15!black}},
xlabel={\dropWe{} (-)},
ymin=0,
ymax=1,
ylabel style={font=\color{white!15!black}},
ylabel={\tcapHmaxmod{} (-)},
axis background/.style={fill=white},
title style={font=\bfseries},
xmajorgrids,
ymajorgrids,
legend style={at={(0.98,0.02)}, anchor=south east, legend cell align=left, align=left, draw=white!15!black},
clip mode=individual,
title={$\text{\wft{}} \approx{} 0.12$}
]

\node at (220,0.9) {\textbf{(a)}};
\addplot [color=black, mark=square*, mark options={solid, fill=blue}]
  table[row sep=crcr]{%
508.471029883192	0.347650300687167\\
508.471029883192	0.347650300687167\\
};

\addplot [color=black, mark=*, mark options={solid, fill=green}]
  table[row sep=crcr]{%
508.48504088168	0.311146518976771\\
508.48504088168	0.311146518976771\\
};

\addplot [color=black, mark=triangle*, mark options={solid, fill=red}]
  table[row sep=crcr]{%
498.867960163284	0.306529460597006\\
498.867960163284	0.233770868555999\\
};

\addplot [color=black, mark=square*, mark options={solid, fill=blue}, forget plot]
  table[row sep=crcr]{%
440.840954053392	0.243557043233828\\
440.840954053392	0.243557043233828\\
};
\addplot [color=black, mark=*, mark options={solid, fill=green}, forget plot]
  table[row sep=crcr]{%
445.036209274399	0.312211089297681\\
445.036209274399	0.312211089297681\\
};
\addplot [color=black, mark=triangle*, mark options={solid, fill=red}, forget plot]
  table[row sep=crcr]{%
434.940594142564	0.338718013037755\\
434.940594142564	0.261214093064824\\
};
\addplot [color=black, mark=square*, mark options={solid, fill=blue}, forget plot]
  table[row sep=crcr]{%
358.310179270558	0.390428172183411\\
358.310179270558	0.390428172183411\\
};
\addplot [color=black, mark=*, mark options={solid, fill=green}, forget plot]
  table[row sep=crcr]{%
368.354660425129	0.412138014042126\\
368.354660425129	0.412138014042126\\
};
\addplot [color=black, mark=triangle*, mark options={solid, fill=red}, forget plot]
  table[row sep=crcr]{%
362.804222243713	0.42540396691287\\
362.804222243713	0.310096075393817\\
};
\addplot [color=black, mark=square*, mark options={solid, fill=blue}, forget plot]
  table[row sep=crcr]{%
314.864350862924	0.383199756018109\\
314.864350862924	0.383199756018109\\
};
\addplot [color=black, mark=*, mark options={solid, fill=green}, forget plot]
  table[row sep=crcr]{%
320.545648797683	0.322114606465932\\
320.545648797683	0.322114606465932\\
};
\addplot [color=black, mark=triangle*, mark options={solid, fill=red}, forget plot]
  table[row sep=crcr]{%
316.28831821484	0.425322864784322\\
316.28831821484	0.292369662650618\\
};
\addplot [color=black, mark=square*, mark options={solid, fill=blue}, forget plot]
  table[row sep=crcr]{%
266.853404078276	0.369512737219381\\
266.853404078276	0.369512737219381\\
};
\addplot [color=black, mark=*, mark options={solid, fill=green}, forget plot]
  table[row sep=crcr]{%
271.76583073723	0.370239355410219\\
271.76583073723	0.370239355410219\\
};
\addplot [color=black, mark=triangle*, mark options={solid, fill=red}, forget plot]
  table[row sep=crcr]{%
269.170235106682	0.352313981603575\\
269.170235106682	0.352313981603575\\
};
\addplot [color=black, mark=square*, mark options={solid, fill=blue}, forget plot]
  table[row sep=crcr]{%
220.427335475403	0.346447274732377\\
220.427335475403	0.346447274732377\\
};
\addplot [color=black, mark=*, mark options={solid, fill=green}, forget plot]
  table[row sep=crcr]{%
215.741494675615	0.342417009964285\\
215.741494675615	0.342417009964285\\
};
\addplot [color=black, mark=triangle*, mark options={solid, fill=red}, forget plot]
  table[row sep=crcr]{%
218.583938591942	0.314592467169209\\
218.583938591942	0.314592467169209\\
};
\end{axis}
\end{tikzpicture}%

%% file: Figures/TimeHmax_B2W_We_time7delta2.tex
\begin{tikzpicture}

\begin{axis}[%
width=0.245\linewidth,
height=0.245\linewidth,
scale only axis,
xmin=200,
xmax=400,
xlabel style={font=\color{white!15!black}},
xlabel={\dropWe{} (-)},
ymin=0,
ymax=1,
ylabel style={font=\color{white!15!black}},
ylabel={\tcapHmaxmod{} (-)},
axis background/.style={fill=white},
title style={font=\bfseries},
xmajorgrids,
ymajorgrids,
legend style={at={(0.98,0.98)}, anchor=north east, legend cell align=left, align=left, draw=white!15!black},
clip mode=individual,
title={$\text{\wft{}} \approx{} 0.26$}
]

\node at (220,0.9) {\textbf{(b)}};
]
\addplot [color=black, mark=square*, mark options={solid, fill=blue}, only marks]
  table[row sep=crcr]{%
502.006156921766	0.374761278117067\\
502.006156921766	0.374761278117067\\
};
\addlegendentry{\Hexa{}}

\addplot [color=black, mark=*, mark options={solid, fill=green}, only marks]
  table[row sep=crcr]{%
501.547976394186	0.320052237923076\\
501.547976394186	0.320052237923076\\
};
\addlegendentry{\SO{}}

\addplot [color=black, mark=triangle*, mark options={solid, fill=red}, only marks]
  table[row sep=crcr]{%
495.658218686723	0.407096980339934\\
495.658218686723	0.247153581695602\\
};
\addlegendentry{\EtOh{}}

\addplot [color=black, mark=square*, mark options={solid, fill=blue}, forget plot]
  table[row sep=crcr]{%
434.268229939372	0.37476532312291\\
434.268229939372	0.37476532312291\\
};
\addplot [color=black, mark=*, mark options={solid, fill=green}, forget plot]
  table[row sep=crcr]{%
434.063366381827	0.346624905077405\\
434.063366381827	0.346624905077405\\
};
\addplot [color=black, mark=triangle*, mark options={solid, fill=red}, forget plot]
  table[row sep=crcr]{%
434.092265396643	0.416537273398344\\
434.092265396643	0.304562729137517\\
};
\addplot [color=black, mark=square*, mark options={solid, fill=blue}, forget plot]
  table[row sep=crcr]{%
363.494692562247	0.398205523517638\\
363.494692562247	0.398205523517638\\
};
\addplot [color=black, mark=*, mark options={solid, fill=green}, forget plot]
  table[row sep=crcr]{%
369.324643699557	0.370323412931675\\
369.324643699557	0.370323412931675\\
};
\addplot [color=black, mark=triangle*, mark options={solid, fill=red}, forget plot]
  table[row sep=crcr]{%
355.919634927362	0.34235380643054\\
355.919634927362	0.34235380643054\\
};
\addplot [color=black, mark=square*, mark options={solid, fill=blue}, forget plot]
  table[row sep=crcr]{%
321.655135921171	0.455823775217537\\
321.655135921171	0.455823775217537\\
};
\addplot [color=black, mark=*, mark options={solid, fill=green}, forget plot]
  table[row sep=crcr]{%
318.321194011279	0.404406129655369\\
318.321194011279	0.404406129655369\\
};
\addplot [color=black, mark=triangle*, mark options={solid, fill=red}, forget plot]
  table[row sep=crcr]{%
318.36112148599	0.365013895814202\\
318.36112148599	0.365013895814202\\
};
\addplot [color=black, mark=square*, mark options={solid, fill=blue}, forget plot]
  table[row sep=crcr]{%
267.654800883446	0.372440153693866\\
267.654800883446	0.372440153693866\\
};
\addplot [color=black, mark=*, mark options={solid, fill=green}, forget plot]
  table[row sep=crcr]{%
269.636317537986	0.367801899243231\\
269.636317537986	0.367801899243231\\
};
\addplot [color=black, mark=triangle*, mark options={solid, fill=red}, forget plot]
  table[row sep=crcr]{%
266.494273655914	0.318326455501765\\
266.494273655914	0.318326455501765\\
};
\addplot [color=black, mark=square*, mark options={solid, fill=blue}, forget plot]
  table[row sep=crcr]{%
216.514512112825	0.381515305603823\\
216.514512112825	0.381515305603823\\
};
\addplot [color=black, mark=*, mark options={solid, fill=green}, forget plot]
  table[row sep=crcr]{%
216.660937992126	0.39052347922615\\
216.660937992126	0.39052347922615\\
};
\addplot [color=black, mark=triangle*, mark options={solid, fill=red}, forget plot]
  table[row sep=crcr]{%
217.038084289605	0.310447443797544\\
217.038084289605	0.310447443797544\\
};
\end{axis}
\end{tikzpicture}%

%% file: Conclusion.tex
\section{Conclusions}\label{sec:Conclusion}

The influence of miscibility and wettability in terms of spreading parameter is investigated for droplet impact onto thin wall-films: immiscible with partial wetting, with full wetting or miscible, respectively. At similar liquid properties, this corresponds to a variation of the interfacial tension between the droplet and the wall-film. 
First, shifts in the splashing limit as well as in the crown extensions are observed. Lower interfacial tension promotes splashing and larger crown extensions. These discrepancies vanish by taking into account the energy stored in the interface (if any) between the droplet and the wall film. 
An energy balance at the end of crown extension taking into account all surface and interfacial tensions also highlights that almost half of the initial kinetic energy is lost for thin wall-films.
Second, a shift in the duration of the crown ascending phase is observed. Smaller interfacial tensions result in longer duration. These differences vanish by scaling time with a capillary time scale taking into account both surface and interfacial tensions, indicating an inertio-capillary driven system.
The dynamics is well captured by the capillary time for different wall-film thicknesses if accounting for the variations of the liquid masses in movement. Shear at the crown base during extension might also prolong the crown ascending phase by influencing the damping rate.
To conclude, the interfacial tension leads to a non-negligible energy storage and recoiling force for droplet impact onto thin wall films where the extension is important. The interfacial tension, which is linked to miscibility and wettability, needs to be taken into account to understand the impact dynamics and to model it accurately.

%% file: Figures/Appendix/Exp_Setup.tex
\begin{tikzpicture}
\node[anchor=south west,inner sep=0] (image) at (0,0) {\includegraphics[scale=0.2,trim={8cm 0cm 8cm 1cm},clip]{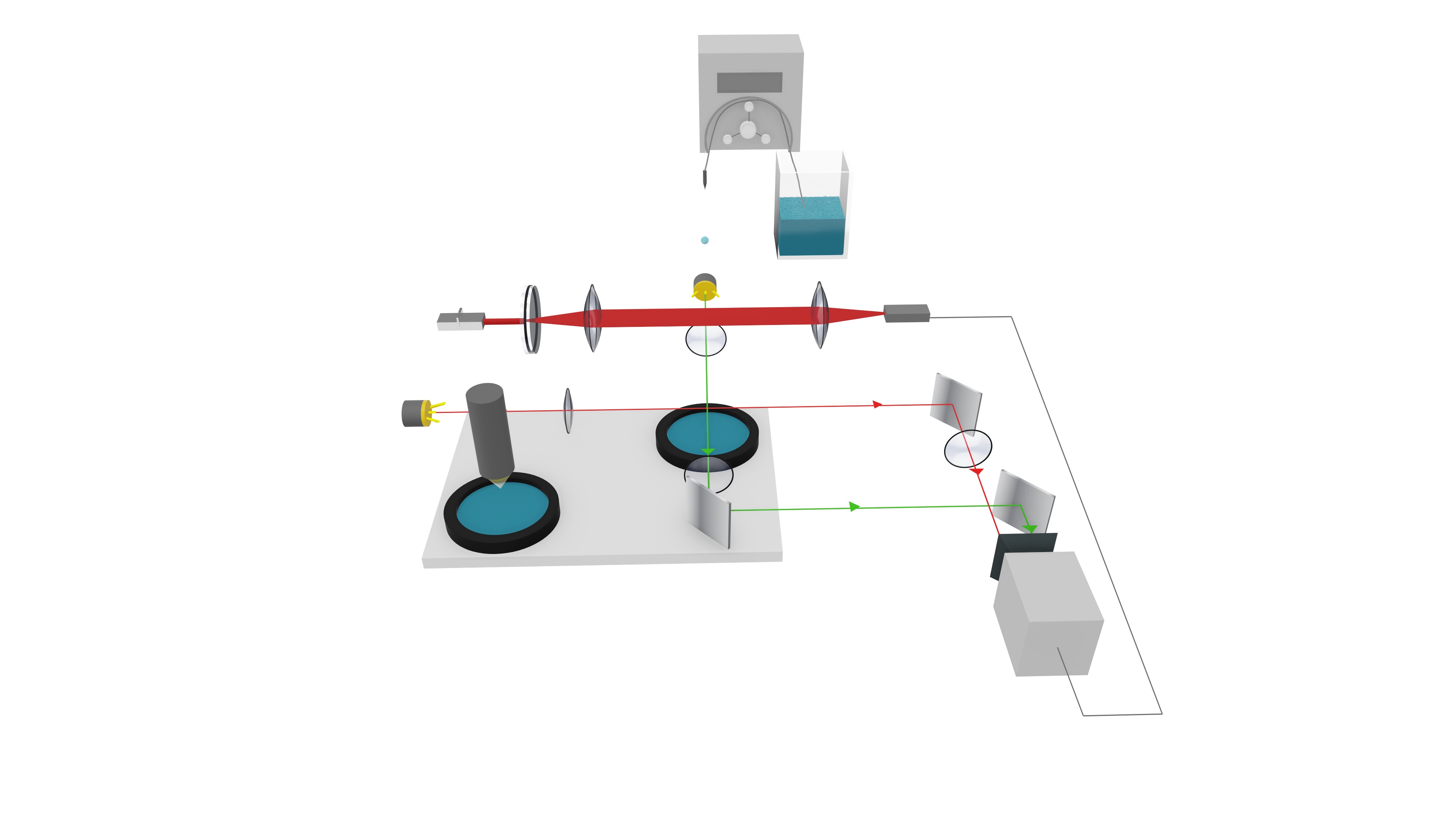}};
    \begin{scope}[x={(image.south east)},y={(image.north west)}]
\coordinate (IM) at (0.55,0.1);
\draw[->,draw=green!70!black, color=green!70!black, line width=1.1pt] (IM)--(0.71,0.26) ;
\node[color=green!70!black,fill=white, anchor = mid,align=center] at (IM) {High-Speed\\Camera};

\coordinate (T) at (0.8,0.85);
\draw[->,draw=red, color=red, line width=1.1pt] (T)--(0.65,0.66) ;
\node[color=red,fill=white,anchor=mid] at (T) {Trigger};

\coordinate (CCI) at (0.15,0.1);
\draw[->,draw=purple, color=purple,thick] (CCI)--(0.265,0.45) ;
\node[color=purple,anchor=mid,fill=white,align=center] at (CCI) {Film thickness\\measurement device (CCI)}; 

\coordinate (IA) at (0.4,0.25);
\draw[->, line width=1.1pt, blue] (IA) --(0.45,0.43) ;
\node[color=blue, fill=white] at (IA) {Impact area};

\coordinate (D) at (0.2,0.85);
\draw[->, line width=1.1pt, blue] (D) --(0.45,0.8) ;
\node[color=blue, fill=white,anchor=mid] at (D) {Dropper};

\coordinate (LED) at (0.2,0.7);
\draw[->, line width=1.1pt, yellow!70!black] (LED) --(0.22,0.54) ; 
\draw[->, line width=1.1pt, yellow!70!black] (LED) --(0.46,0.68) ; 
\node[color=yellow!70!black, fill=white,anchor=mid] at (LED) {LED};

      \end{scope}    
\end{tikzpicture}

%% file: App_surf_nrj_delta2.tex
\begin{figure}
\centering
\input{Figures/Appendix/MaxCrownSurf_B2W_We_d_delta_2_Dim.tex}%
\input{Figures/Appendix/MaxSurfNrj_FW_KinEn_Wed_delta_2.tex}
  \caption{(a) Maximum crown surface \CSmax{} at different Weber numbers \dropWe{}. (b) Maximum crown surface energy \CSEmax{} normalized with the initial droplet kinetic energy \DKE{} at different Weber numbers \dropWe{}.}
\label{fig:App_surf_nrj_delta2}
\end{figure}
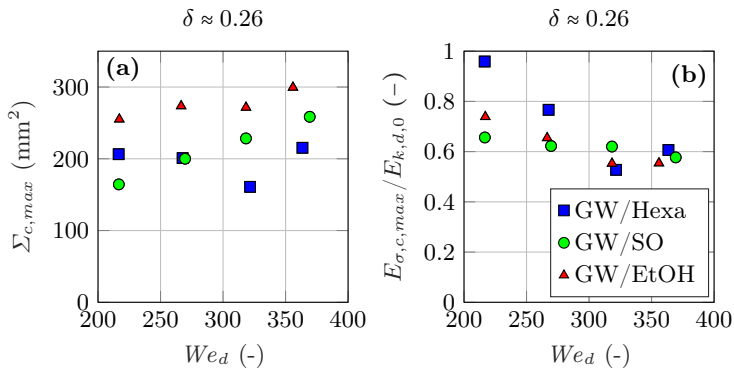
\begin{figure}
	\centering
\begin{tikzpicture}
\node[anchor=south west,inner sep=0] (image) at (0,0) {\includegraphics[width=0.5\linewidth]{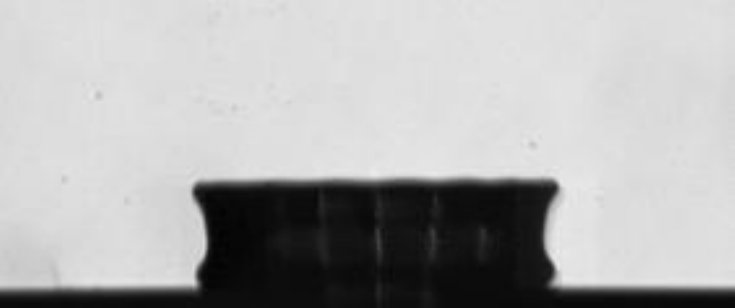}};    
    \begin{scope}[x={(image.south east)},y={(image.north west)}]            
       \draw[thick] (0.1,0.9)--(0.190909,0.9);
     \end{scope}  
\end{tikzpicture}
\input{Figures/Appendix/appendix_crown_curved}
	\caption{Top: Typical high-speed image (without background substraction) of \Hexa{} at \dropWe{}=320 and $\text{\wft{}}\approx{}0.259$ (see full experimental condition in Table \ref{tab:c3s1_exprangeavg}). The scale bar measures \SI{2.00}{\milli\meter}. Bottom: schematic of a cylindrical crown with curved wall.}\label{fig:concavity}
\end{figure}
\begin{figure}
\centering
\input{Figures/Appendix/appendix_underestimation2}
\caption{Normalized crown surface in the case where the crown wall is bended inward in function of the crown aspect ratio \Hrim{}/\Rrim{}, for different values of the minor axis $b$ of the ellipse as shown in the schematic of Fig.~\ref{fig:concavity} (bottom). This quantifies the underestimation of the measured crown surface \CS{} when the crown wall is assumed to be a straight line between base and rim radii.} \label{fig:underestimation}
\end{figure}
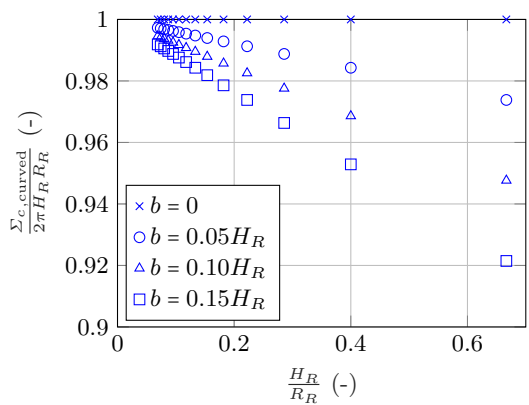

The maximum crown surfaces \CSmax{} and normalized crow surface energies \CSEmax{}/\DKE{} are given in Fig.~\ref{fig:App_surf_nrj_delta2} (a) and (b), respectively, in function of \dropWe{} for the three different liquid pairs at $\text{\wft{}}\approx{} 0.26$.
The trends observed are in general similar to those of $\text{\wft{}}\approx{} 0.12$: the points of the different liquid pairs come together if the interfacial energy between droplet and wall film is taken into account, the crown surface \CS{} tends to increase with growing \dropWe{}, and \CSEmax{}/\DKE{} decreases with growing \dropWe{}.
Considering the crown surface energies (Fig.~\ref{fig:App_surf_nrj_delta2}(b)), the values are higher than those of $\text{\wft{}}\approx{} 0.12$ as explained in Sec.~\ref{sec:nrj_storage}.
However, some discrepancies are observed for small \dropWe{}, where the crown surfaces are much higher than expected. This is particularly pronounced for \Hexa{}, whose trend with decreasing \dropWe{} is reversed, leading to crown surfaces even higher than those of \SO{}. Note that a similar trend is also observed for \Hexa{} at $\text{\wft{}}\approx{} 0.12$ for the lowest \dropWe{}, but in much smaller amplitude.
This sudden increase is peculiar since it corresponds to higher values of normalized crown surface energies approaching 1, which would indicate extremely low losses.
The possible reasons for these discrepancies are listed hereafter.\\
First, it is possible that these experimental conditions lead to a much better energy transfer from the incoming droplet kinetic energy to surface energy. Low Weber/Reynolds numbers and high wall-film thickness promote a smooth energy transfer, corresponding to a clear deposition case (or \textit{full} deposition case), as it is confirmed by the unperturbed rim of Fig.~\ref{fig:concavity}. Furthermore, no energy is lost to form the corrugations on the crown rim, and there is no mass losses. Since the interfacial tension of \Hexa{} is higher, this effect could happen at higher \dropWe{} and stronger than for the other liquid pairs, as it has been observed for the splashing limit and explained in Sec.~\ref{sec:ShiftSplashing}.\\
Second, the particularly strong increase of \Hexa{} could be due to contaminations that would decrease the value of the interfacial tension, and thus increase the crown extension. However, this effect alone is unlikely since it has been observed at several impact conditions, and also for all liquid pairs (\SO{} and \EtOh{}) although in smaller amplitude.\\
Last, part of this increasing trend could be due to measurement errors of the crown surface due to its concavity for these cases, especially for \Hexa{} as illustrated in Fig.~\ref{fig:concavity}. Similar crown contractions have already been observed in the literature for n-hexadecane droplet/wall-film experiments \citep{Geppert2017}, but they were rather located at the top of the crown. In the present cases, this curvature of the crown wall could be due to the interfacial tension acting on it, since it is the most varying parameter between the three liquid pairs.
It could also result from the gradient of the crown thickness and/or liquid properties inside the crown wall. To better understand these effects, the spatial distribution of the liquids need to be resolved which goes beyond the scope of this study, but could be for example further investigated with Direct Numerical Simulations.
Coming back to the evaluation of the crown surface, this crown wall bended inward can lead to an overestimation of the real crown surface. Indeed, the crown surface is calculated by assuming a straight line between the rim and base radii (the crown wall being considered as a conical frustum, as in Fig.~\ref{fig:CrownSchematic}).
In order to get a rough idea of this overestimation, one can compare the surface of a simplified cylindrical crown with the same crown where the side wall is curved inward as illustrated in Fig.~\ref{fig:concavity} (bottom). In this simplified case, the envelop of the crown wall is assumed to follow an ellipse centered at (\Rrim{},\Hrim{}/2) and a vertical major axis $a$ equal to $\text{\Hrim{}/2}$. The minor axis $b$ can be varied to see the influence of the crown wall curvature.
In this configuration, the crown radius $r$ is dependent on the height $z$ as follows :
$$r=\text{\Rrim{}}-b\sqrt{1-\frac{\left(\text{\Hrim{}}/2-z\right)^2}{a^2}}$$
The crown surface can be calculated as :
$$\text{\CS{}}=\int_{0}^{\text{\Hrim{}}} 2 \pi r(z) dz$$
The solution of this integration for $a=\text{\Hrim{}}/2$, normalized with the cylindrical surface of reference $2\pi\text{\Rrim{}}\text{\Hrim{}}$ is provided in function of the crown aspect ratio \Hrim{}/\Rrim{} in Fig.~\ref{fig:underestimation} for different values of the minor axis $b$. For $b=0$, the coefficient is equal to one, the surfaces are equivalents. With increasing curvature, the ratio decreases, and it decreases also with increasing aspect ratio. The highest aspect ratio observed in the present database is \Hrim{}/\Rrim{}=0.65. The highest value of b can be estimated from the high-speed images. In Fig.~\ref{tab:c3s1_exprangeavg}, the value of $b$ is 	approximately 0.13\Hrim{}, which is a particularly pronounced case. Hence, in the worst case scenario we consider the configuration of $b=0.15\text{\Hrim{}}$ with $\text{\Hrim{}}/\text{\Rrim{}}=0.65$, which leads to an underestimation of almost 8\%.

%% file: Figures/Appendix/MaxCrownSurf_B2W_We_d_delta_2_Dim.tex
\begin{tikzpicture}

\begin{axis}[%
width=0.245\linewidth,
height=0.245\linewidth,
scale only axis,
xmin=200,
xmax=400,
xlabel style={font=\color{white!15!black}},
xlabel={\dropWe{} (-)},
ymin=0,
ymax=350,
ylabel style={font=\color{white!15!black}},
ylabel={\CSmax{} (\si{\milli\meter\squared})},
axis background/.style={fill=white},
title style={font=\bfseries},
xmajorgrids,
ymajorgrids,
legend style={at={(0.98,0.02)}, anchor=south east, legend cell align=left, align=left, draw=white!15!black},
clip mode=individual,
title={$\text{\wft{}} \approx{} 0.26$}
]

\node at (220,325) {\textbf{(a)}};

\addplot [color=black, mark=square*, mark options={solid, fill=blue}]
  table[row sep=crcr]{%
502.006156921766	199.348952570615\\
502.006156921766	199.348952570615\\
};

\addplot [color=black, mark=*, mark options={solid, fill=green}]
  table[row sep=crcr]{%
501.547976394186	251.259155756974\\
501.547976394186	251.259155756974\\
};

\addplot [color=black, mark=triangle*, mark options={solid, fill=red}]
  table[row sep=crcr]{%
495.658218686723	536.623921014746\\
495.658218686723	362.557914829791\\
};

\addplot [color=black, mark=square*, mark options={solid, fill=blue}, forget plot]
  table[row sep=crcr]{%
434.268229939372	215.857229786572\\
434.268229939372	215.857229786572\\
};
\addplot [color=black, mark=*, mark options={solid, fill=green}, forget plot]
  table[row sep=crcr]{%
434.063366381827	242.009544320678\\
434.063366381827	242.009544320678\\
};
\addplot [color=black, mark=triangle*, mark options={solid, fill=red}, forget plot]
  table[row sep=crcr]{%
434.092265396643	414.737693775947\\
434.092265396643	375.102465040874\\
};
\addplot [color=black, mark=square*, mark options={solid, fill=blue}, forget plot]
  table[row sep=crcr]{%
363.494692562247	215.346333686126\\
363.494692562247	215.346333686126\\
};
\addplot [color=black, mark=*, mark options={solid, fill=green}, forget plot]
  table[row sep=crcr]{%
369.324643699557	258.531826502423\\
369.324643699557	258.531826502423\\
};
\addplot [color=black, mark=triangle*, mark options={solid, fill=red}, forget plot]
  table[row sep=crcr]{%
355.919634927362	299.423350896893\\
355.919634927362	299.423350896893\\
};
\addplot [color=black, mark=square*, mark options={solid, fill=blue}, forget plot]
  table[row sep=crcr]{%
321.655135921171	160.929053865554\\
321.655135921171	160.929053865554\\
};
\addplot [color=black, mark=*, mark options={solid, fill=green}, forget plot]
  table[row sep=crcr]{%
318.321194011279	228.39095059159\\
318.321194011279	228.39095059159\\
};
\addplot [color=black, mark=triangle*, mark options={solid, fill=red}, forget plot]
  table[row sep=crcr]{%
318.36112148599	271.533672357354\\
318.36112148599	271.533672357354\\
};
\addplot [color=black, mark=square*, mark options={solid, fill=blue}, forget plot]
  table[row sep=crcr]{%
267.654800883446	200.908720921075\\
267.654800883446	200.908720921075\\
};
\addplot [color=black, mark=*, mark options={solid, fill=green}, forget plot]
  table[row sep=crcr]{%
269.636317537986	200.082576811993\\
269.636317537986	200.082576811993\\
};
\addplot [color=black, mark=triangle*, mark options={solid, fill=red}, forget plot]
  table[row sep=crcr]{%
266.494273655914	273.518021830381\\
266.494273655914	273.518021830381\\
};
\addplot [color=black, mark=square*, mark options={solid, fill=blue}, forget plot]
  table[row sep=crcr]{%
216.514512112825	206.391940854686\\
216.514512112825	206.391940854686\\
};
\addplot [color=black, mark=*, mark options={solid, fill=green}, forget plot]
  table[row sep=crcr]{%
216.660937992126	164.335406978491\\
216.660937992126	164.335406978491\\
};
\addplot [color=black, mark=triangle*, mark options={solid, fill=red}, forget plot]
  table[row sep=crcr]{%
217.038084289605	255.07501630472\\
217.038084289605	255.07501630472\\
};
\end{axis}
\end{tikzpicture}%

%% file: Figures/Appendix/MaxSurfNrj_FW_KinEn_Wed_delta_2.tex
\begin{tikzpicture}

\begin{axis}[%
width=0.245\linewidth,
height=0.245\linewidth,
scale only axis,
xmin=200,
xmax=400,
xlabel style={font=\color{white!15!black}},
xlabel={\dropWe{} (-)},
ymin=0,
ymax=1,
ylabel style={font=\color{white!15!black}},
ylabel={\CSEmax{}/\DKE{} $(-)$},
axis background/.style={fill=white},
title style={font=\bfseries},
xmajorgrids,
ymajorgrids,
legend style={at={(0.98,0.02)}, anchor=south east, legend cell align=left, align=left, draw=white!15!black},
clip mode=individual,
title={$\text{\wft{}} \approx{} 0.26$}
]

\node at (380,0.9) {\textbf{(b)}};

\addplot [color=black, mark=square*, mark options={solid, fill=blue}, only marks]
  table[row sep=crcr]{%
502.006156921766	0.373298040371674\\
502.006156921766	0.373298040371674\\
};
\addlegendentry{\Hexa{}}

\addplot [color=black, mark=*, mark options={solid, fill=green}, only marks]
  table[row sep=crcr]{%
501.547976394186	0.395342185019883\\
501.547976394186	0.395342185019883\\
};
\addlegendentry{\SO{}}

\addplot [color=black, mark=triangle*, mark options={solid, fill=red}, only marks]
  table[row sep=crcr]{%
495.658218686723	0.623711848603942\\
495.658218686723	0.428861707742733\\
};
\addlegendentry{\EtOh{}}

\addplot [color=black, mark=square*, mark options={solid, fill=blue}, forget plot]
  table[row sep=crcr]{%
434.268229939372	0.480568730995052\\
434.268229939372	0.480568730995052\\
};
\addplot [color=black, mark=*, mark options={solid, fill=green}, forget plot]
  table[row sep=crcr]{%
434.063366381827	0.455903084706349\\
434.063366381827	0.455903084706349\\
};
\addplot [color=black, mark=triangle*, mark options={solid, fill=red}, forget plot]
  table[row sep=crcr]{%
434.092265396643	0.566973622884393\\
434.092265396643	0.51388321082991\\
};
\addplot [color=black, mark=square*, mark options={solid, fill=blue}, forget plot]
  table[row sep=crcr]{%
363.494692562247	0.606268313697049\\
363.494692562247	0.606268313697049\\
};
\addplot [color=black, mark=*, mark options={solid, fill=green}, forget plot]
  table[row sep=crcr]{%
369.324643699557	0.577029858825046\\
369.324643699557	0.577029858825046\\
};
\addplot [color=black, mark=triangle*, mark options={solid, fill=red}, forget plot]
  table[row sep=crcr]{%
355.919634927362	0.554105136293999\\
355.919634927362	0.554105136293999\\
};
\addplot [color=black, mark=square*, mark options={solid, fill=blue}, forget plot]
  table[row sep=crcr]{%
321.655135921171	0.527579857955007\\
321.655135921171	0.527579857955007\\
};
\addplot [color=black, mark=*, mark options={solid, fill=green}, forget plot]
  table[row sep=crcr]{%
318.321194011279	0.620195406870722\\
318.321194011279	0.620195406870722\\
};
\addplot [color=black, mark=triangle*, mark options={solid, fill=red}, forget plot]
  table[row sep=crcr]{%
318.36112148599	0.552796383810965\\
318.36112148599	0.552796383810965\\
};
\addplot [color=black, mark=square*, mark options={solid, fill=blue}, forget plot]
  table[row sep=crcr]{%
267.654800883446	0.766585450586008\\
267.654800883446	0.766585450586008\\
};
\addplot [color=black, mark=*, mark options={solid, fill=green}, forget plot]
  table[row sep=crcr]{%
269.636317537986	0.622297167542162\\
269.636317537986	0.622297167542162\\
};
\addplot [color=black, mark=triangle*, mark options={solid, fill=red}, forget plot]
  table[row sep=crcr]{%
266.494273655914	0.654387050188622\\
266.494273655914	0.654387050188622\\
};
\addplot [color=black, mark=square*, mark options={solid, fill=blue}, forget plot]
  table[row sep=crcr]{%
216.514512112825	0.958821324960032\\
216.514512112825	0.958821324960032\\
};
\addplot [color=black, mark=*, mark options={solid, fill=green}, forget plot]
  table[row sep=crcr]{%
216.660937992126	0.656396671933113\\
216.660937992126	0.656396671933113\\
};
\addplot [color=black, mark=triangle*, mark options={solid, fill=red}, forget plot]
  table[row sep=crcr]{%
217.038084289605	0.738466360864482\\
217.038084289605	0.738466360864482\\
};
\end{axis}
\end{tikzpicture}%

%% file: Figures/Appendix/appendix_crown_curved.tex
\definecolor{c0000ff}{RGB}{0,0,255}
\definecolor{c44aa00}{RGB}{68,170,0}
\definecolor{c668500}{RGB}{102,133,0}
\definecolor{c55d400}{RGB}{85,212,0} 
\definecolor{cff0000}{RGB}{255,0,0} 
\definecolor{cff8300}{RGB}{255,131,0} 
\newcommand{\WidthDashed}{0.5pt}

\begin{tikzpicture}    
	 \node[anchor=south west,inner sep=0] (image) at (0,0) {\includegraphics[scale=0.4,trim={3cm 12cm 3cm 9.5cm},clip]{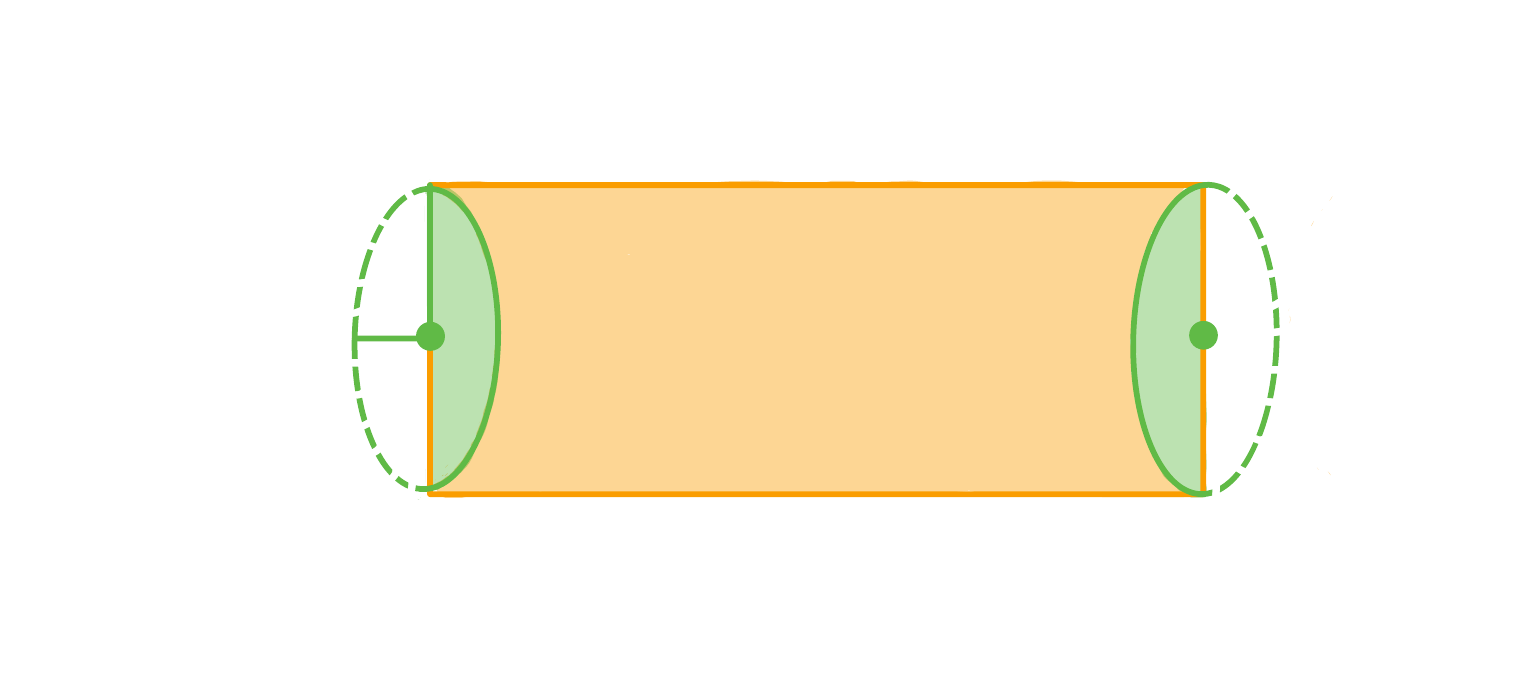}};  
		
    \begin{scope}[x={(image.south east)},y={(image.north west)}] 				
\draw[-,thick, c668500] (0.2,0.9) node[anchor=south east]{$a$} -- (0.28,0.6) ;;	
\draw[-,thick, c668500] (0.1,0.76) node[anchor=south east]{$b$} -- (0.26,0.5) ;;	
		
\draw[-,thick, black] (0,0.27)-- (1,0.27);

\draw[->, black] (0.53,0.27) -- (0.53,0.5) node[anchor=east]{$z$};
\draw[->, black] (0.53,0.27)  -- (0.4,0.27) node[anchor=south]{$r$};

        \node[orange] (R1) at (0.28,0.72) {$\bullet$}; 
        \node[orange] (R2) at (0.78,0.72) {$\bullet$}; 
        
        \node[orange] (B1) at (0.28,0.27) {$\bullet$}; 
        \node[orange] (B2) at (0.78,0.27) {$\bullet$}; 
				\draw[orange, dashed](R1)--(0.08,0.72); 
				\draw[orange, dashed](B1)--(0.08,0.27); 
        \draw[orange, thick, arrows={Triangle[angle=90:5pt,orange] - Triangle[angle=90:5pt,orange]}](0.2,0.27)--(0.2,0.72) node[midway,sloped,left,rotate=270,orange] {\Hrim};;
				    
        \draw[orange, dashed](R1)--(R1|-,0.95); 
        \draw[orange, dashed](R2)--(R2|-,0.95); 
        \draw[orange,thick, arrows={Triangle[angle=90:5pt,orange] - Triangle[angle=90:5pt,orange]}](R1|-,0.85)--(R2|-,0.85) node[midway,sloped,above,orange] {2\Rrim};;
 
        \draw[orange, dashed](B1|-,0.035)--(B1);  
        \draw[orange, dashed](B2|-,0.035)--(B2);
        \draw[orange, thick, arrows={Triangle[angle=90:5pt,orange] - Triangle[angle=90:5pt,orange]}](B1|-,0.15)--(B2|-,0.15) node[midway,sloped,below,orange] {2\Rbase};;		
     \end{scope}  
\end{tikzpicture}

%% file: Figures/Appendix/appendix_underestimation2.tex
\begin{tikzpicture}

\begin{axis}[%
width=0.4\linewidth,
height=0.3\linewidth,
at={(0,0)},
scale only axis,
xmin=0,
xmax=0.7,
xlabel style={font=\color{white!15!black}},
xlabel={$\frac{\text{\Hrim{}}}{\text{\Rrim{}}}$ (-)},
ylabel style={font=\color{white!15!black}},
ylabel={$\frac{\text{\CS{}}_{,\text{curved}}}{2\pi\text{\Hrim{}\Rrim{}}}$ (-)},
ymin=0.9,
ymax=1,
axis background/.style={fill=white},
xmajorgrids,
ymajorgrids,
legend style={at={(0.02,0.02)}, anchor=south west, legend cell align=left, align=left, draw=black},
]
\addplot [color=blue, mark=x, mark options={solid, blue}, only marks]
  table[row sep=crcr]{%
0.666666666666667	1\\
};
\addlegendentry{$b=0$};

\addplot [color=blue, mark=x, mark options={solid, blue}, forget plot]
  table[row sep=crcr]{%
0.4	1\\
};
\addplot [color=blue, mark=x, mark options={solid, blue}, forget plot]
  table[row sep=crcr]{%
0.285714285714286	1\\
};
\addplot [color=blue, mark=x, mark options={solid, blue}, forget plot]
  table[row sep=crcr]{%
0.222222222222222	1\\
};
\addplot [color=blue, mark=x, mark options={solid, blue}, forget plot]
  table[row sep=crcr]{%
0.181818181818182	1\\
};
\addplot [color=blue, mark=x, mark options={solid, blue}, forget plot]
  table[row sep=crcr]{%
0.153846153846154	1\\
};
\addplot [color=blue, mark=x, mark options={solid, blue}, forget plot]
  table[row sep=crcr]{%
0.133333333333333	1\\
};
\addplot [color=blue, mark=x, mark options={solid, blue}, forget plot]
  table[row sep=crcr]{%
0.117647058823529	1\\
};
\addplot [color=blue, mark=x, mark options={solid, blue}, forget plot]
  table[row sep=crcr]{%
0.105263157894737	1\\
};
\addplot [color=blue, mark=x, mark options={solid, blue}, forget plot]
  table[row sep=crcr]{%
0.0952380952380952	1\\
};
\addplot [color=blue, mark=x, mark options={solid, blue}, forget plot]
  table[row sep=crcr]{%
0.0869565217391304	1\\
};
\addplot [color=blue, mark=x, mark options={solid, blue}, forget plot]
  table[row sep=crcr]{%
0.08	1\\
};
\addplot [color=blue, mark=x, mark options={solid, blue}, forget plot]
  table[row sep=crcr]{%
0.0740740740740741	1\\
};
\addplot [color=blue, mark=x, mark options={solid, blue}, forget plot]
  table[row sep=crcr]{%
0.0689655172413793	1\\
};
\addplot [color=blue, mark=o, mark options={solid, blue}, only marks]
  table[row sep=crcr]{%
0.666666666666667	0.973820061220085\\
};
\addlegendentry{$b=0.05\text{\Hrim{}}$};

\addplot [color=blue, mark=o, mark options={solid, blue}, forget plot]
  table[row sep=crcr]{%
0.4	0.984292036732051\\
};
\addplot [color=blue, mark=o, mark options={solid, blue}, forget plot]
  table[row sep=crcr]{%
0.285714285714286	0.988780026237179\\
};
\addplot [color=blue, mark=o, mark options={solid, blue}, forget plot]
  table[row sep=crcr]{%
0.222222222222222	0.991273353740028\\
};
\addplot [color=blue, mark=o, mark options={solid, blue}, forget plot]
  table[row sep=crcr]{%
0.181818181818182	0.992860016696387\\
};
\addplot [color=blue, mark=o, mark options={solid, blue}, forget plot]
  table[row sep=crcr]{%
0.153846153846154	0.993958475666173\\
};
\addplot [color=blue, mark=o, mark options={solid, blue}, forget plot]
  table[row sep=crcr]{%
0.133333333333333	0.994764012244017\\
};
\addplot [color=blue, mark=o, mark options={solid, blue}, forget plot]
  table[row sep=crcr]{%
0.117647058823529	0.995380010803544\\
};
\addplot [color=blue, mark=o, mark options={solid, blue}, forget plot]
  table[row sep=crcr]{%
0.105263157894737	0.995866325455803\\
};
\addplot [color=blue, mark=o, mark options={solid, blue}, forget plot]
  table[row sep=crcr]{%
0.0952380952380952	0.996260008745726\\
};
\addplot [color=blue, mark=o, mark options={solid, blue}, forget plot]
  table[row sep=crcr]{%
0.0869565217391304	0.996585225376533\\
};
\addplot [color=blue, mark=o, mark options={solid, blue}, forget plot]
  table[row sep=crcr]{%
0.08	0.99685840734641\\
};
\addplot [color=blue, mark=o, mark options={solid, blue}, forget plot]
  table[row sep=crcr]{%
0.0740740740740741	0.997091117913343\\
};
\addplot [color=blue, mark=o, mark options={solid, blue}, forget plot]
  table[row sep=crcr]{%
0.0689655172413793	0.997291730471043\\
};
\addplot [color=blue, mark=triangle, mark options={solid, blue}, only marks]
  table[row sep=crcr]{%
0.666666666666667	0.94764012244017\\
};
\addlegendentry{$b=0.10\text{\Hrim{}}$};
\addplot [color=blue, mark=triangle, mark options={solid, blue}, forget plot]
  table[row sep=crcr]{%
0.4	0.968584073464102\\
};
\addplot [color=blue, mark=triangle, mark options={solid, blue}, forget plot]
  table[row sep=crcr]{%
0.285714285714286	0.977560052474359\\
};
\addplot [color=blue, mark=triangle, mark options={solid, blue}, forget plot]
  table[row sep=crcr]{%
0.222222222222222	0.982546707480057\\
};
\addplot [color=blue, mark=triangle, mark options={solid, blue}, forget plot]
  table[row sep=crcr]{%
0.181818181818182	0.985720033392774\\
};
\addplot [color=blue, mark=triangle, mark options={solid, blue}, forget plot]
  table[row sep=crcr]{%
0.153846153846154	0.987916951332347\\
};
\addplot [color=blue, mark=triangle, mark options={solid, blue}, forget plot]
  table[row sep=crcr]{%
0.133333333333333	0.989528024488034\\
};
\addplot [color=blue, mark=triangle, mark options={solid, blue}, forget plot]
  table[row sep=crcr]{%
0.117647058823529	0.990760021607089\\
};
\addplot [color=blue, mark=triangle, mark options={solid, blue}, forget plot]
  table[row sep=crcr]{%
0.105263157894737	0.991732650911606\\
};
\addplot [color=blue, mark=triangle, mark options={solid, blue}, forget plot]
  table[row sep=crcr]{%
0.0952380952380952	0.992520017491453\\
};
\addplot [color=blue, mark=triangle, mark options={solid, blue}, forget plot]
  table[row sep=crcr]{%
0.0869565217391304	0.993170450753066\\
};
\addplot [color=blue, mark=triangle, mark options={solid, blue}, forget plot]
  table[row sep=crcr]{%
0.08	0.99371681469282\\
};
\addplot [color=blue, mark=triangle, mark options={solid, blue}, forget plot]
  table[row sep=crcr]{%
0.0740740740740741	0.994182235826686\\
};
\addplot [color=blue, mark=triangle, mark options={solid, blue}, forget plot]
  table[row sep=crcr]{%
0.0689655172413793	0.994583460942087\\
};
\addplot [color=blue, mark=square, mark options={solid, blue},  only marks]
  table[row sep=crcr]{%
0.666666666666667	0.921460183660255\\
};
\addlegendentry{$b=0.15\text{\Hrim{}}$};
\addplot [color=blue, mark=square, mark options={solid, blue}, forget plot]
  table[row sep=crcr]{%
0.4	0.952876110196153\\
};
\addplot [color=blue, mark=square, mark options={solid, blue}, forget plot]
  table[row sep=crcr]{%
0.285714285714286	0.966340078711538\\
};
\addplot [color=blue, mark=square, mark options={solid, blue}, forget plot]
  table[row sep=crcr]{%
0.222222222222222	0.973820061220085\\
};
\addplot [color=blue, mark=square, mark options={solid, blue}, forget plot]
  table[row sep=crcr]{%
0.181818181818182	0.97858005008916\\
};
\addplot [color=blue, mark=square, mark options={solid, blue}, forget plot]
  table[row sep=crcr]{%
0.153846153846154	0.98187542699852\\
};
\addplot [color=blue, mark=square, mark options={solid, blue}, forget plot]
  table[row sep=crcr]{%
0.133333333333333	0.984292036732051\\
};
\addplot [color=blue, mark=square, mark options={solid, blue}, forget plot]
  table[row sep=crcr]{%
0.117647058823529	0.986140032410633\\
};
\addplot [color=blue, mark=square, mark options={solid, blue}, forget plot]
  table[row sep=crcr]{%
0.105263157894737	0.987598976367409\\
};
\addplot [color=blue, mark=square, mark options={solid, blue}, forget plot]
  table[row sep=crcr]{%
0.0952380952380952	0.988780026237179\\
};
\addplot [color=blue, mark=square, mark options={solid, blue}, forget plot]
  table[row sep=crcr]{%
0.0869565217391304	0.989755676129598\\
};
\addplot [color=blue, mark=square, mark options={solid, blue}, forget plot]
  table[row sep=crcr]{%
0.08	0.990575222039231\\
};
\addplot [color=blue, mark=square, mark options={solid, blue}, forget plot]
  table[row sep=crcr]{%
0.0740740740740741	0.991273353740028\\
};
\addplot [color=blue, mark=square, mark options={solid, blue}, forget plot]
  table[row sep=crcr]{%
0.0689655172413793	0.99187519141313\\
};
\end{axis}
\end{tikzpicture}

%% file: Main.bbl
\begin{thebibliography}{50}
\expandafter\ifx\csname natexlab\endcsname\relax\def\natexlab#1{#1}\fi
\def\au#1{#1} \def\ed#1{#1} \def\yr#1{#1}\def\at#1{#1}\def\jt#1{\textit{#1}}
  \def\bt#1{#1}\def\bvol#1{\textbf{#1}} \def\vol#1{#1} \def\pg#1{#1}
  \def\publ#1{#1}\def\arxiv#1{#1}\def\org#1{#1}\def\st#1{\textit{#1}}

\bibitem[Aljedaani {\em et~al.\/}(2018)Aljedaani, Wang, Jetly \&
  Thoroddsen]{Aljedaani2018}
{\sc \au{Aljedaani, A.~B.}, \au{Wang, C.}, \au{Jetly, A.} \& \au{Thoroddsen,
  S.~T.}} \yr{2018}  \at{Experiments on the breakup of drop-impact crowns by
  {M}arangoni holes}.  \jt{J. Fluid Mech.}  \bvol{844},  \pg{162--186}.

\bibitem[Banks {\em et~al.\/}(2013)Banks, Ajawara, Sanchez, Surti \&
  Aguilar]{Banks2013}
{\sc \au{Banks, D.}, \au{Ajawara, C.}, \au{Sanchez, R.}, \au{Surti, H.} \&
  \au{Aguilar, G.}} \yr{2013}  \at{Effects of drop and film viscosity on drop
  impacts onto thin films}.  \jt{Atomization and Sprays}  \bvol{23(6)},
  \pg{525--540}.

\bibitem[Baumgartner {\em et~al.\/}(2020)Baumgartner, Bernard, Weigand,
  Lamanna, Brenn \& Planchette]{Baumgartner2019}
{\sc \au{Baumgartner, D.}, \au{Bernard, R.}, \au{Weigand, B.}, \au{Lamanna,
  G.}, \au{Brenn, G.} \& \au{Planchette, C.}} \yr{2020}  \at{Influence of
  liquid miscibility and wettability on the structures produced by drop-jet
  collisions}.  \jt{Journal of fluid mechanics}  \bvol{885},  \pg{A23}.

\bibitem[Bernard {\em et~al.\/}(2017)Bernard, Foltyn, Geppert, Lamanna \&
  Weigand]{Bernard2017}
{\sc \au{Bernard, R.}, \au{Foltyn, P.}, \au{Geppert, A.}, \au{Lamanna, G.} \&
  \au{Weigand, B.}} \yr{2017}  \bt{Generalized analysis of the
  deposition/splashing limit for one- and two-component droplet impacts upon
  thin films}.

\bibitem[Bernard {\em et~al.\/}(2018)Bernard, Geppert, Vaikuntanathan, Lamanna
  \& Weigand]{Bernard2018}
{\sc \au{Bernard, R.}, \au{Geppert, A.}, \au{Vaikuntanathan, V.}, \au{Lamanna,
  G.} \& \au{Weigand, B.}} \yr{2018}  \at{On the scaling of crown rim diameter
  during droplet impact on thin wall-films}.  \jt{ICLASS 2018} .

\bibitem[Bernard {\em et~al.\/}(2020)Bernard, Vaikuntanathan, Weigand \&
  Lamanna]{Bernard2020b}
{\sc \au{Bernard, R.}, \au{Vaikuntanathan, V.}, \au{Weigand, B.} \&
  \au{Lamanna, G.}} \yr{2020}  \at{On the crown rim expansion kinematics during
  droplet impact on wall-films}.  \jt{Experimental Thermal and Fluid Science}
  \bvol{118},  \pg{110168}.

\bibitem[Che \& Matar(2018)]{Che2018}
{\sc \au{Che, Zhizhao} \& \au{Matar, Omar~K.}} \yr{2018}  \at{Impact of
  droplets on immiscible liquid films}.  \jt{Soft Matter}  \bvol{14}~(9),
  \pg{1540--1551}.

\bibitem[Chen {\em et~al.\/}(2017)Chen, Chen \& Amirfazli]{Chen2017}
{\sc \au{Chen, N.}, \au{Chen, H.} \& \au{Amirfazli, A.}} \yr{2017}  \at{Drop
  impact onto a thin film: Miscibility effect}.  \jt{Physics of Fluids}
  \bvol{29}~(9),  \pg{092106}.

\bibitem[Cossali {\em et~al.\/}(1997)Cossali, Coghe \& Marengo]{Cossali1997}
{\sc \au{Cossali, G.~E.}, \au{Coghe, A.} \& \au{Marengo, M.}} \yr{1997}
  \at{The impact of a single drop on a wetted solid surface}.  \jt{Experiments
  in Fluids}  \bvol{22}~(6),  \pg{463--472}.

\bibitem[Cossali {\em et~al.\/}(2004)Cossali, Marengo, Coghe \&
  Zhdanov]{Cossali2004}
{\sc \au{Cossali, G.~E.}, \au{Marengo, M.}, \au{Coghe, A.} \& \au{Zhdanov, S.}}
  \yr{2004}  \at{The role of time in single drop splash on thin film}.
  \jt{Experiments in Fluids}  \bvol{36}~(6),  \pg{888--900}.

\bibitem[De~Gennes {\em et~al.\/}(2013)De~Gennes, Brochard-Wyart \&
  Qu{\'e}r{\'e}]{DeGennes2013}
{\sc \au{De~Gennes, P.-G.}, \au{Brochard-Wyart, F.} \& \au{Qu{\'e}r{\'e}, D.}}
  \yr{2013} {\em Capillarity and wetting phenomena: drops, bubbles, pearls,
  waves\/}.  \publ{Springer Science \& Business Media}.

\bibitem[Deegan {\em et~al.\/}(2007)Deegan, Brunet \& Eggers]{Deegan2007}
{\sc \au{Deegan, R~D}, \au{Brunet, P} \& \au{Eggers, J}} \yr{2007}
  \at{Complexities of splashing}.  \jt{Nonlinearity}  \bvol{21}~(1),
  \pg{C1--C11}.

\bibitem[Enders \& Kahl(2008)]{Enders2008}
{\sc \au{Enders, S.} \& \au{Kahl, H.}} \yr{2008}  \at{Interfacial properties of
  water+ alcohol mixtures}.  \jt{Fluid Phase Equilibria}  \bvol{263}~(2),
  \pg{160--167}.

\bibitem[Fulcher \& Davis(1976)]{Fulcher1976}
{\sc \au{Fulcher, L.~P.} \& \au{Davis, B.~F.}} \yr{1976}  \at{Theoretical and
  experimental study of the motion of the simple pendulum}.  \jt{American
  Journal of Physics}  \bvol{44}~(1),  \pg{51--55}.

\bibitem[Gao \& Li(2015)]{Gao2015}
{\sc \au{Gao, X.} \& \au{Li, R.}} \yr{2015}  \at{Impact of a single drop on a
  flowing liquid film}.  \jt{Phys. Rev. E}  \bvol{92},  \pg{053005}.

\bibitem[Geppert {\em et~al.\/}(2020)Geppert, Bernard, Weigand \&
  Lamanna]{Geppert2020}
{\sc \au{Geppert, A.}, \au{Bernard, R.}, \au{Weigand, B.} \& \au{Lamanna, G.}}
  \yr{2020} {\em Analytical Model for Crown Spreading During Drop Impact onto
  Wetted Walls: Effect of Liquids Viscosity on Momentum Transfer\/},  \pg{pp.
  177--190}.  \publ{Springer International Publishing}.

\bibitem[Geppert {\em et~al.\/}(2016)Geppert, Chatzianagnostou, Mesiter, Gomaa,
  Lamanna \& Weigand]{Geppert2016}
{\sc \au{Geppert, A.}, \au{Chatzianagnostou, D.}, \au{Mesiter, C.}, \au{Gomaa,
  H.}, \au{Lamanna, G.} \& \au{Weigand, B.}} \yr{2016}  \at{Classification of
  impact morphology and splashing/deposition limit for n-hexadecane}.
  \jt{Atomization and Sprays}  \bvol{26},  \pg{983--1007}.

\bibitem[Geppert {\em et~al.\/}(2014)Geppert, Gomaa, Meister, Lamanna \&
  Weigand]{geppert2014}
{\sc \au{Geppert, A}, \au{Gomaa, H}, \au{Meister, C}, \au{Lamanna, G} \&
  \au{Weigand, B}} \yr{2014} Droplet wall-film-interaction: Impact morphology
  and splashing/deposition boundary of hyspin/n-hexadecane two-component
  system.  \bt{In {\em Proceedings of the 26th Annual Conference on Liquid
  Atomization and Spray Systems, ILASS-Americas\/}}.

\bibitem[Geppert {\em et~al.\/}(2017)Geppert, Terzis, Lamanna, Marengo \&
  Weigand]{Geppert2017}
{\sc \au{Geppert, A.}, \au{Terzis, A.}, \au{Lamanna, G.}, \au{Marengo, M.} \&
  \au{Weigand, B.}} \yr{2017}  \at{A benchmark study for the crown‑type
  splashing dynamics of one and two‑component droplet wall–film
  interactions}.  \jt{Exp. in Fluids}  \bvol{58},  \pg{172(1--27)}.

\bibitem[Geppert(2019)]{Geppert2019}
{\sc \au{Geppert, A.~K.}} \yr{2019}  \at{Experimental investigation of droplet
  wall-film interaction of binary systems}. PhD thesis.

\bibitem[Hasegawa \& Nara(2019)]{Hasegawa2019}
{\sc \au{Hasegawa, K.} \& \au{Nara, T.}} \yr{2019}  \at{Energy conservation
  during single droplet impact on deep liquid pool and jet formation}.
  \jt{{AIP} Advances}  \bvol{9}~(8),  \pg{085218}.

\bibitem[van Hinsberg {\em et~al.\/}(2010)van Hinsberg, Budakli, Göhler,
  Berberovi{\'{c}}, Roisman, Gambaryan-Roisman, Tropea \&
  Stephan]{vanHinsberg2010}
{\sc \au{van Hinsberg, Nils~Paul}, \au{Budakli, Mete}, \au{Göhler, Sebastian},
  \au{Berberovi{\'{c}}, Edin}, \au{Roisman, Ilia~V.}, \au{Gambaryan-Roisman,
  Tatiana}, \au{Tropea, Cameron} \& \au{Stephan, Peter}} \yr{2010}
  \at{Dynamics of the cavity and the surface film for impingements of single
  drops on liquid films of various thicknesses}.  \jt{Journal of Colloid and
  Interface Science}  \bvol{350}~(1),  \pg{336--343}.

\bibitem[Huang \& Chen(2018)]{Huang2018}
{\sc \au{Huang, Hai-Meng} \& \au{Chen, Xiao-Peng}} \yr{2018}  \at{Energetic
  analysis of drop’s maximum spreading on solid surface with low impact
  speed}.  \jt{Physics of Fluids}  \bvol{30}~(2),  \pg{022106}.

\bibitem[Kittel(2019)]{Kittel2019}
{\sc \au{Kittel, H.~M.}} \yr{2019}  \at{Drop impact onto a wall wetted by a
  thin film of another liquid}. PhD thesis, Technische Universit{\"a}t
  Darmstadt.

\bibitem[Kittel {\em et~al.\/}(2017)Kittel, Roisman \& Tropea]{Kittel2017}
{\sc \au{Kittel, Hannah~M}, \au{Roisman, Ilia~V} \& \au{Tropea, Cameron}}
  \yr{2017} Splashing of a very viscous liquid drop impacting onto a solid wall
  wetted by another liquid.  \bt{In {\em Proceedings {ILASS}{\textendash}Europe
  2017. 28th Conference on Liquid Atomization and Spray Systems\/}}.
  \publ{Universitat Polit{\`{e}}cnica Val{\`{e}}ncia}.

\bibitem[Kittel {\em et~al.\/}(2018{\natexlab{{\em a\/}}})Kittel, Roisman \&
  Tropea]{Kittel2018a}
{\sc \au{Kittel, H.~M.}, \au{Roisman, I.~V.} \& \au{Tropea, C.}}
  \yr{2018{\natexlab{{\em a\/}}}}  \at{Content of secondary droplets formed by
  drop impact onto a solid wall wetted by another liquid}.  \jt{ICLASS 2018} .

\bibitem[Kittel {\em et~al.\/}(2018{\natexlab{{\em b\/}}})Kittel, Roisman \&
  Tropea]{Kittel2018}
{\sc \au{Kittel, H.~M.}, \au{Roisman, I.~V.} \& \au{Tropea, C.}}
  \yr{2018{\natexlab{{\em b\/}}}}  \at{Splash of a drop impacting onto a solid
  substrate wetted by a thin film of another liquid}.  \jt{Phys. Rev. Fluids}
  \bvol{3},  \pg{073601(1--17)}.

\bibitem[Lamanna {\em et~al.\/}(2019)Lamanna, Geppert \& Weigand]{Lamanna2019}
{\sc \au{Lamanna, G}, \au{Geppert, A} \& \au{Weigand, B}} \yr{2019}  \bt{On the
  effect of a thin liquid film on the crown propagation in drop impact
  studies}.

\bibitem[Lel {\em et~al.\/}(2008)Lel, Kellermann, Dietze, Kneer \&
  Pavlenko]{Lel2008}
{\sc \au{Lel, V.V.}, \au{Kellermann, A.}, \au{Dietze, G.}, \au{Kneer, R.} \&
  \au{Pavlenko, A.N.}} \yr{2008}  \at{Investigations of the {M}arangoni effect
  on the regular structures in heated wavy liquid films}.  \jt{Experiments in
  Fluids}  \bvol{44}~(2),  \pg{341--354}.

\bibitem[Liang \& Mudawar(2016)]{Liang2016}
{\sc \au{Liang, G.} \& \au{Mudawar, I.}} \yr{2016}  \at{Review of mass and
  momentum interactions during drop impact on a liquid film}.  \jt{Int. J. Heat
  Mass Transfer}  \bvol{101},  \pg{577--599}.

\bibitem[Marcotte {\em et~al.\/}(2019)Marcotte, Michon, S{\'{e}}on \&
  Josserand]{Marcotte2019}
{\sc \au{Marcotte, Florence}, \au{Michon, Guy-Jean}, \au{S{\'{e}}on, Thomas} \&
  \au{Josserand, Christophe}} \yr{2019}  \at{Ejecta, corolla, and splashes from
  drop impacts on viscous fluids}.  \jt{Physical Review Letters}
  \bvol{122}~(1).

\bibitem[Motzkus {\em et~al.\/}(2009)Motzkus, Gensdarmes \&
  G{\'{e}}hin]{Motzkus2009}
{\sc \au{Motzkus, C.}, \au{Gensdarmes, F.} \& \au{G{\'{e}}hin, E.}} \yr{2009}
  \at{Parameter study of microdroplet formation by impact of millimetre-size
  droplets onto a liquid film}.  \jt{Journal of Aerosol Science}
  \bvol{40}~(8),  \pg{680--692}.

\bibitem[Okumura {\em et~al.\/}(2003)Okumura, Chevy, Richard, Qu{\'{e}}r{\'{e}}
  \& Clanet]{Okumura2003}
{\sc \au{Okumura, K.}, \au{Chevy, F.}, \au{Richard, D.}, \au{Qu{\'{e}}r{\'{e}},
  D.} \& \au{Clanet, C.}} \yr{2003}  \at{Water spring: A model for bouncing
  drops}.  \jt{EPL}  \bvol{62}~(2),  \pg{237--243}.

\bibitem[Petitjeans \& Maxworthy(1996)]{Petitjeans1996}
{\sc \au{Petitjeans, P.} \& \au{Maxworthy, T.}} \yr{1996}  \at{Miscible
  displacements in capillary tubes. part 1. experiments}.  \jt{Journal of Fluid
  Mechanics}  \bvol{326},  \pg{37--56}.

\bibitem[Planchette {\em et~al.\/}(2017)Planchette, Hinterbichler, Liu, Bothe
  \& Brenn]{Planchette2017}
{\sc \au{Planchette, C.}, \au{Hinterbichler, H.}, \au{Liu, M.}, \au{Bothe, D.}
  \& \au{Brenn, G.}} \yr{2017}  \at{Colliding drops as coalescing and
  fragmenting liquid springs}.  \jt{Journal of fluid mechanics}  \bvol{814},
  \pg{277--300}.

\bibitem[Roisman {\em et~al.\/}(2008)Roisman, van Hinsberg \&
  Tropea]{Roisman2008}
{\sc \au{Roisman, I.~V.}, \au{van Hinsberg, N.~P.} \& \au{Tropea, C.}}
  \yr{2008}  \at{Propagation of a kinematic instability in a liquid layer:
  Capillary and gravity effects}.  \jt{Phys. Rev. E}  \bvol{77},  \pg{046305}.

\bibitem[Roisman {\em et~al.\/}(2012)Roisman, Planchette, Lorenceau \&
  Brenn]{Roisman2012}
{\sc \au{Roisman, I.~V.}, \au{Planchette, C.}, \au{Lorenceau, E.} \& \au{Brenn,
  G.}} \yr{2012}  \at{Binary collisions of drops of immiscible liquids}.
  \jt{J. Fluid Mech.}  \bvol{690},  \pg{512–535}.

\bibitem[Ross \& Becher(1992)]{Ross1992}
{\sc \au{Ross, Sydney} \& \au{Becher, Paul}} \yr{1992}  \at{The history of the
  spreading coefficient}.  \jt{Journal of Colloid and Interface Science}
  \bvol{149}~(2),  \pg{575--579}.

\bibitem[Shaikh {\em et~al.\/}(2018)Shaikh, Toyofuku, Hoang \&
  Marston]{Shaikh2018}
{\sc \au{Shaikh, S.}, \au{Toyofuku, G.}, \au{Hoang, R.} \& \au{Marston, J.O.}}
  \yr{2018}  \at{Immiscible impact dynamics of droplets onto millimetric
  films}.  \jt{Experiments in Fluids}  \bvol{59}~(1),  \pg{7}.

\bibitem[Tang {\em et~al.\/}(2019)Tang, Saha, Sun \& Law]{Tang2019}
{\sc \au{Tang, X.}, \au{Saha, A.}, \au{Sun, C.} \& \au{Law, C.~K.}} \yr{2019}
  \at{Spreading and oscillation dynamics of drop impacting liquid film}.
  \jt{Journal of Fluid Mechanics}  \bvol{881},  \pg{859--871}.

\bibitem[Thoroddsen(2002)]{Thoroddsen2002}
{\sc \au{Thoroddsen, S.~T.}} \yr{2002}  \at{The ejecta sheet generated by the
  impact of a drop}.  \jt{Journal of Fluid Mechanics}  \bvol{451},
  \pg{373--381}.

\bibitem[Thoroddsen {\em et~al.\/}(2006)Thoroddsen, Etoh \&
  Takehara]{Thoroddsen2006}
{\sc \au{Thoroddsen, S.~T.}, \au{Etoh, T.~G.} \& \au{Takehara, K.}} \yr{2006}
  \at{Crown breakup by {M}arangoni instability}.  \jt{J. Fluid Mech.}
  \bvol{557},  \pg{63--72}.

\bibitem[Thoroddsen {\em et~al.\/}(2011)Thoroddsen, Thoraval, Takehara \&
  Etoh]{Thoroddsen2011}
{\sc \au{Thoroddsen, S.~T.}, \au{Thoraval, M.-J.}, \au{Takehara, K.} \&
  \au{Etoh, T.~G.}} \yr{2011}  \at{Droplet splashing by a slingshot mechanism}.
   \jt{Physical Review Letters}  \bvol{106}~(3).

\bibitem[Truzzolillo \& Cipelletti(2017)]{truzzolillo2017}
{\sc \au{Truzzolillo, D.} \& \au{Cipelletti, L.}} \yr{2017}
  \at{Off-equilibrium surface tension in miscible fluids}.  \jt{Soft matter}
  \bvol{13}~(1),  \pg{13--21}.

\bibitem[Vaikuntanathan \& Sivakumar(2016)]{Vaikuntanathan2016}
{\sc \au{Vaikuntanathan, V.} \& \au{Sivakumar, D.}} \yr{2016}  \at{Maximum
  spreading of liquid drops impacting on groove-textured surfaces: effect of
  surface texture}.  \jt{Langmuir}  \bvol{32}~(10),  \pg{2399--2409}.

\bibitem[Wang {\em et~al.\/}(2020)Wang, Wang, Yu \& Chen]{Wang2020}
{\sc \au{Wang, Bo}, \au{Wang, Chenyu}, \au{Yu, Yude} \& \au{Chen, Xiaodong}}
  \yr{2020}  \at{Spreading and penetration of a micro-sized water droplet
  impacting onto oil layers}.  \jt{Physics of Fluids}  \bvol{32}~(1),
  \pg{012003}.

\bibitem[Wildeman {\em et~al.\/}(2016)Wildeman, Visser, Sun \&
  Lohse]{Wildeman2016}
{\sc \au{Wildeman, S.}, \au{Visser, C.~W.}, \au{Sun, C.} \& \au{Lohse, D.}}
  \yr{2016}  \at{On the spreading of impacting drops}.  \jt{Journal of fluid
  mechanics}  \bvol{805},  \pg{636--655}.

\bibitem[Worthington \& Cole(1897)]{Worthington1897}
{\sc \au{Worthington, A.~M.} \& \au{Cole, R.~S.}} \yr{1897}  \at{Impact with a
  liquid surface, studied by the aid of instantaneous photography}.
  \jt{Philosoph. Trans. Royal Soc. London Ser. A}  \bvol{189},  \pg{137--148}.

\bibitem[Zhang {\em et~al.\/}(2012)Zhang, Toole, Fezzaa \& Deegan]{Zhang2012}
{\sc \au{Zhang, L.~V.}, \au{Toole, J.}, \au{Fezzaa, K.} \& \au{Deegan, R.~D.}}
  \yr{2012}  \at{Splashing from drop impact into a deep pool: multiplicity of
  jets and the failure of conventional scaling}.  \jt{Journal of Fluid
  Mechanics}  \bvol{703},  \pg{402--413}.

\bibitem[Zhang {\em et~al.\/}(2019)Zhang, Liu, Qu \& Hu]{Zhang2019}
{\sc \au{Zhang, Y.}, \au{Liu, P.}, \au{Qu, Q.} \& \au{Hu, T.}} \yr{2019}
  \at{Energy conversion during the crown evolution of the drop impact upon
  films}.  \jt{International Journal of Multiphase Flow}  \bvol{115},
  \pg{40--61}.

\end{thebibliography}
